\documentclass[12pt,twoside,english]{report}

\PassOptionsToPackage{unicode=true, pdfusetitle, bookmarks=true, colorlinks=false}{hyperref}

\usepackage[T1]{fontenc}
\usepackage[utf8]{inputenc}

\usepackage[main=english]{babel}

\usepackage{geometry}
\usepackage{fancyhdr}
\usepackage{calc}
\usepackage{setspace}
\usepackage{tikz}

\usepackage{slashed}
\usepackage{nextpage}
\newcommand\myclearpage{\cleartooddpage[\thispagestyle{empty}]}
\usepackage[fit]{truncate}

\fancypagestyle{plain}{
  \fancyhf{}
  \fancyhead[LE]{\truncate{.95\headwidth}{\leftmark}}
  \fancyhead[RO]{\truncate{.95\headwidth}{\rightmark}}
  \fancyfoot[LE,RO]{\thepage}
}

\newcommand{\aver}[1]{\langle #1 \rangle}
\newcommand{\ket}[1]{|#1\rangle}
\newcommand{\bra}[1]{\langle#1|}
\newcommand{\tr}{\text{tr}}
\newcommand{\Tr}{\text{tr}}
\newcommand{\e}{\text{e}}
\newcommand{\cov}{\text{cov}}

\newcommand{\mb}{{\bf m}}
\newcommand{\nb}{{\bf n}}
\newcommand{\red}[1]{\textcolor{red}{#1}}

\newcommand{\blambda}{{\bm \lambda}}
\newcommand{\me}[1]{\textcolor{black}{#1}}

\usepackage{amsmath}
\usepackage{bm}
\usepackage{amssymb}
\usepackage{graphicx}
\usepackage{mathtools}
\usepackage[dvipsnames]{xcolor}
\usepackage{tikz}
\usepackage{orcidlink}
\usepackage{mathrsfs}
\usepackage{cite}
\usepackage[most]{tcolorbox}

\usepackage{array}
\newcolumntype{P}[1]{>{\centering\arraybackslash}p{#1}}
\usepackage{makecell}
\usepackage{float}
\usepackage{multirow}
\usepackage{dcolumn}
\usepackage{booktabs}

\usepackage{hyperref}

\usepackage{pdfpages}

\begin{document}

%---------------------------------------------------------------------%

\begin{center}
{\large{}\thispagestyle{empty}}{\large\par}
\par
\end{center}

\begin{center}
Matheus Eiji Ohno Bezerra
\par\end{center}

\vspace{3cm}

\begin{center}
{\Large{} From the Hong-Ou-Mandel Effect to Quantum Sensing: Interference of Nonclassical Light with Partial Distinguishability and Noise}{\huge\par}
\par\end{center}

\begin{center}
\vspace{4cm}
\par\end{center}

\begin{flushright}
\begin{minipage}[t]{0.6\textwidth}
Thesis submitted to the Program in Physics of the Universidade Federal do ABC, as a partial requirement for the degree of Doctor of Philosophy in Physics.
\end{minipage}
\end{flushright}

\begin{center}
\vspace{2cm}
Supervisor: Prof. Dr. Valery Shchesnovich \\ [0.5cm]
Research Internship Supervisor: Prof. Dr. Rafał Demkowicz-Dobrzański
\par\end{center}

\begin{center}
\vspace{3cm}
Santo André,  State of São Paulo,  Brazil - 2025
\par\end{center}

%-------------------------------------------------------%

\newpage

\begin{center}
{\large{}\thispagestyle{empty}}{\large\par}
\par\end{center}

\selectlanguage{american}%
\begin{center}
\textbf{\large{}Thesis Defense Committee}{\large\par}
\par\end{center}

\vspace{0.5cm}

The doctoral thesis defense took place on 25 November 2025 and the examining committee was composed of the following members:

\vspace{0.5cm}

\begin{itemize}
    \item Valery Shchesnovich, Universidade Federal do ABC (Chair)
    \item Celso Jorge Villas Boas, Universidade Federal de São Carlos
    \item Daniel Jost Brod, Universidade Federal Fluminense
    \item Jelmer Jan Renema, University of Twente
    \item Luciano Soares da Cruz, Universidade Federal do ABC
\end{itemize}

\begin{center}
\myclearpage
\par\end{center}

%-------------------------------------------------------%

\newpage

\begin{center}
{\large{}\thispagestyle{empty}}{\large\par}
\par\end{center}

\begin{center}
\textbf{\large{}Acknowledgments}{\large\par}
\par\end{center}

I am immensely grateful to my supervisor Valery for the continuous support of my research since the end of my undergraduate studies in physics to my PhD, as well as for the essential discussions and numerous insights over all these years. I am also very grateful to my supervisor Rafał for all the support and helpfulness provided during my research internship at the University of Warsaw. I would also like to thank Francesco for all the great discussions and always collaborative spirit during our work in quantum parameter estimation. Finally, I acknowledge the Universidade Federal do ABC for the support provided throughout my undergraduate, master’s, and doctoral studies, and the University of Warsaw for hosting me during my research internship.

\vspace{0.5cm}

This study was financed, in part, by the São Paulo Research Foundation (FAPESP), Brasil, Process Numbers 2022/13635-3 and 2021/03251-0; and by the Coordenação de Aperfeiçoamento de Pessoal de Nível Superior - Brasil (CAPES) - Finance Code 001.

%-------------------------------------------------------%

\newpage

\begin{center}
{\large{}\thispagestyle{empty}}{\large\par}
\par\end{center}

\begin{center}
\textbf{\large{}Resumo}{\large\par}
\par\end{center}

%\vspace{0.5cm}

Esta tese explora a interferência de estados não clássicos da luz, em particular os estados de Fock e os estados Gaussianos, em interferômetros lineares sujeitos a ruído, com aplicações em informação e metrologia quântica. Utilizando o formalismo do espaço de fase, são desenvolvidas ferramentas analíticas baseadas em funções geradoras para descrever de forma unificada a interferência óptica quântica. Para estados de Fock com múltiplos fótons, novos eventos de probabilidade nula (leis de supressão) são identificados além do princípio de permutação de simetria previamente conhecido, revelando estruturas de interferência mais ricas que se degradam com a distinguibilidade dos fótons. No caso dos estados Gaussianos, a descrição baseada no Hafnian do Gaussian Boson Sampling é estendida para incluir distinguibilidade parcial por meio da matriz de sobreposição dos estados internos dos fótons. Por fim, é examinada a conexão entre esses fenômenos de interferência e a estimação quântica de multi-parâmetros, considerando a estimação simultânea de fase e perda de fótons. O estudo mostra que, embora a incompatibilidade do estado de prova possa desaparecer para estados não Gaussianos otimizados e estados Gaussianos de dois modos, em regimes de alto número de fótons, a incompatibilidade de medição permanece como uma limitação fundamental mesmo nesse limite.

\vspace{0.5cm}

\textbf{Palavras chave:} Interferência quântica; Boson sampling; estados Gaussianos; metrologia quântica.

%-------------------------------------------------------%

\newpage

\begin{center}
{\large{}\thispagestyle{empty}}{\large\par}
\par\end{center}

\selectlanguage{american}%
\begin{center}
\textbf{\large{}Abstract}{\large\par}
\par\end{center}

%\selectlanguage{american}%
This thesis explores the interference of nonclassical states of light, particularly Fock and Gaussian states, in noisy linear interferometers, with applications to quantum information and quantum sensing. Using the phase-space formalism, analytical tools based on generating functions are developed to describe quantum optical interference in a unified way. For multiphoton Fock states, new zero probability events (suppression laws) are identified beyond the previously derived symmetry permutation principle, revealing rich interference structures that is degraded with photon distinguishability. For Gaussian states, the Hafnian-based description of Gaussian Boson Sampling is extended to include partial distinguishability via the overlap matrix of the internal state of the photons. Finally, the link between these interference effects and quantum multiparameter estimation is examined for the simultaneous estimation of phase and loss. This study shows that while probe incompatibility can vanish for optimized non-Gaussian states and some two-mode Gaussian states, at high photon number, measurement incompatibility remains a fundamental constraint even in this limit.

\vspace{0.5cm}

\selectlanguage{american}%

%-------------------------------------------------------%

\newpage

\tableofcontents{}

\begin{center}
{\large{}\thispagestyle{empty}}{\large\par}
\par
\end{center}

%%%%%%%%%%%%%%%%%%%%%%%%%%%%%%%%%%%%%%%%%%%%%%%%%%%%%%%%%%%%%%%%%%%%%%%%%%%%%%%%%%%%%%%%%%%%%%%%%%%%%%%%%%%%%%%%%%%%%%%%%%%%%%%%%%%%%%%%%%%%%%%%%%%%%%%%%%%%%%%%%%%%%%%%%%%%%%%%%%%%%%%%%%%%%%%%%%%%%%%%%%%%%%%%%%%%%%%%
\begin{center}
\myclearpage
\par\end{center}

\chapter{Introduction}

Multiphoton interference is a fundamental resource for photonic quantum technologies, with direct applications in quantum information processing and quantum-enhanced metrology. A basic manifestation of two-photon interference is the \textit{Hong-Ou-Mandel (HOM) effect}~\cite{HOM1987}, in which a destructive interference at a balanced beamsplitter suppresses coincidence events  for indistinguishable photons.  This interference is gradually degraded as distinguishability is introduced, 
due to an imperfect overlap of the internal state of the photons, 
such as their temporal delay, spectral profiles, and polarization~\cite{HOM1987,Loudon1989,Mandel1991,OuTempDist, 4photon}.  In  multiphoton interferences, such zero-probability events have also been investigated, 
mainly within a permutation-symmetry principle~\cite{Dittel1,Dittel2}.  In addition, for three or more photons, partial distinguishability exhibits a richer structure, 
including the effects of collective phases~\cite{CollectivePhase2,CollectivePhase1,CollectivePhase3}.

In the context of demonstrating quantum advantage, one of the most influential models in non-universal photonic quantum computation, is the \textit{Boson Sampling} (BS) protocol, proposed by Aaronson and Arkhipov~\cite{BSAA}. 
In this protocol, $N$ indistinguishable single photons are injected into a $M$-mode linear interferometer, 
followed by photon-number-resolving (PNR) detection at the output.
The resulting probabilities reflect the interference effects that arising from the genuine bosonic signature of the multiphoton state. In particular, the transition amplitudes between input and output configurations are governed by the matrix operation  known as the permanent~\cite{origBS,BSAA}. The permanent accounts for all possible permutations of $N$ photons paths, which makes its evaluation combinatorially hard and causes the computational cost to scale exponentially with the photon number $N$~\cite{HardnessPermanent}. This hardness is what gives the potential of BS to demonstrate the quantum advantage.

Besides other alternatives to the original BS proposal, the
Boson Sampling with Gaussian states~\cite{BSGaussianLund} employs two-mode squeezed vacuum sources together with heralding 
to prepare single-photon Fock inputs in random modes before the interferometer.  Building on this Gaussian-source perspective, \textit{Gaussian Boson Sampling} (GBS)~\cite{GBS,GBSdetailed} 
samples from the full Gaussian output statistics obtained by injecting single-mode squeezed vacuum states 
into a linear interferometer and performing PNR  detection at the output. From a practical perspective, GBS is particularly attractive because it avoids one of the main challenges of BS, 
namely the requirement of $N$ heralded single photons in well-defined input spatial modes. 
Squeezed states, by contrast, can be generated deterministically using well-established nonlinear optical processes 
based on second-order or third-order nonlinearities~\cite{GenerationSMSS1,GenerationSMSS2,GenerationSMSS3}. This allows experiments to scale to a larger number of modes and higher average photon numbers, as demonstrated in several recent outstanding realizations~\cite{Pan2020,Pan2023,Pan2025,Madsen2022,Paesani2019}. Beyond its experimental feasibility, GBS also features a distinct mathematical structure in its output statistics: 
while BS probabilities are governed by matrix permanents, GBS probabilities are determined by matrix hafnians. The computation of the hafnian also scales exponentially with the matrix size, implying that GBS is also classically intractable for large systems\cite{GBS,GBSdetailed}. Indeed, GBS has some applications beyond quantum advantage demonstrations, including problems in graph theory~\cite{GBSGraph2018,GBSGraph2018_2} and molecular vibronic spectra simulations~\cite{GBSMolecule2015,GBSMolecule2020}.

In such protocols, the most relevant sources of noise are the distinguishability of the photons~\cite{ValeryBS1,BSAlg2018,ValeryPRA2019,SimulabilityMoylett2020,RenemaSimulabilityGBS2020,Shi2020,RenemaSimulability2025,QuesadaLoopHafnian} and photon losses~\cite{BrodLoss2016,OszmaniecLoss2018,ValeryQuantum2019,PatronSimulabilityGBS,Shi2020,BressaniGBS2024,QuesadaLoopHafnian}. 
These imperfections degrade the quantum interference, progressively turning this problem into regimes that can be efficiently simulated on classical computers~\cite{Kalai2014,SimulabilityCaves2016,ValeryPRA2019}. 
While partial distinguishability in Fock-state interference is well characterized~\cite{ValeryBS1,TichyBS}, its effects in interference with Gaussian states are not fully explored.
In the case of squeezed states, the distinguishability may arise from spectral correlations within photon pairs produced by a single source, quantified by the Schmidt number of the joint spectral function~\cite{Christ_2011,4photon}, or from differences between photons originating from distinct sources~\cite{Shi2020,ValeryGBS}. 
Understanding how partial distinguishability affects Gaussian-state interference is important not only for modelling it as a noise mechanism in Gaussian photonic protocols, but also for elucidating a fundamental aspect of quantum interference.

In quantum metrology, optical implementations are among the most prominent, as notably demonstrated by the use of squeezed states for enhancing the phase estimation in gravitational-wave detectors~\cite{Schnabel2016,LIGO2019,Virgo2019,LIGO2023}. More generally, appropriately engineered quantum optical states can surpass the standard quantum limit and, in ideal conditions, achieve Heisenberg-limited scaling in parameter estimation~\cite{YurkeSU2,Giovanneetti2006}. However, in optical systems, the photon losses are a fundamental source of noise and can drastically reduce the precision achievable in phase estimation~\cite{Escher2011,Demkowicz2012,Demkowicz-Dobrzanski2015a}. In addition, estimating losses themselves is crucial in applications such as absorption imaging and spectroscopy~\cite{Taylor2016,Shi2020}. 
Since both phase and loss encode essential information in optical fields~\cite{Polino2020}, their joint estimation is central to \textit{multiparameter quantum metrology}~\cite{Crowley2014,Albarelli2019c,Demkowicz-Dobrzanski2020,Pezze2025}. Multiparameter quantum metrology addresses the simultaneous estimation of several parameters, establishing the ultimate precision limits and the trade-offs that arise when optimizing them together. 
These trade-offs arise from two different sources of incompatibility: at the level of the probe and at the level of the measurement. 
In general, the probe state that is optimal for one parameter is not optimal for another~\cite{ProbeIncomp2022}, and the measurements that individually maximize the sensitivity to different parameters may not be jointly implementable~\cite{Ragy2016,SzczykulskaMultiparameter,Albarelli2022}, typically because the corresponding optimal observables do not commute.

This thesis investigates the interference of nonclassical light, in particular Fock and Gaussian states, in linear optical interferometers subject to realistic noise effects, with implications for both quantum computing and quantum sensing. This text is organized as follows. To begin, Chapter~\ref{chapter:preliminaries} introduces the theoretical background of quantum optics and quantum mechanics, including the description of nonclassical states, the phase-space formalism and the basics of quantum metrology that will be adopted throughout the work. In sequence, Chapter~\ref{chapter:suppression_laws} focuses on multiphoton interference with Fock states, beginning with the Hong–Ou–Mandel effect and extending to general zero-output probabilities (suppression laws) derived through generating-function techniques. We identify entire families of such suppression laws that cannot be explained for by previously known symmetry-based approaches. In Chapter~\ref{chapter:GBS}, we focus on the interference of single-mode squeezed states, employing a phase-space formalism to describe their interference in the presence of partial distinguishability, which is quantified by a general overlap matrix. Within this framework, we investigate partial distinguishability as a source of noise in the GBS protocol and analyse its effect on zero-output probabilities. Finally, in Chapter~\ref{chapter:parameter_estimation}, we investigate quantum multiparameter estimation within the quantum optical interference paradigm, focusing on the simultaneous estimation of phase and loss. In this context, the incompatibilities arising from both the choice of probe states and the measurement strategies are analysed.

%%%%%%%%%%%%%%%%%%%%%%%%%%%%%%%%%%%%%%%%%%%%%%%%%%%%%%%%%%%%%%%%%%%%%%%%%%%%%%%%%%%%%%%%%%%%%%%%%%%%%%%%%%%%%%%%%%%%%%%%%%%%%%%%%%%%%%%%%%%%%%%%%%%%%%%%%%%%%%%%%%%%%%%%%%%%%%%%%%%%%%%%%%%%%%%%%%%%%%%%%%%%%%%%%%%%%%%%

\begin{center}
\myclearpage
\par
\end{center}

\chapter{Preliminaries}
\label{chapter:preliminaries}

%--------------------------------------------------------------------------------------------------------------------------------------------------------------------------%

\section{Quantization of the electromagnetic field}
\label{sec:quantization_EM}

We begin by revisiting the main concepts of quantum mechanics and quantum optics, 
starting with the quantization of the electromagnetic field.
A particularly important physical system in this context is the harmonic oscillator, 
as it models a wide range of phenomena: small oscillations around a stable equilibrium, 
particles confined in a quadratic potential, circular motion, and, most importantly for our purposes, the quantized modes of the electromagnetic field. 
Several approaches to the quantization of the electromagnetic field exist in the literature; in this work, we follow the concise and elegant formulation presented in Ref.~\cite{Ballentine1998}.
Consider a set of particles with unit mass, $m = 1$, each in an harmonic motion with frequency $\omega_k$ (in radians per unit time). 
The classical Hamiltonian describing this system is given by:
\begin{equation}
    H= \frac{1}{2} \sum_k \left( p^2_k +  \omega^2_k q^2_k \right)
    \label{hamiltonian_SHO}
\end{equation}
where $\{ q_k \}$ and $\{ p_k \}$ are the canonical coordinates, respectively the spatial coordinates and the momentum. In this system, the Jacobi equation gives~\cite{GoldsteinMechanics}:
\begin{equation}
    \frac{d q_k}{dt} = \frac{\partial H}{\partial p_k} =  p_k 
    , \qquad
    \frac{d p_k}{dt} = \frac{\partial H}{\partial q_k} = - \omega^2 q_k
    \label{Jacobi_osc}
\end{equation}

Since position and momentum are observables, in quantum mechanics they are promoted to the hermitian operators $q_k \rightarrow \hat{q}_k$ and $q_k \rightarrow \hat{p}_k$, and the Poisson brackets are promoted to a commutation relation, according to:
\begin{equation}
\{ q_k , p_{k'} \} = \delta_{k k'}
\ \mapsto\
[ \hat q_k , \hat p_{k'} ] = i\hbar\,\delta_{k k'} .
\end{equation}

This procedure, known as \textit{canonical quantization}, establishes the bridge between classical and quantum descriptions. 
The operators $\hat{q}_k$ and $\hat{p}_k$ no longer represent definite values of position and momentum, 
but rather quantum observables whose measurement outcomes are governed by probability distributions 
defined by the quantum states.

The next step is to determine the eigenvalues and eigenvectors of the Hamiltonian operator in Eq.~\ref{hamiltonian_SHO}, using an algebraic approach. 
These eigenvalues correspond to the quantized energy levels of the system, 
while the eigenvectors represent the energy eigenstates of the quantum harmonic oscillator. We first define the {\it creation and annihilation operators} respectively by:
\begin{align}
    \hat{a}^\dagger_k = \sqrt{ \frac{ \omega_k}{2 \hbar}} \left( \hat{q}_k + \frac{i}{ \omega_k} \hat{p}_k \right)
    , \qquad
    \hat{a}_k= \sqrt{\frac{ \omega_k}{2 \hbar}} \left( \hat{q}_k - \frac{i}{ \omega_k} \hat{p}_k \right)
    \label{creation_annihilation_def_pos_mom}
\end{align}
that satisfy the bosonic commutation relations:
\begin{align}
    & [\hat{a}^\dagger_k, \hat{a}^\dagger_j] = [\hat{a}_k, \hat{a}_j] = 0 ,
    \nonumber\\
    & [\hat{a}_k, \hat{a}^\dagger_j] = \delta_{kj} .
    \label{creation_annihilation_def}
\end{align}

By expressing the position $\hat{q}_k$ and momentum $\hat{p}_k$ operators in terms of the creation and annihilation operators, 
and performing some straightforward algebraic manipulations on Eq.~(\ref{hamiltonian_SHO}), 
we obtain the following expression for the quantized Hamiltonian:
\begin{equation}
    \hat{H} = \sum_k \hbar \omega_k \left( \hat{n}_k + \frac{1}{2} \right),
    \ \qquad
    \hat{n}_k = \hat{a}^\dagger_k \hat{a}_k
    ,
    \label{PhotonNumberOp_def}
\end{equation}
where $\hat{n}_k$ is called the \textit{number operator}. 
Physically, it measures the number of energy quanta, interpreted as the photons in the field mode $k$.
Therefore, the problem is reduced to finding the eigenvalues and eigenvectors of the number operator.

We now define the eigenstates of the number operator, which will serve as the basis of the quantized field description. Denoting its orthonormal eigenvectors by
\begin{equation}
    \hat{n}_k \ket{n_k} = n_k \ket{n_k},
    \qquad n_k \in \mathbb{N},
    \label{Fock_def}
\end{equation}
where, throughout this text, $\mathbb{N}$ denotes the set of non-negative integers,  $\mathbb{N} = \{0, 1, 2, \ldots\}$. 
The states $\ket{n_k}$ are known as \textit{Fock states} (or \textit{number states}) and represent configurations with exactly $n_k$ quanta (e.g., photons) in mode $k$. For an $M$-mode system, the joint state is denoted by
\begin{equation}
    \ket{\mathbf{n}} = \ket{n_1, n_2, \ldots, n_M},
    \label{multimode_Fock}
\end{equation}
which forms a complete orthonormal basis of the Hilbert space of the field modes. Each Fock state is also an eigenstate of the Hamiltonian operator:
\begin{equation}
    \hat{H} \ket{n_k} = \sum_k \hbar \omega_k \left( n_k + \frac{1}{2} \right) \ket{n_k},
\end{equation}
where the term $\hbar \omega_k / 2$ represents the zero-point energy of the quantum harmonic oscillator.

The creation and annihilation operators, $\hat{a}_k^\dagger$ and $\hat{a}_k$, 
act on the multimode Fock states according to:
\begin{equation}
    \hat{a}_k^\dagger \ket{n_1, n_2, \ldots, n_k, \ldots} 
    = \sqrt{n_k + 1} \, \ket{n_1, n_2, \ldots, n_k + 1, \ldots},
    \label{creation_action}
\end{equation}
\begin{equation}
    \hat{a}_k \ket{n_1, n_2, \ldots, n_k, \ldots} 
    = \sqrt{n_k} \, \ket{n_1, n_2, \ldots, n_k - 1, \ldots}.
    \label{annihilation_action}
\end{equation}
Thus, we can recast Eq.~(\ref{multimode_Fock}) in the following way:
\begin{equation}
     | \nb \rangle_{\bf a} =  \prod_{k=1}^M \frac{(\hat{a}^\dagger_k)^{n_k}}{\sqrt{n_k!}}|0\rangle .
     \label{Fock_input}
\end{equation}

Having established the quantization of the harmonic oscillator, we now connect this formalism to classical electrodynamics, which provides the physical basis for the quantization of the electromagnetic field. The electric field ${\bf E}$ and the magnetic field ${\bf B}$ satisfy the Maxwell’s equations on the vacuum~\cite{JacksonlElectrodynamics}:
\begin{eqnarray}
    && \bm{\nabla} \cdot {\bf E} = 0, 
    \label{M1} \\[2mm]
    && \bm{\nabla} \times {\bf E} = - \frac{1}{c} \frac{\partial {\bf B}}{\partial t}, 
    \label{M2} \\[2mm]
    && \bm{\nabla} \cdot {\bf B} = 0, 
    \label{M3} \\[2mm]
    && \bm{\nabla} \times {\bf B} =  \frac{1}{c} \frac{\partial {\bf E}}{\partial t}
    ,
    \label{M4}
\end{eqnarray}
where $\bm{\nabla}$ is the usual nabla operator and $c$ is the speed of light in the vacuum. 
We assume that the fields are confined within a finite volume $V$ bounded by a perfectly conducting surface $S$. 
Therefore, the boundary conditions at the surface are given by
\begin{eqnarray}
    {\bf n} \times {\bf E}({\bf x},t) = 0, 
    \qquad
    {\bf n} \cdot {\bf B}({\bf x},t) = 0,
    \qquad
    {\bf x} \in S,
    \label{boundary}
\end{eqnarray}
where ${\bf n}$ is the unit vector normal to the surface $S$. From the Maxwell’s equations~(\ref{M1}) and~(\ref{M2}), 
we obtain the wave equation for the electric field:
\begin{equation}
    \nabla^2 {\bf E} - \frac{1}{c^2} \frac{\partial^2 {\bf E}}{\partial t^2} = 0.
\end{equation}
Similarly, combining Eqs.~(\ref{M3}) and~(\ref{M4}) we obtain the corresponding wave equation for the magnetic field:
\begin{equation}
    \nabla^2 {\bf B} - \frac{1}{c^2} \frac{\partial^2 {\bf B}}{\partial t^2} = 0.
\end{equation}

Now, the objective is to find the solutions of the wave equation for the electric and magnetic fields. Consider a set of real functions $\{ {\bf u}_k({\bf x}) \}$ with null divergence $ \bm{\nabla} {\bf u}_k({\bf x}) = 0 $, that satisfies the Helmholtz equation in a volume $V$:
\begin{equation}
    \nabla^2 {\bf u}_k({\bf x}) = - \lambda^2_k \hspace{1mm} {\bf u}_k({\bf x}) , \qquad \lambda_k \in \mathbb{R} ,
\end{equation}
and the boundary condition on the surface $S$ of this volume:
\begin{equation}
    {\bf n} \times {\bf u}_k({\bf x})  = 0 
    , \qquad
    {\bf x} \in S,
\end{equation}

This set of functions forms an orthogonal basis and therefore we can expand the electric and magnetic fields in a way that the wave equations are satisfied~\cite{JacksonlElectrodynamics,Ballentine1998}:
\begin{eqnarray}
    && {\bf E}({\bf x},t) = \sum_k f_k (t) \hspace{1mm} {\bf u}_k({\bf x}), \\
    && {\bf B}({\bf x},t) = \sum_k g_k (t)  \hspace{1mm}  \bm{\nabla} \times {\bf u}_k({\bf x}),
    \label{EMfields}
\end{eqnarray} 
where the temporal coefficients satisfy the equation of motion of a harmonic oscillator with frequency $\omega_k = c k_k$:
\begin{equation}
     \frac{d^2 f_k (t)}{dt^2} = - \omega^2_k \hspace{1mm} f_k (t) , \qquad
     \frac{d^2 g_k (t)}{dt^2} = - \omega^2_k \hspace{1mm} g_k (t)
\end{equation}

Since the basis functions have null divergence, the Maxwell equations Eq.~(\ref{M1}) and Eq.~(\ref{M3}) are trivially satisfied. From the boundary condition imposed for the functions ${\bf u}_k({\bf x})$, the boundary condition for the electric and magnetic fields of Eq.(\ref{boundary}) is satisfied. So, remains to find the relation between the temporal coefficients $f_k (t)$ and $g_k (t)$ in a way that the Maxwell equations Eq.~(\ref{M2}) and Eq.~(\ref{M4}) are also satisfied. Substituting then, the expansions of ${\bf E}$ and ${\bf B}$ in the remaining Maxwell equation, we have the following relations:

\begin{equation}
\frac{d f_k (t)}{dt} = \frac{\omega^2_k}{c} \hspace{1mm} g_k (t) , \qquad \frac{d g_k (t)}{dt} = -c \hspace{1mm} f_k (t)
\end{equation}
that have the same structure as the Jacobi equations for the harmonic oscillator, according to Eq.(\ref{Jacobi_osc}). Therefore, due to these similarities, we set the following convenient correspondence between the field modes $\{ f_k , g_k \}$ and the canonical variables $\{ q_k , p_k \}$:
\begin{equation}
     f_k \mapsto 2 \omega_k \sqrt{\pi} \, q_k \, 
     , \qquad
     g_k \mapsto \frac{2 \sqrt{\pi}}{k_k} \hspace{1mm} p_k
\end{equation}

Using the orthogonality of the mode functions $\{ {\bf u}_k({\bf x}) \}$ and 
$\{ \bm{\nabla} \times {\bf u}_k({\bf x}) \}$, together with the correspondences introduced above, 
the Hamiltonian of the electromagnetic field can be written as
\begin{equation}
H = \frac{1}{8 \pi} \int d^3 x \left( {\bf E}^2 + {\bf B}^2 \right)
  = \frac{1}{8 \pi} \sum_k \left( f_k^2 + k_k^2 g_k^2 \right) .
\end{equation}
Making use of the identifications between the field amplitudes $\{ f_k , g_k \}$ and the canonical variables 
$\{ q_k , p_k \}$, this Hamiltonian assumes the standard form of a collection of independent harmonic oscillators,
\begin{equation}
H = \frac{1}{2} \sum_k \left( p_k^2 + \omega_k^2 q_k^2 \right).
\end{equation}

Finally, for the quantization of the electromagnetic fields, the modes $f_k$ and $g_k$ are promoted as operators that satisfy the previous correspondence. In terms of the creation and annihilation operators defined in Eq.~(\ref{creation_annihilation_def_pos_mom}), we have the following ~\cite{Ballentine1998}:
\begin{equation}
 \hat{f}_k \longleftrightarrow \sqrt{2 \pi \hbar \omega_k} \hspace{1mm} \left( \hat{a}^\dagger_k + \hat{a}_k  \right) 
 , \qquad
 \hat{g}_k \longleftrightarrow i \left( \frac{2 \pi \hbar c^2}{\omega_k} \right) \hspace{1mm}  \left( \hat{a}^\dagger_k - \hat{a}_k  \right)
\end{equation}
and therefore, in the Heisenberg picture, the electric and magnetic field operators are obtained by Eq.(\ref{EMfields}), 
\begin{eqnarray}
&& \hat{{\bf E}}({\bf x},t) = \sum_k  \sqrt{ 2 \pi \hbar \omega_k} \hspace{1mm} \left( \hat{a}^\dagger_k (t) + \hat{a}_k (t) \right) \hspace{1mm} {\bf u}_k({\bf x}) \\
&& \hat{{\bf B}}({\bf x},t) = i c \sum_k  \left( \frac{2 \pi \hbar }{\omega_k} \right) \hspace{1mm}  \left( \hat{a}^\dagger_k (t) - \hat{a}_k (t)  \right) \hspace{1mm}  \bm{\nabla} \times {\bf u}_k({\bf x})
\end{eqnarray}

%---------------------------------------------------------------------------------------------------------------------------------------------------------------------%

\section{Quantum state evolution in linear optical interferometers}
\label{sec:state_evolution}

This section establishes the theoretical framework, together with the notation and
conventions adopted to describe the state evolution. Let the initial state be denoted
by $\hat{\rho}_{\mathrm{in}}$, which evolves into an output state
$\hat{\rho}_{\mathrm{out}}$. The most general form of such an evolution is described, in the Schrödinger picture,
by a \textit{quantum channel}, that is, a completely positive and trace-preserving (CPTP) map~\cite{NielsenChuang, BreuerPetruccione}, 
\begin{equation}
    \hat{\rho}_{\text{in}} \mapsto \hat{\rho}_{\text{out}} = \Lambda(\hat{\rho}_{\text{in}})
    .
    \label{quantum_channel}
\end{equation}
The map $\Lambda$ is linear,
\begin{equation}
    \Lambda\!\left(\alpha_1 \hat{\rho}^{(1)}_{\mathrm{in}}
    + \alpha_2 \hat{\rho}^{(2)}_{\mathrm{in}}\right)
    =
    \alpha_1 \Lambda\!\left(\hat{\rho}^{(1)}_{\mathrm{in}}\right)
    + \alpha_2 \Lambda\!\left(\hat{\rho}^{(2)}_{\mathrm{in}}\right),
\end{equation}
for arbitrary $\alpha_1,\alpha_2 \in \mathbb{C}$, trace preserving,
\begin{equation}
    \tr\!\left[\Lambda(\hat{\rho}_{\mathrm{in}})\right]
    =
    \tr(\hat{\rho}_{\mathrm{in}})=1,
\end{equation}
and completely positive, meaning that
\begin{equation}
    \left(\Lambda \otimes \mathbb{I}_n\right)(\hat{\rho}) \geq 0
\end{equation}
for any positive operator $\hat{\rho}$ acting on an extended Hilbert space and for any auxiliary dimension $n$. 

Denoting by $M$ the number of spatial modes under consideration, in the entire text we restrict attention to input states of the general form
\begin{equation}
    \hat{\rho}_{\text{in}} = \sum_{{\bf n}, {\bf n}'}  c_{{\bf n}, {\bf n}'} \, \prod^M_{k=1} \frac{(\hat{a}^\dagger_k)^{n_k}}{n_k!} \ket{0} \bra{0} \prod^M_{k=1} \frac{\hat{a}_k^{n'_k}}{n'_k!} ,
    \label{input_state_GeneralForm}
\end{equation}
where ${\bf n} = (n_1,\ldots,n_M)$ and $\hat{a}^\dagger_1,\ldots,\hat{a}^\dagger_M$
denote the creation operators acting on the input modes. For our purposes, the state evolution is fully characterized by the transformation
of the creation and annihilation operators acting on the system modes.

We first consider \textit{unitary evolutions}, in which the initial state evolves according to a linear unitary operator $\hat{U}$,
\begin{equation}
    \hat{\rho}_{\mathrm{in}} \mapsto
    \hat{\rho}_{\mathrm{out}} =
    \hat{U}\,\hat{\rho}_{\mathrm{in}}\,\hat{U}^\dagger .
\end{equation}
Such evolutions define quantum channels, since they are linear, completely positive,
and trace preserving as a consequence of the unitarity condition
$\hat{U}^\dagger\hat{U}=\mathbb{I}$. From a group-theoretical perspective, unitary operators form a Lie group, whose elements can be expressed as~\cite{BookSzekeres}
\begin{equation}
    \hat{U} =
    \exp\!\left(i\sum_k g_k \hat{G}_k\right),
    \label{unitary_generator}
\end{equation}
where the Hermitian operators $\hat{G}_k$ are the generators of the
associated Lie algebra and $g_k \in \mathbb{R}$ are the corresponding group
parameters. In physical settings, these generators typically correspond to
observables such as:
\begin{itemize}
    \item The Hamiltonian $\hat{H}$, generating a one-parameter unitary subgroup with the group parameter given by the evolution time $t$.
    \item The photon number operator $\hat{n}$, generating optical phase shifts $\varphi$.
    \item The momentum and position operators $\{\hat{P},\hat{X}\}$, generating
    translations in position $x$ and momentum $p$, respectively.
    \item The angular momentum operators
    $\{\hat{L}_x,\hat{L}_y,\hat{L}_z\}$, generating rotations $\theta_k$ about the corresponding axes.
\end{itemize}

%Unitary transformations generated by operators that are at most quadratic in thecreation and annihilation operators are referred to as \textit{Gaussian unitaries}, since they map Gaussian states onto Gaussian states (see Sec.~\ref{sec:Phase_Space}). Gaussian unitaries can be divided into \textit{passive unitaries}, which preserve the total photon number, and \textit{active unitaries}, which change the photon number and therefore do not preserve the energy.

The \textit{linear interferometers} implement transformations that conserve the total photon number, commonly referred to as passive unitaries. In linear optics, such transformations arise from unitary evolutions of the form given in Eq.~(\ref{unitary_generator}), generated by quadratic Hamiltonians 
\begin{equation}
    \hat{H} = \sum_{i,j} h_{ij} \hat{a}^\dagger_i \hat{a}_j .
    \label{passive_unitaries}
\end{equation}
Denoting by $M$ the number of spatial modes of the interferometer, the corresponding transformations act on the $M$-dimensional mode space and form the unitary group $\mathbb{U}(M)$, whose elements are represented by $M \times M$ unitary matrices. In the following, we highlight two fundamental transformations belonging to this group:
\begin{itemize}
    \item The simplest element is a \textit{phase shift}, which is an element of the group $\mathbb{U}(1)$  and introduces a phase $\varphi$ on a spatial mode $k$. It is  generated by the photon number operator $\hat{n}_k$ acting on the corresponding spatial mode, according to the following unitary
    \begin{equation}
        \hat{U}_{\varphi} = \e^{i \varphi \hat{n}_k} 
        ,
        \label{phase_encoding}
    \end{equation}
    and transforms the corresponding creation operator as
    \begin{equation}
        \hat{a}^\dagger_k \mapsto \hat{U}_{\varphi} \, \hat{a}^\dagger_k \, \hat{U}^\dagger_{\varphi} = \e^{i \varphi} \, \hat{a}^\dagger_k
    \end{equation}

    \item A second fundamental element is a \textit{beamsplitter}, which is an element of the group $\mathbb{SU}(2)$~\footnote{Here, an global phase factor $\e^{i\phi_0}$ has been omitted, since it has no physical effect on the interference. If such a phase is included, the transformation is promoted from $\mathbb{SU}(2)$ to $\mathbb{U}(2)$, with determinant $\e^{i\phi_0}$.} that mixes two different optical modes $k$ and $j$. It is represented by the unitary~\cite{YurkeSU2,CamposSU2}:
    \begin{equation}
    \hat{U}_{\mathrm{BS}}
    =
    \exp\!\left[-i(\phi_t-\phi_r)\hat{L}_z\right]\,
    \exp\!\left[-2i\,\arccos\!\big(\sqrt{\tau}\big)\,\hat{L}_y\right]\,
    \exp\!\left[-i(\phi_t+\phi_r)\hat{L}_z\right]
    ,
    \end{equation}
    where $\tau$ is the transmissivity, $\phi_t$ the transmission phase and $\phi_r$ the reflectivity phase; and the generators $\hat{L}_2$ and $\hat{L}_3$ are the angular momentum operators in the Swinger's representation
    \begin{equation}
        \hat{L}_y = \frac{1}{2i} \left( \hat{a}^\dagger_1 \hat{a}_2 - \hat{a}^\dagger_2 \hat{a}_1 \right)
        , \qquad
        \hat{L}_z = \frac{1}{2} \left(\hat{a}^\dagger_1 \hat{a}_1 - \hat{a}^\dagger_2 \hat{a}_2 \right)
        .
    \end{equation}
    The beamsplitter evolves the state mixing the two optical modes as follows
    \begin{equation}
            \begin{pmatrix}
            \hat{a}_k \\
            \hat{a}_j
        \end{pmatrix}
        \mapsto
        \,
        \hat{U}_{\mathrm{BS}}
        \begin{pmatrix}
            \hat{a}_k \\
            \hat{a}_j
        \end{pmatrix}
        \,
        \hat{U}^\dagger_{\mathrm{BS}}
        =
        \begin{pmatrix}
            \sqrt{\tau} \, \e^{-i \phi_t} & \sqrt{1-\tau} \, \e^{-i \phi_r} \\
            -\sqrt{1-\tau} \, \e^{i \phi_r} & \sqrt{\tau} \, \e^{-i \phi_t} 
        \end{pmatrix} 
        \begin{pmatrix}
            \hat{a}_k \\
            \hat{a}_j
        \end{pmatrix}
        \label{beamsplitter_general}
    \end{equation}
    During the text, a common choice for the phases is $\phi_r=\pi/2$ and $\phi_t=0$ . 
\end{itemize}

\begin{figure}[t]
    \centering
\includegraphics[width=0.5 \columnwidth]{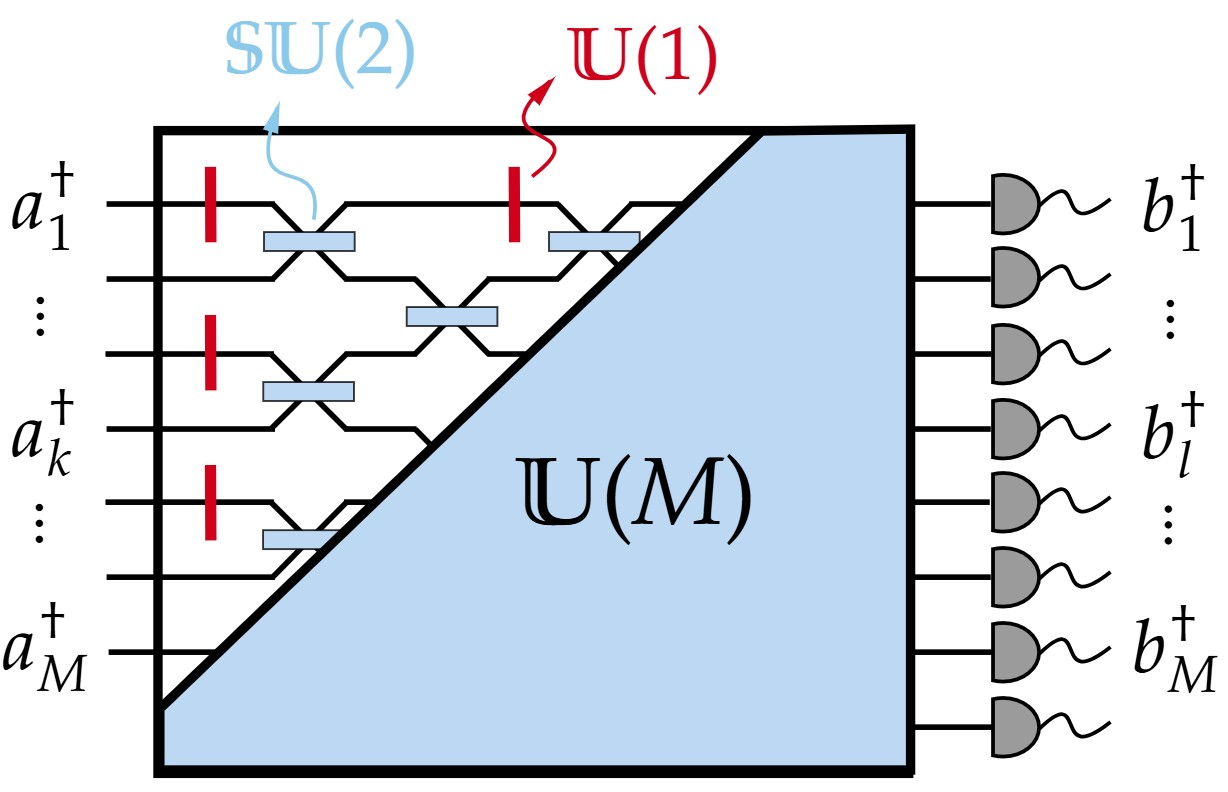}
\caption{Representation of the construction of a $M$-mode interferometer in terms of elements of linear optics: beamsplitters (blue) and phase shifters (red). Indeed, the entire interferometer is represented by a matrix that is an element of the group $\mathbb{U}(M)$, while the elements are matrices belonging to the groups $\mathbb{SU}(2)$ and $\mathbb{U}(1)$ respectively.}
\label{interferometer_main_scheme}
\end{figure}

To summarize the above construction, a linear optical interferometer with $M$ spatial modes implements a passive, photon-number–preserving unitary transformation described by an $M\times M$ matrix $U \in \mathbb{U}(M)$. A fundamental result is that any element of $\mathbb{U}(M)$ can be decomposed into a sequence of $\mathbb{SU}(2)$ transformations
(two-mode beam splitters) and $\mathbb{U}(1)$ phase shifts. As a consequence, any $M$-mode unitary transformation admits a physical realization as a network of beamsplitters and phase shifters acting on individual spatial modes \cite{InterferometerRealization1994,InterferometerRealization1997}, as illustrated in Fig.~\ref{interferometer_main_scheme}. Let $\hat{\mathbf{a}}^\dagger = (\hat{a}^\dagger_1,\ldots,\hat{a}^\dagger_M)$ the
the vector of input creation operators, and
$\hat{\mathbf{b}}^\dagger = (\hat{b}^\dagger_1,\ldots,\hat{b}^\dagger_M)$ the
corresponding output operators. The interferometer evolves the state according to the unitary transform
\begin{equation}
    \hat{\bf b}^\dagger = U \, \hat{\bf a}^\dagger
    , \qquad
    \hat{b}^\dagger_l = \sum^M_{k=1} U_{lk} \hat{a}^\dagger_k. 
    \label{multiport_evolution}
\end{equation}

%In addition, the active unitaries correspond to the unitary transformations in Eq.~(\ref{unitary_generator}) with generators being a Hamiltonian of the form
%\begin{equation}
%    \hat{\mathcal{H}} = \sum_{i,j} h_{ij} \left( \hat{a}^\dagger_i \hat{a}^\dagger_j + \hat{a}_i \hat{a}_j \right) .
%    \label{active_unitaries}
%\end{equation}
%The main operators belonging to this class are the displacement operators, which are crucial in the formulation of the quantum phase space, discussed in the following at Sec.~\ref{sec:Phase_Space}; and the single-mode and two-mode squeezing operators, which will be extensively explored in Chapters~\ref{chapter:GBS} and ~\ref{chapter:parameter_estimation}.

In the previous discussion, we restricted attention to the evolution of quantum states in a closed-system interferometer, fully described by a unitary transformation. However, such a description cannot fully capture the state evolution in the interferometer in the presence of any dissipative process. In the most general setting, the evolution
of the input state into the output state is described in terms of \textit{Kraus operators} $\hat{K}_m$, according to~\cite{NielsenChuang}
\begin{equation}
    \hat{\rho}_{\text{in}} \mapsto  \hat{\rho}_{\text{out}} = \sum_m \hat{K}_m \, \hat{\rho}_{\text{in}} \, \hat{K}^\dagger_m
    , \qquad
    \sum_m \hat{K}^\dagger_m \, \hat{K}_m = \mathbb{I} .
    \label{general_evolution_Kraus}
\end{equation}

One of the most relevant dissipative processes in quantum optics is \textit{photon loss}, which constitutes a fundamental source of noise affecting  any optical devices. Its impact on quantum-enhanced protocols will be discussed in detail in Chapter~\ref{chapter:parameter_estimation}. A standard and
physically transparent model for photon loss is obtained by coupling the optical mode of interest to an unobserved vacuum mode through a fictitious beam splitter, which redirects part of the photons into the environment, where it is not detected.
From Eq.~(\ref{beamsplitter_general}) we can describe this model in the following way
\begin{equation}
    \hat{a}^\dagger \mapsto \sqrt{\eta} \, \hat{a}^\dagger + \sqrt{1-\eta} \, \hat{e}^\dagger ,
    \label{loss_def}
\end{equation}
where $\hat{e}^\dagger$ creates photons on the virtual mode. The transmissivity of this fictitious beamsplitter gives the loss rate $\eta \in [0,1]$, where $\eta=1$ means that no photon is lost, while $\eta=0$ means that all photons are lost. To investigate explicitly the effect of losses, let us consider their action on a Fock state:
\begin{equation}
    \ket{n}_a = \frac{(\hat{a}^\dagger)^n}{\sqrt{n!}} \ket{0}
    \mapsto 
    \ket{\widetilde{n}}_a = \frac{1}{n!} \left( \sqrt{\eta} \, \hat{a}^\dagger + \sqrt{1-\eta} \, \hat{e}^\dagger \right)^n | 0 \rangle ,
    \label{action_loss_FockState}
\end{equation}
where $\ket{\widetilde{n}}_a$ is the Fock state affected by this fictitious beamsplitter~\cite{Dorner2009},
\begin{align}
    \ket{\widetilde{n}}_a = & ~ \frac{1}{\sqrt{n!}}  \sum^n_{m=0} \binom{n}{m} \left( \sqrt{\eta} \, \hat{a}^\dagger \right)^{n-m} \left( \sqrt{1-\eta} \, \hat{e}^\dagger \right)^m \ket{0} \nonumber\\
     = & ~ \sum^n_{m=0} \sqrt{B^{n}_m} \, \ket{n-m}_a \ket{m}_e.
     \label{FockState_wLoss}
\end{align}
where on $ \ket{n-m}_a $ denotes the true Fock state with $m$ photons lost, $ \ket{m}_e $ denotes a virtual Fock state which we do not have access to. Thus, the previous evolution corresponds to a binomial process, with coefficients given by
\begin{equation}
    B^{n}_m = \binom{n}{m} \eta^{n-m}(1-\eta)^m .
    \label{binomial_loss}
\end{equation}
Finally, the output state is obtained by tracing out over this virtual mode, resulting in the following mixed state:
\begin{equation}
    \hat{\rho}_{\text{out}} = \tr_e \big( \ket{\widetilde{n}}_a \bra{\widetilde{n}}_a \big) = \sum^n_{m=0} B^{n}_m \ket{n-m}_a \bra{n-m}_a .
\end{equation}

%---------------------------------------------------------------------------------------------------------------------------------------------------------------------%

\section{Phase space formalism}
\label{sec:Phase_Space}

We begin this chapter by introducing \textit{coherent states}, which play a central role in the phase-space formalism of quantum optics. They are defined as the eigenstates of the annihilation operator~\cite{Glauber1963},
\begin{equation}
    \hat{a}_k \, |\alpha_k\rangle = \alpha_k \, |\alpha_k\rangle
    ,
    \label{coherent_eigen_def}
\end{equation}
where $\alpha_k \in \mathbb{C}$ is a complex amplitude associated with the spatial mode $k$. The coherent  states minimize the Heisenberg uncertainty relation and are commonly regarded as the quantum states that most closely reproduce the statistical properties of classical electromagnetic fields. Coherent states form an overcomplete basis in the Hilbert space, providing a resolution of the identity,
\begin{equation}
    \frac{1}{\pi} \int d^2 \alpha_k \, \ket{\alpha_k} \bra{\alpha_k} = \mathbb{I},
\end{equation}
with $\int d^2 \alpha_k = \int_{\mathbb{R}} d\,\Re(\alpha_k)\int_{\mathbb{R}} d\,\Im(\alpha_k)$. However, they are not mutually orthogonal,
\begin{equation}
    \big| \langle \alpha_k | \alpha'_k \rangle \big|^2
    =
    \exp\!\left(-|\alpha_k - \alpha'_k|^2\right).
\end{equation}
Additionally, on the Fock state basis, a coherent state is written as:
\begin{equation}
    \ket{\alpha_k} = \e^{-|\alpha_k|^2/2} \sum^\infty_{n=0} \frac{\alpha^{n_k}_k}{n_k!} \ket{n_k}
    \label{coherent_fock_basis}
    .
\end{equation}

\begin{figure}[H]
    \centering
\includegraphics[width=0.5 \columnwidth]{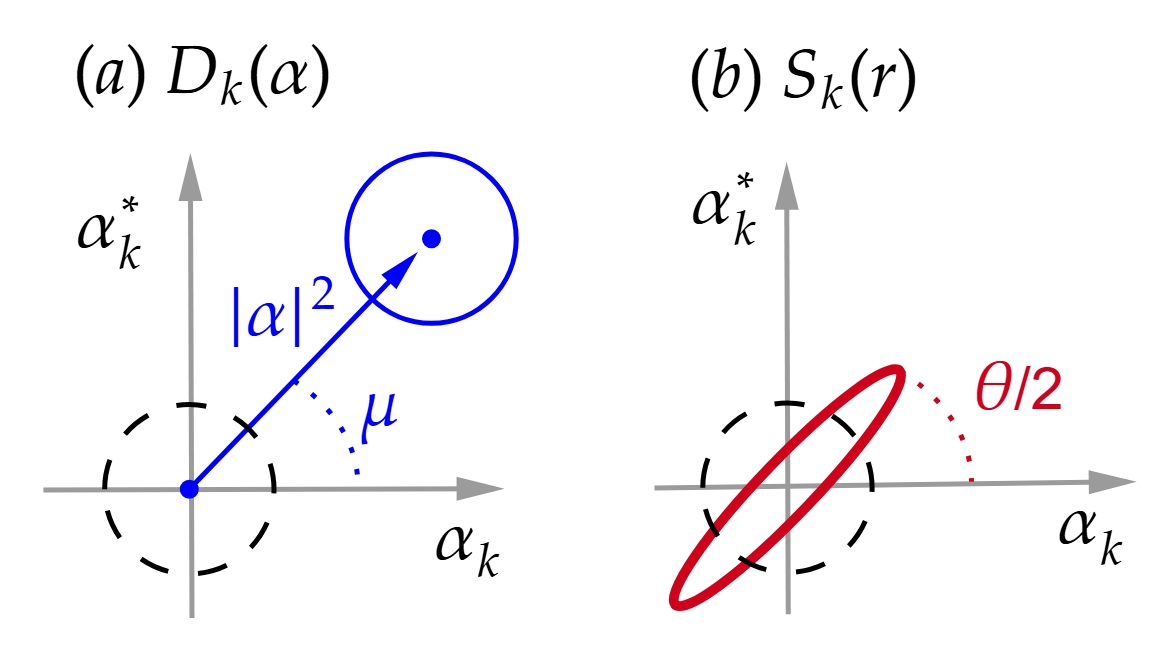}
\caption{Illustration of the operators acting on the phase space (a) Displacement operator $\hat{D}_k (\alpha_k)$, defined in Eq.~(\ref{displacement_def}), (b) Single-mode squeezing $\hat{S}_k (r)$, defined in Eq.~(\ref{sm_operator_def}).}
\label{displacement_squeezing}
\end{figure}

An equivalent and operationally convenient definition of coherent states is obtained through the \textit{displacement operator}, which is a unitary operator on the same for of given in  Eq.~(\ref{unitary_generator}), as
\begin{equation}
    \hat{D}_k(\alpha_k)
    =
    \exp\!\left(
        \alpha_k \hat{a}^\dagger_k
        -
        \alpha_k^{*} \hat{a}_k
    \right)
    .
    \label{displacement_op_def}
\end{equation}
In this way, the coherent state is obtained from the action of the displacement operator on the vacuum~\footnote{Using the Baker-Campbell-Hausdorff theorem, we can rewrite the displacement operator in a convenient way,
\begin{equation}
    \hat{D}_k (\alpha_k) = \e^{-|\alpha_k|^2/2} ~ \e^{\alpha_k \hat{a}^\dagger_k} ~ \e^{-\alpha^*_k \hat{a}_k}  ,
\end{equation}
and then, from Eq.~(\ref{coherent_state_def}) we recover the expression for the coherent state written in the Fock basis, as shown in Eq.~(\ref{coherent_fock_basis}).},
\begin{equation}
    |\alpha_k\rangle
    =
    \hat{D}_k(\alpha_k)\,|0\rangle
    .
    \label{coherent_state_def}
\end{equation}
The displacement operator acts by translating the field operators in phase space according to
\begin{equation}
    \hat{D}_k(\alpha_k)\,
    \begin{pmatrix}
        \hat{a}^\dagger_k \\
        \hat{a}_k
    \end{pmatrix}
    \hat{D}_k^\dagger(\alpha_k)
    =
    \begin{pmatrix}
        \hat{a}^\dagger_k + \alpha^*_k \\
        \hat{a}_k + \alpha_k
    \end{pmatrix},
    \label{displacement_action}
\end{equation}
which corresponds to a translation in optical phase space. As a direct consequence, the expectation value of the photon-number operator is shifted as illustrated in Fig.~\ref{displacement_squeezing}.(a),
\begin{equation}
    \langle \hat{n}_k \rangle
    \;\mapsto\;
    \langle \hat{n}_k \rangle + |\alpha_k|^2
    .
\end{equation}

The \textit{phase-space} formulation of quantum mechanics offers an alternative description of quantum states. In this approach, quantum states are described by quasiprobability distributions associated with different operator orderings, each defining a distinct phase-space representation for computing expectation values~\cite{BookBarnett,BookSerafini}. Introducing the multimode complex amplitude vector $\bm{\alpha} = (\alpha_1, \ldots, \alpha_M)$, we can express expectation values of observables in terms of normally or anti normally ordered operator functions, as explained in details in the following.
\begin{itemize}

    \item \textit{Normal ordering:} all creation operators $\hat{a}^\dagger_i$ are placed to the left of all annihilation operators $\hat{a}_i$. 
    The expectation value of a normally ordered operator function is given by
    \begin{equation}
        \langle \, \mathcal{N} \{ f(\hat{\bf a},\hat{\bf a}^\dagger) \} \, \rangle_\rho 
        =  \int d^2{\bm \alpha} \, f({\bm \alpha},\bm \alpha^*) \, \mathcal{P} ({\bm \alpha}),
        \label{expectation_NormalOrdering}
    \end{equation}
    where $\mathcal{P}({\bm \alpha})$ is the \textit{Glauber–Sudarshan} $P$ function, defined as a diagonal representation of the density operator in the coherent-state basis:
    \begin{equation}
        \hat{\rho} 
        = \int \frac{d^2{\bm \alpha}}{\pi^M} \, \mathcal{P}({\bm \alpha}) \, 
          | {\bm \alpha} \rangle \langle {\bm \alpha} |.
        \label{Glauber_def}
    \end{equation}
    This representation is particularly useful for describing classical-like optical fields, 
    since $\mathcal{P}$ plays the role of a quasiprobability distribution in phase space, which is written as a mixture of coherent states.
    
    \item \textit{Antinormal ordering:} this ordering places all annihilation operators $\hat{a}_i$ to the left of the creation operators $\hat{a}^\dagger_i$. 
    The corresponding expectation value is written as
    \begin{equation}
        \langle \, \mathcal{A} \{ f(\hat{\bf a},\hat{\bf a}^\dagger) \} \, \rangle_\rho 
        =  \int d^2{\bm \alpha} \, f({\bm \alpha},\bm \alpha^*) \, \mathcal{Q} ({\bm \alpha}),
        \label{expectation_AntiNormalOrdering}
    \end{equation}
    where $\mathcal{Q}({\bm \alpha})$ is the \textit{Husimi} $Q$ function, defined by
    \begin{equation}
        \mathcal{Q} ({\bm \alpha}) = \frac{1}{\pi^M} 
        \langle {\bm \alpha} | \hat{\rho} | {\bm \alpha} \rangle.
        \label{Husimi_def}
    \end{equation}
    The $\mathcal{Q}$ function is always positive and smooth, making it a convenient tool for representing 
    quantum states that are close to classical ones, although it cannot exhibit negativity.

    \item \textit{Symmetric ordering:} in this ordering, creation and annihilation operators are symmetrically ordered, such that all operator products are averaged over all possible permutations of $\hat{a}_i$ and $\hat{a}^\dagger_i$. 
    The expectation value of a symmetrically ordered operator function is given by
    \begin{equation}
        \langle \, \mathcal{S} \{ f(\hat{\bf a},\hat{\bf a}^\dagger) \} \, \rangle_\rho 
        =  \int d^2{\bm \alpha} \, f({\bm \alpha},\bm \alpha^*) \, \mathcal{W} ({\bm \alpha}),
        \label{expectation_SymmetricOrdering}
    \end{equation}
    where $\mathcal{W}({\bm \alpha})$ is the \textit{Wigner} function, defined as the phase-space quasiprobability distribution associated with symmetric operator ordering. 
    The Wigner function provides a complete representation of the quantum state and allows for a direct correspondence with classical phase-space dynamics. 
    Unlike the $\mathcal{P}$ and $\mathcal{Q}$ functions, $\mathcal{W}$ can take negative values, which is commonly interpreted as a signature of nonclassical behavior.

\end{itemize}

One important identity is the expectation value of two operators, let us say $\hat{A}_1$ and $\hat{A}_2$, that can be written in terms of the $\mathcal{P}$ and the $\mathcal{Q}$ function as follows:
\begin{align}
    \Tr \left(\hat{A}_1 \, \hat{A}_2 \right) = & ~ \text{tr} \left[ \left( \int \frac{d^2{\bm \alpha}}{\pi^M} \, \mathcal{P}_{\hat{A}_1} ({\bm \alpha}) \, | {\bm \alpha} \rangle \langle {\bm \alpha} | \right) \hat{A}_2 \, \right] \nonumber\\
    = & ~ \int d^2{\bm \alpha} \, \mathcal{P}_{\hat{A}_1} ({\bm \alpha}) \left(\frac{1}{\pi^M} \langle {\bm \alpha} | \hat{A}_2 | {\bm \alpha} \rangle \right) \nonumber\\
    = & ~ \int d^2{\bm \alpha} \, \mathcal{P}_{\hat{A}_1} ({\bm \alpha}) \, \mathcal{Q}_{\hat{A}_2} ({\bm \alpha}).
    \label{trace_PhaseSpace}
\end{align}

An important identity connecting the normal and antinormal orderings 
is obtained by considering the expectation value of the exponential of the creation and annihilation operators:~\cite{BookBarnett}:
\begin{equation}
    \langle \, \mathcal{N} \big\{ \exp \left( (\e^\lambda - 1) \, \hat{a}^\dagger_k \hat{a}_k \right) \big\} \, \rangle
    =
    \exp \left( \lambda \hat{a}^\dagger_k \hat{a}_k  \right)
    =
    \langle \, \mathcal{A} \big\{ \exp \left( (1-\e^{-\lambda}) \hat{a}_k \hat{a}^\dagger_k \right) \big\} \, \rangle .
    \label{normal_antinormal_relation}
\end{equation}

Any Gaussian state can be fully described by the its first and second moments of the creation and annihilation operators~\cite{BookSerafini}. The first moments are encoded on the \textit{displacement vector}, which is a $2M$ dimensional complex vector, defined by
\begin{equation}
    {\bf d} = 
    \begin{pmatrix}
        \langle \hat{a}_1 \rangle &
        \cdots &
        \langle \hat{a}_M \rangle &
        \langle \hat{a}^\dagger_1 \rangle &
        \cdots &
        \langle \hat{a}^\dagger_M \rangle
    \end{pmatrix}^t .
    \label{displacement_def}
\end{equation}
The second moments are encoded on the \textit{covariance matrix}, which is a $2M \times 2M$ Hermitian matrix with the following block structure:
\begin{equation}
    \sigma = 
    \begin{pmatrix}
        \sigma_N & \sigma_S \\
        \sigma^\dagger_S & \sigma_N ,
    \end{pmatrix},
    \label{covariance_def}
\end{equation}
where each block of the covariance matrix encodes a different type of second-order correlation
between the field operators,
\begin{align}
    & [\sigma_N]_{ij} = \aver{\{\hat{a}_i , \hat{a}^\dagger_j \}} - 2 \aver{\hat{a}_i} \aver{\hat{a}^\dagger_j} , \nonumber\\
    & [\sigma_S]_{ij} = \aver{\{\hat{a}_i , \hat{a}_j \}} - 2 \aver{\hat{a}_i} \aver{\hat{a}_j} .
\end{align}
The block $\sigma_N$ contains correlations involving one annihilation and one creation
operator and therefore describes fluctuations in the mode occupations, while the block $\sigma_S$ contains correlations between pairs of annihilation operators and
quantifies phase-sensitive, nonclassical correlations such as squeezing.
The bosonic algebra of the creation and annihilation operators is preserved in the phase-space through the \textit{symplectic matrix}, which is a $2M \times 2M$ matrix defined from the corresponding commutation relation,
\begin{equation}
    \Omega
    =
    \begin{pmatrix}
    [\hat{a}_i , \hat{a}^\dagger_j] & [\hat{a}_i , \hat{a}_j] \\   
    [\hat{a}^\dagger_i , \hat{a}^\dagger_j] & [\hat{a}^\dagger_i , \hat{a}_j]
    \end{pmatrix}
    = 
    \begin{pmatrix}
    I_M & 0_M \\   
    0_M& - I_M
    \end{pmatrix}
    \label{symplectic_matrix} ,
\end{equation}
where the covariance matrix obeys the following inequality:
\begin{equation}
    \sigma + \Omega  \geq  0 .
\end{equation}

Any Gaussian state admits a $p$-ordered quasiprobability distribution given by~\cite{BookBarnett,BookSerafini}
\begin{equation}
    \mathcal{W}({\bm \alpha}, p) = \frac{1}{\sqrt{\pi^M \det\left[ \sigma(p) \right]}}\exp \left[ - (\mathcal{A}-{\bf d})^\dagger \sigma(p)^{-1} (\mathcal{A}-{\bf d})  \right],
    \label{GaussianState_general_form}
\end{equation}
where the $p$-ordered covariance matrix is given by
\begin{equation}
    \sigma(p) = \sigma - \frac{p}{2} I_{2M}.
    \label{CovMatrix_ordering}
\end{equation}
For $p=0$, Eq.~(\ref{GaussianState_general_form}) reduces to the symmetrically ordered Wigner function,
with $\sigma(0)$ coinciding with the covariance matrix defined in Eq.~(\ref{covariance_def}).
The choice $p=1$ corresponds to normal ordering and gives the Glauber-Sudarshan $\mathcal{P}$ function, introduced in Eq.~(\ref{Glauber_def});
while $p=-1$ corresponds to antinormal ordering and gives the Husimi $\mathcal{Q}$ function
introduced in Eq.~(\ref{Husimi_def}).

For Gaussian states, the state evolution can be fully characterized by the corresponding transformations
of the displacement vector and the covariance matrix. 
In the case of a unitary evolution, these transformations are described by the following linear map ~\cite{BookSerafini,Safranek2016}:
\begin{align}
    & \sigma_{\text{in}} \mapsto \sigma_{\text{out}} = S \sigma_{\text{in}} S^\dagger ,
    \\
    & {\bf d}_{\text{in}} \mapsto {\bf d}_{\text{out}} = S {\bf d}_{\text{in}} + {\bf d}.
    \label{Gaussian_evolution_general}
\end{align}
In the second line of Eq.~(\ref{Gaussian_evolution_general}), the shift induced by the vector $\bm{d}$ corresponds to the action of the displacement operator given in Eq.~(\ref{displacement_action}).
The matrix $S$ represents a \textit{symplectic transformation} acting on phase space and belongs to the complex symplectic group
$\mathrm{Sp}(2M,\mathbb{C})$, where $2M$ is the dimension of the full phase space.
Such transformations are generated by Hamiltonians that are at most quadratic in the creation and annihilation operators,
as given in Eq.~(\ref{unitary_generator}).
In block form, the symplectic matrix $S$ can be written as
\begin{equation}
    S =
    \begin{pmatrix}
        S_{11} & S_{12} \\
        S^*_{12} & S^*_{11}
    \label{symplectic_general}
    \end{pmatrix} ,
\end{equation}
where $S_{11}$ and $S_{12}$ are complex $M \times M$ matrices; and preserve the symmetry of the symplectic matrix $\Omega$ defined in Eq.~(\ref{symplectic_matrix}),
\begin{equation}
    S \Omega S^\dagger = \Omega .
\end{equation}
In the following, the symplectic transformations that will be used throughout the present text are highlighted.
\begin{itemize}
    \item Any evolution given by a \textit{linear interferometer} belongs to such group, since they trivially satisfy the conditions imposed by the previous two equation. In the phase-space, the evolution is simply recast in terms of the coordinates ${\bm \alpha}=(\alpha_1,...,\alpha_M)$. Using Eq.~(\ref{multiport_evolution}) and identifying that $\hat{a}_k|\alpha_k \rangle = \alpha_k|\alpha_k \rangle$ and $\hat{b}_k| \beta_k \rangle = \beta_k| \beta_k \rangle$. Thus we have the corresponding transformation ${\bm \beta} = U^* {\bm \alpha}$, which can be written explicitly as
    \begin{equation}
    \beta_l = \sum^M_{l=1} U^*_{kl} \, \alpha_k .
    \label{PhaseSpace_evolution}
    \end{equation}
    This transformation is precisely the same as the one defined in Eq.~(\ref{multiport_evolution}) replacing each set of creation and annihilation operators $\{ \hat{a}_k, \hat{a}^\dagger_k \}$ by the corresponding coordinate of the phase space $\{ \alpha_k, \alpha^*_k \}$.
    Thus, the action of a multiport $U$ corresponds to the following transformation in the complete phase-space  coordinates $\mathcal{A} =(\alpha_1,...,\alpha_M,\alpha^*_1,...,\alpha^*_M)$:
    \begin{equation}
        \mathcal{A}
        \mapsto
        \begin{pmatrix}
            U & 0_M \\
            0_M & U^*
        \end{pmatrix}
        \mathcal{A}
    \label{multiport_evolution_PhaseSpace}
    \end{equation}
    In Eq.~(\ref{multiport_evolution_PhaseSpace}), the $2M \times 2M$ matrix transforming the coordinates $\mathcal{A}$ is exactly the corresponding symplectic transformation in Eq.~(\ref{symplectic_general}).
    
    \item Following, we consider the \textit{single-mode squeezing operator}, denoted by $\hat{S}_k (r)$, which acts on a single spatial mode $k$ as follows:
    \begin{equation}
        \hat{S}_k (r) = \exp \left[ \frac{r \, \e^{i \theta}}{2} (\hat{a}^\dagger_k)^2 - \frac{r \, \e^{-i \theta}}{2} \hat{a}^2_k \, \right] ,
        \label{sm_operator_def}
    \end{equation}
    and acts in the creation operators of the corresponding mode as follows
    \begin{align}
        \hat{S}_k (r) \, \hat{a}^\dagger_k \, \hat{S}^\dagger_k(r) = \text{cosh} \, r \, \hat{a}^\dagger_k + \text{sinh} \, r \, \e^{i \theta} \, \hat{a}_k .
    \end{align}
    The coefficient $r$ is the squeezing strength and shifts the average photon number on the corresponding mode as $\langle \hat{n}_k \rangle \mapsto \langle \hat{n}_k \rangle + \text{sinh}^2 \, r$, while $\theta$ is the squeezing phase and determines the direction of the squeezing on the phase space, as illustrated in Fig.~\ref{displacement_squeezing}.(b). Such transformation can be written in terms of the transformation defined in Eq.~(\ref{Gaussian_evolution_general}) from its corresponding matrix $S_k$ that acts on the coordinates $(\alpha_k, \alpha^*_k)$,
    \begin{equation}
        \begin{pmatrix}
            \alpha_k \\
            \alpha^*_k
        \end{pmatrix}
        \mapsto
        \begin{pmatrix}
            \text{cosh} \, r & \text{sinh} \, r \, \e^{i \theta} \\
            \text{sinh} \, r \, \e^{-i \theta} & \text{cosh} \, r
        \end{pmatrix} 
        \begin{pmatrix}
            \alpha_k \\
            \alpha^*_k
        \end{pmatrix}
        .
        \label{sm_squeezing_matrix}
    \end{equation}
    \item Finally, we also consider the \textit{two-mode squeezing operator}, denoted by $\hat{S}_{k,j} (r)$, which couples two modes $k$ and $j$ as follows:
        \begin{equation}
        \hat{S}_{k,j} (r) = \exp \left( r \, \e^{i \theta} \, \hat{a}^\dagger_k \hat{a}^\dagger_j - r \, \e^{-i \theta} \, \hat{a}_k \hat{a}_j \right) ,
        \label{tm_operator_def}
    \end{equation}
    and acts in the creation operators of the corresponding modes as follows
    \begin{align}
        & \hat{S}_{k,j} (r) \, \hat{a}^\dagger_k \, \hat{S}^\dagger_{k,j}(r) = \text{cosh} \, r \, \hat{a}^\dagger_k + \text{sinh} \, r \, \e^{i \theta} \, \hat{a}_j , \nonumber\\
        & \hat{S}_{k,j} (r) \, \hat{a}^\dagger_j \, \hat{S}^\dagger_{k,j}(r) = \text{cosh} \, r \, \hat{a}^\dagger_j + \text{sinh} \, r \, \e^{i \theta} \, \hat{a}_k .
    \end{align}
    The coefficient $r$ is the squeezing strength and shifts the average photon number in both modes as $\langle \hat{n}_k \rangle \mapsto \langle \hat{n}_k \rangle + \text{sinh}^2 \, r$ and $\langle \hat{n}_j \rangle \mapsto \langle \hat{n}_j \rangle + \text{sinh}^2 \, r$, while $\theta$ remains the squeezing. Note that, at the full system,  two-mode squeezing  creates twice as many photons (thus, twice the energy) of a single-mode squeezing. Such transformation can be written in terms of the transformation defined in Eq.~(\ref{Gaussian_evolution_general}) from its corresponding matrix $S_{k,j}$ that acts on the coordinates $(\alpha_k, \alpha_j, \alpha^*_k, \alpha^*_j)$,
    \begin{equation}
        \begin{pmatrix}
            \alpha_k \\
            \alpha_j \\
            \alpha^*_k \\
            \alpha^*_j
        \end{pmatrix}
        \mapsto
        \begin{pmatrix}
            \text{cosh} \, r & 0 & 0 & \text{sinh} \, r \, \e^{i \theta} \\
            0 & \text{cosh} \, r & \text{sinh} \, r \, \e^{i \theta} & 0  \\
            0 & \text{sinh} \, r \, \e^{-i \theta} & \text{cosh} \, r & 0 \\
            \text{sinh} \, r \, \e^{-i \theta} & 0 & 0 & \text{cosh} \, r &
        \end{pmatrix} 
        \begin{pmatrix}
            \alpha_k \\
            \alpha_j \\
            \alpha^*_k \\
            \alpha^*_j
        \end{pmatrix}
        .
    \label{tm_squeezing_matrix}
    \end{equation}
\end{itemize}

%-------------------------------------------------------------------------------------------------------------------------------------------------------------------%

\section{Quantum measurement and estimation theory}

Let us consider an input state $\hat{\rho}_{\mathrm{in}}$ entering an $M$-mode interferometer, as described in Sec.~\ref{sec:state_evolution}. 
We are interested in the output statistics obtained by measuring an observable $X_l$ at each output mode $l$. 
In quantum mechanics, any observable $X_l$ is represented by a Hermitian operator $\hat{X}_l$ with spectral decomposition that can be denoted as ~\cite{NielsenChuang}
\begin{equation}
    \hat{X}_l = \sum_{x_l \in \text{sp}(\hat{X}_l)} x_l \, \hat{\Pi}_{x_l}
    , \qquad
    \hat{\Pi}_{x_l} = \ket{x_l} \bra{x_l}
    .
    \label{observabel_spectral}
\end{equation}
In Eq.~(\ref{observabel_spectral}) $ \text{sp}(\hat{X}_l)$ denotes the spectrum of the operator, which is the set of all eigenvalues $x_l$, and $ \ket{x_i}$ the corresponding eigenvectors. Indeed, such a type of measurement is called a \textit{projective measurement}, because it is performed in terms of the projectors $\hat{\Pi}_{x_l}$, which are Hermitian operators that satisfy
\begin{equation}
    \hat{\Pi}_{x_l} \hat{\Pi}_{x'_l} = \delta_{x_l, x'_l} \, \hat{\Pi}_{x_l} .
\end{equation}
The probability of obtaining the joint outcome ${\bf x}=(x_1,\ldots,x_M)$ is given by the Born rule
\begin{equation}
    P({\bf x}) = \Tr \left( \hat{\rho}_{\text{out}} \, \hat{\Pi}_{\bf x} \right) 
    , \qquad
    \hat{\Pi}_{\bf x} = \prod^M_{l=1} \hat{\Pi}_{x_l} ,
\end{equation}
where the product is well-defined because operators acting on different modes commute.
More generally, a measurement is described by a complete set set of positive operators $\{\hat{\Pi}_{\bf x}\}$,
\begin{equation}
    \hat{\Pi}_{\bf x}\ge 0,
    \qquad
    \sum_{x_1}\cdots\sum_{x_M}\hat{\Pi}_{\bf x}=\mathbb{I},
\end{equation}
which defines a \textit{positive operator-valued measure} (POVM), where the projective measurements are a particular case of this general framework~\cite{NielsenChuang,BookHolevo,BookHelstrom}.
Two measurements used throughout this work are:

\begin{itemize}

    \item \textit{Photon-number-resolving} (PNR) detection. It consists of measuring the total number of photons in a given output mode $k$, described by the photon-number operator in the output basis
    \begin{equation}
        \hat{N}_k = \hat{b}^\dagger_k \hat{b}_k
        .
    \end{equation}
    The corresponding projective measurement in a $M$-mode interferometer is given in terms of the Fock states in the output basis $\ket{{\bf n}}_{\bf b}$, through the following POVM:
    \begin{equation}
        \hat{\Pi}_{\bf n} = \ket{{\bf n}}_{\bf b} \bra{{\bf n}}_{\bf b}
        , \qquad
        | \nb \rangle_{\bf b} \equiv  \prod_{k=1}^M \frac{(\hat{b}^\dagger_k)^{n_k}}{\sqrt{n_k!}} \ket{0}
        .
        \label{born_rule}
    \end{equation}
    The measurement outcomes ${\bf n} = (n_1,\ldots,n_M)$, with $n_k \in \mathbb{N}$, correspond to the photon numbers detected in each output mode, and throughout the text we denote the joint photon-counting probability distribution by $P({\bf n})$.

    \item \textit{Homodyne detection}. It consists of measuring a field quadrature in a given output mode $k$, described by the operator
    \begin{equation}
         \hat{Q}_k(\xi) = \frac{1}{\sqrt{2}} \left(\hat{b}^\dagger_k e^{i\xi} +  \hat{b}_k e^{-i \xi}  \right),
         \label{quadrature_op}
    \end{equation}
        where $\xi$ denotes the phase of the local oscillator that defines the measured quadrature. The local oscillator is a strong coherent reference field that interferes with the signal mode and fixes the quadrature measured in the homodyne detection scheme. The corresponding projective measurement is described by the POVM elements
        \begin{equation}
            \hat{\Pi}_{q_k} = \ket{q_k;\xi}_{b_k}\bra{q_k;\xi}_{b_k},
        \end{equation}
        where $\ket{q_k;\xi}_{b_k}$ denotes the eigenstate of the quadrature operator $\hat{Q}_k(\xi)$ with eigenvalue $q_k \in \mathbb{R}$. Homodyne detection therefore provides access to continuous-valued measurement outcomes, widely used to characterize optical states in phase space. 

\end{itemize}

In many relevant scenarios, the quantity of interest is not an observable, but a parameter $\lambda$, such as a phase, frequency, time delay, or photon loss. In that situations, there is no Hermitian operator associated with the parameter that can be directly measured, as discussed in the theory above. In that case, $\lambda$ must be inferred indirectly from measurement outcomes, which is a problem consolidated in the quantum parameter estimation theory~\cite{BookHelstrom,BookHolevo}.  The analogy is the following: an observable $X$ is measured by performing a measurement associated with the corresponding Hermitian operator $\hat{X}$; whereas a parameter $\lambda$ is estimated through an estimator, denoted by $\hat{\lambda}$, which is constructed from the outcomes of an indirect measurement.  The uncertainty of an estimator $\hat{\lambda}$ is quantified by its variance
\begin{equation}
    \Delta^2 \lambda = \langle (\hat{\lambda} - \langle \hat{\lambda} \rangle)^2 \rangle .
    \label{def_variance_lambda}
\end{equation}
Let us consider, for instance, an observable $\hat{X}$ whose expectation value depends on the parameter $\lambda$ as $\langle \hat{X} \rangle = f(\lambda)$. In this case, a common choice in evaluating the error in estimating $\lambda$ is from the standard error propagation formula,
\begin{equation}
    \Delta^2 \lambda
    =
    \frac{\Delta^2 \hat{X}}
    {\left| \partial_\lambda \langle \hat{X} \rangle \right|^2}
    +
    \mathcal{O}\!\left(\langle (\delta \hat{X})^3 \rangle\right),
    \label{error_prop}
\end{equation}

We now briefly recall some fundamental results from classical estimation theory.
Let $\lambda$ denote an unknown parameter, and suppose that, for each value of $\lambda$, a measurement produces outcomes $\{x\}$ distributed according to a probability density function $P(x|\lambda)$. All the information that can be extracted about $\lambda$ from this measurement is encoded in the dependence of $P(x|\lambda)$ on the parameter. 
A central goal of estimation theory is to determine how precisely $\lambda$ can be inferred from such data.
In particular, one seeks lower bounds on the variance of any estimator of $\lambda$. A central objective of estimation theory is to quantify how precisely $\lambda$ can be inferred from such data by deriving lower bounds on the variance of any estimator of $\lambda$. In classical statistics, a fundamental result in this direction is the \textit{Cramér-Rao bound} (CRB), which establishes a lower bound on the variance of any unbiased estimator of $\lambda$~\cite{BookHelstrom},
\begin{equation}
    \Delta^2 \lambda \geq \frac{1}{F_C\left[P(x|\lambda)\right]},
\end{equation}
where we introduce the classical Fisher information (FI) as follows:
\begin{equation}
    F_C\left[P(x|\lambda)\right] = \sum_x \, P(x|\lambda) 
    \left[ \frac{\partial}{\partial \lambda} \ln P(x|\lambda) \right]^2 ,
    \label{FI_def}
\end{equation}
which quantifies how much information the data $P(x|\lambda)$ carries about the parameter $\lambda$. In the previous equations, ``classical'' simply means that these results were developed in the context of statistics, without any assumptions about quantum mechanics. 

We now incorporate the discussion of quantum detection into the FI framework.
Suppose that the parameter $\lambda$ encoded in an input quantum state $\hat{\rho}_{\mathrm{in}}$ (also refered to as probe state)
through a parameter-dependent quantum channel $\Lambda_\lambda$, as introduced in Eq.~(\ref{quantum_channel}).
The resulting output state can be denoted as
\begin{equation}
    \hat{\rho}_\lambda = \Lambda_\lambda(\hat{\rho}_{\mathrm{in}}),
\end{equation}
where the explicit dependence on $\lambda$ emphasizes that the parameter is encoded via the quantum evolution,
which is the standard setting in quantum metrology. A measurement is subsequently performed on the output state by means of a generalized measurement,
described by a POVM $\{\hat{\Pi}_x\}$,
\begin{equation}
    P(x|\lambda) = \tr \left(\hat{\rho}_\lambda \, \hat{\Pi}_x \right).
\end{equation}

Starting from the classical FI defined in Eq.~(\ref{FI_def}),
one may optimize over all possible measurements in order to quantify the
maximum amount of information that can be extracted from a quantum state
about the parameter $\lambda$.
Following standard results from quantum estimation theory, the \textit{quantum Fisher
information} (QFI) is defined as the supremum of the FI information over
all possible POVMs~\cite{BookHelstrom, BookHolevo},
\begin{equation}
    F (\hat{\rho}_\lambda) = \sup_{\hat{\Pi}_x} \big\{ F_C\left[P(x|\lambda)\right] \big\},
\end{equation}
where the supremum is taken over all POVMs $\{\hat{\Pi}_x\}$. This optimization reflects the fact that, among all possible measurement
strategies, there exists at least one that extracts the largest possible
amount of information about the parameter $\lambda$ encoded in the quantum
state $\hat{\rho}_\lambda$. In this way, the maximization of the FI over all possible POVMs results in the QFI,which is a quantity that depends only on the  state $\hat{\rho}_\lambda$,
\begin{equation}
    F (\hat{\rho}_\lambda) =  \tr \left( \hat{\rho}_\lambda L^2_\lambda  \right),
    \label{qfi_def}
\end{equation}
where $L_\lambda$ is a Hermitian operator called the
symmetric logarithmic derivative (SLD), defined as follows
\begin{equation}
    \frac{\partial \hat{\rho}_\lambda}{\partial \lambda} = \frac{1}{2} \{ \hat{\rho}_\lambda , L_\lambda \} .
    \label{SLD_def}
\end{equation}

Therefore, for a given probe state $\hat{\rho}_\lambda$, the quantum Fisher information
sets a fundamental lower bound on the uncertainty of estimating the parameter $\lambda$,
known as the \emph{quantum Cramér--Rao bound} (quantum CRB),
\begin{equation}
    \Delta^2 \lambda \geq \frac{1}{F(\hat{\rho}_\lambda)}.
    \label{CRB}
\end{equation}
It is important to emphasize that, while the QFI is independent of the specific measurement performed,
it depends crucially on the choice of the probe state.
Different quantum states can exhibit highly different QFI values,
and thus enable distinct levels of estimation precision.
As a result, a central theme in quantum metrology is the identification and characterization
of probe states $\hat{\rho}_\lambda$ that maximize the QFI for a given parameter $\lambda$ and then, the corresponding measurement that saturates the bound in Eq.~(\ref{CRB}).
This topic will be investigated in detail in Chapter~\ref{chapter:parameter_estimation}.

%%%%%%%%%%%%%%%%%%%%%%%%%%%%%%%%%%%%%%%%%%%%%%%%%%%%%%%%%%%%%%%%%%%%%%%%%%%%%%%%%%%%%%%%%%%%%%%%%%%%%%%%%%%%%%%%%%%%%%%%%%%%%%%%%%%%%%%%%%%%%%%%%%%%%%%%%%%%%%%%%%%%%%%%%%%%%%%%%%%%%%%%%%%%%%%%%%%%%%%%%%%%%%%%%%%%%%%%

\begin{center}
\myclearpage
\par
\end{center}

\chapter{Zero probability events beyond the permutation symmetry principle}
\label{chapter:suppression_laws} 

\begin{tcolorbox}[colback=gray!5,colframe=black,title={This chapter is based on the following publication:}]
\underline{M.~E.~O.~Bezerra} and V.~S.~Shchesnovich,  
``Families of bosonic suppression laws beyond the permutation symmetry principle'',   
New Journal of Physics, \textbf{25}, 093047 (2023).  
\href{https://doi.org/10.1088/1367-2630/acfa1e}{DOI: 10.1088/1367-2630/acfa1e} \, 
(\href{https://arxiv.org/abs/2301.02192}{arXiv:2305.12345})
\end{tcolorbox}

\section{Background}

One of the most distinctive signatures of quantum theory is the superposition principle, which under appropriate conditions, leads to the existence of destructive interference in multi-path scenario, with the probability of certain outcomes being exactly zero. The simplest manifestation of this quantum signature is the \textit{Hong-Ou-Mandel effect} (HOM)~\cite{HOM1987}, which consists of the interference of two single photons at a balanced beam splitter, represented by the following matrix\footnote{This matrix can be obtained from Eq.~(\ref{beamsplitter_general}) by identifying the unitary matrix $B = U_2(\tau; \phi_r, \phi_t)$ with transmissivity $\tau = 1/2$, reflection phase $\phi_r = \pi/2$, and transmission phase $\phi_t = 0$.}:
\begin{equation}
    B = 
    \frac{1}{\sqrt{2}}
    \begin{pmatrix}
        1 & i \\
        i & 1
    \end{pmatrix} 
    ,
    \label{beamsplitter_balanced}
\end{equation}
which implements the unitary evolution described in Eq.~(\ref{multiport_evolution}). Denoting by $|1,1\rangle_{\mathbf a}$ the input Fock state corresponding to one photon entering each input mode of the beam splitter, we have the following possible amplitudes
\begin{equation}
    _{\bf b}\langle 2, 0 | 1, 1 \rangle_{\bf a}  = \,  _{\bf b}\langle 0, 2 | 1, 1 \rangle_{\bf a} = \frac{1}{2} ,
    \qquad
    _{\bf b}\langle 1, 1 | 1, 1 \rangle_{\bf a} =  0 
    ,
    \label{amplitude_2photon}
\end{equation}
where $|n_1,n_2\rangle_{\mathbf b}$ denotes the state detected at the output. Therefore, according to Eq.~(\ref{amplitude_2photon}), the amplitude for coincidence detection vanishes, indicating complete destructive interference for this output event, which constitutes a fundamental bosonic signature of two-photon interference. However, this perfect destructive interference occurs only when the photons are completely indistinguishable~\cite{HOM1987}. Distinguishability arises from mode mismatch in the internal states of the photons, which correspond to unresolved degrees of freedom such as their spectral profiles, temporal delays, and polarization~\cite{HOM1987,Loudon1989}.  Denoting by $\ket{\psi_1}$ and $\ket{\psi_2}$ the internal states of the two photons, the coincidence-detection probability becomes~\cite{Loudon1989},
\begin{equation}
    P(1,1) = \frac{1}{2} \Big(1- |\aver{\psi_1 | \psi_2}|^2 \Big)
    .
    \label{prob_11_HOM}
\end{equation}
Therefore, Eq.~(\ref{prob_11_HOM}) gives a continuous statistics between the completely indistinguishable case $\aver{\psi_1 | \psi_2}=1$, recovering Eq.~(\ref{amplitude_2photon}), and the completely distinguishable case $\aver{\psi_1 | \psi_2}=0$. The latter results in the probability $P(1,1)=1/2$, which corresponds to a classical probability obtained by simply multiplying single-photon probabilities, and thus contains no quantum interference. The quantity $|\langle \psi_1 | \psi_2 \rangle|^2$, commonly referred to as the visibility, measures the overlap between the two photonic states and therefore characterizes the experimental quality of the two-photon interference.

\begin{figure}[t]
    \centering
    \includegraphics[width=0.8 \columnwidth]{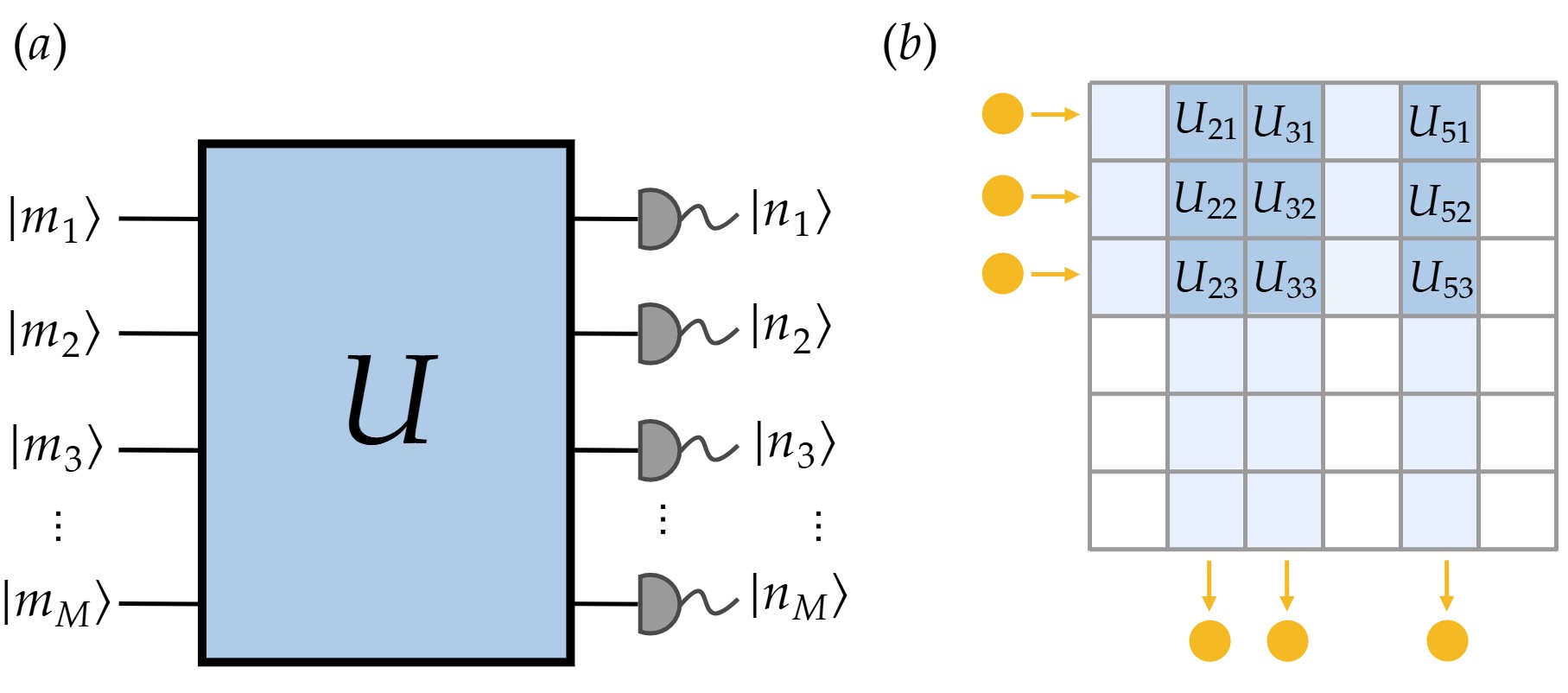}
    \caption{(a) Illustration of the scheme for the interference of $N$ photons prepared in the input Fock state $\ket{\bf m}_{\bf a}$, with ${\bf m}=(m_1,\ldots,m_M)$, and detected in the output Fock state $\ket{\bf n}_{\bf b}$, with ${\bf n}=(n_1,\ldots,n_M)$, after interference at the multiport $U$. (b) Example of three-photon interference in a six-mode interferometer: the photons are prepared in the input configuration ${\bf m}_3=(1,1,1,0,0,0)$ and detected in the output configuration ${\bf n}_3=(0,1,1,0,1,0)$. The associated input and output mode label vectors are ${\bf k}=(1,2,3)$ and ${\bf l}=(2,3,5)$, respectively, where $U_{lk}$ denotes the transition amplitude from input mode $k$ to output mode $l$, as defined in Eq.~(\ref{multiport_evolution}).}
    \label{BS_scheme}
\end{figure}

Moving to a more general interferometric setting, we consider $N$ indistinguishable photons interfering in an $M$-mode interferometer, as illustrated in Fig.~\ref{BS_scheme}(a). Recalling the definition of an $M$-mode Fock state with $N$ photons given in Eq.~(\ref{Fock_input}), we evaluate the transition probability from an input state $|\mathbf m\rangle_{\bf a}$, with ${\bf m} = (m_1,\ldots,m_M)$, to an output state $|\mathbf n\rangle_{\mathbf b}$, with ${\bf n} = (n_1,\ldots,n_M)$, after interfering at an $M$-mode linear interferometer $U$ described by Eq.~(\ref{multiport_evolution}).  Identifying the input state as $\hat{\rho}_{\bf m} = \ket{\bf m}_{\bf a}\bra{\bf m}_{\bf a}$, the photon-counting probability is given by
\begin{equation}
    P({\nb}) = \tr \left( \hat{\rho}_{{\bf m}} \, \hat{\Pi}_{\bf n} \right)
    =
    \big| {_{\bf b}}\langle {\bf n} | {\bf m} \rangle_{\bf a} \big|^2,
    \label{probabili_BS}
\end{equation}
where $\hat{\Pi}_{\bf n} = \ket{{\bf n}}_{\bf b} \bra{{\bf n}}_{\bf b}$ is the PNR operator introduced in Eq.~(\ref{born_rule}). The corresponding transition amplitude is given by~\cite{origBS,BSAA}
\begin{equation}
    _{\bf b} \langle {\bf n} | {\bf m} \rangle_{\bf a}
    =
    \frac{\operatorname{per}\!\left(U_{\mb,\nb}\right)}{\sqrt{{\bf n}! \, {\bf m}!}},
    \label{amplitude_permanent}
\end{equation}
where $\mathbf{m}! = m_1! \cdots m_M!$ and $\mathbf{n}! = n_1! \cdots n_M!$. In Eq.~(\ref{amplitude_permanent}), $\operatorname{per}(U_{\mathbf{m},\mathbf{n}})$ denotes the \emph{permanent} of the $N \times N$ submatrix of $U$, constructed by repeating rows and columns according to the input and output occupation configurations $\mathbf{m}$ and $\mathbf{n}$, respectively. The permanent accounts for all possible permutations of indistinguishable photons across the interferometer paths and is defined as~\cite{BookMinc}
\begin{equation}
    \operatorname{per}\!\left(U_{\mb,\nb}\right)
    =
    \sum_{\sigma \in \mathbb{S}_N}
    \prod_{i=1}^{N}
    U_{k_{\sigma(i)},\,l_i},
    \label{permanent}
\end{equation}
where $k_i$ and $l_i$ label the input and output modes associated with the $i$-th photon, and $\sigma$ denotes a permutation in the symmetric group of $N$ elements, $\mathbb{S}_N$. This labeling makes explicit how each permutation assigns input modes to output modes. The following examples illustrate how to compute the permanent for the selected input-output configurations:

\begin{itemize}

    \item For the interference of two photons in a two-mode interferometer, we focus on the HOM configuration, with the input state prepared in the configuration ${\bf m}_2 = (1,1)$ and detected in the output configuration ${\bf n}_2 = (1,1)$. Therefore, from Eq.~(\ref{permanent}), the coincidence amplitude is given by
    \begin{equation}
        {}_{\bf b}\langle {\bf n}_2 | {\bf m}_2 \rangle_{\bf a}
        =
        \operatorname{per}
        \begin{pmatrix}
            U_{11} & U_{12} \\
            U_{21} & U_{22}
        \end{pmatrix}
        =
        U_{11} U_{22} + U_{12} U_{21}.
    \end{equation}
    and considering the balanced beam-splitter matrix given in Eq.~(\ref{beamsplitter_balanced}), we immediately recover Eq.~(\ref{amplitude_2photon}).

    \item Considering the interference of three photons in a six-mode interferometer, as illustrated in Fig.~\ref{BS_scheme}.(b), the photons are prepared in the input configuration ${\bf m}_3 = (1,1,1,0,0,0)$ and detected in the output configuration ${\bf n}_3 = (0,1,1,0,1,0)$. According to Eq.~(\ref{probabili_BS}), the corresponding transition amplitude is given by
    \begin{equation}
    {}_{\mathbf{b}}\langle \mathbf{n}_3 | \mathbf{m}_3 \rangle_{\mathbf{a}}=
    \operatorname{per}
    \begin{pmatrix}
        U_{21} & U_{22} & U_{23} \\
        U_{31} & U_{32} & U_{33} \\
        U_{51} & U_{52} & U_{53}
    \end{pmatrix} 
    ,
    \end{equation}
    and then, from the definition of the permanent given in Eq.~(\ref{permanent}) we arrive at:
    \begin{equation}
    {}_{\mathbf{b}}\langle \mathbf{n}_3 | \mathbf{m}_3 \rangle_{\mathbf{a}}=
    U_{21}U_{32}U_{53}
    +U_{22}U_{33}U_{51}
    +U_{23}U_{31}U_{52} 
    +U_{21}U_{33}U_{52}
    +U_{22}U_{31}U_{53}
    +U_{23}U_{32}U_{51}
    .
    \end{equation}
\end{itemize}

This scenario naturally connects to the \textit{Boson Sampling} (BS) model of quantum computation, originally proposed by Aaronson and Arkhipov~\cite{BSAA}. The central problem of the BS lies in the evaluation of the matrix permanents given in Eq.~(\ref{amplitude_permanent}), which are believed to be classically intractable for sufficiently large systems with $M > N$, due to their exponential scaling with $N$~\cite{HardnessPermanent}. This computational hardness is a direct consequence of the symmetric structure of bosonic states, which results in genuine quantum interference effects involving many indistinguishable photons. 
However, this multiphoton interference features are progressively degraded in the presence of noise, which suppresses quantum signatures in the output statistics and ultimately diminishes the computational advantage of BS~\cite{Kalai2014,SimulabilityCaves2016,ValeryPRA2019}. Among the various sources of noise, partial distinguishability~\cite{BSAlg2018,SimulabilityMoylett2020,RenemaSimulability2025} and photon losses~\cite{BrodLoss2016, OszmaniecLoss2018,ValeryQuantum2019,SimulabilityMoylett2020} have been extensively studied as the dominant practical limitations. As distinguishability increases, the quantum features of multiphoton interference are progressively suppressed, and the detection statistics approach a classical probabilistic distribution that no longer involves interference.

The same quantum signature responsible for destructive interference in the HOM effect is also manifested in the multiphoton interference, resulting in zero transition amplitudes ${}_{\bf b}\langle {\bf n} | {\bf m} \rangle_{\bf a} = 0$, which are known as \textit{suppression laws}.  The permanent in Eq.~(\ref{permanent}) is a multivariate polynomial in the matrix elements of the interferometer $U_{kl}$. In principle, a direct derivation of suppression laws would require finding the roots of this polynomial, which is highly nontrivial, since the computation of the permanent itself becomes intractable for large values of $M$ and $N$, as discussed above. In previous works, Dittel \textit{et al.} derived a permutation symmetry principle for the derivation of suppression laws~\cite{Dittel1,Dittel2}, which was suggested as being the general mechanism underlying these effects. This method is based on the permutation symmetries of the interferometer, as well as of the input and output configurations and unifies several suppression laws previously derived for specific classes of interferometers:  discrete Fourier transform~\cite{LimFourier2005, TichyFourier2010,TichyFourier2012,CrespiFourier2016}, Sylvester matrices~\cite{CrespiSylvester2015,SciarrinoSylvester2018} and hypercube unitaries~\cite{DittelHypercube2017}. Let $\mathbb{S}_M$ the symmetric group of permutations of $M$ elements.  Let an input configuration ${\bf m}$ symmetric under a permutation $\sigma \in \mathbb{S}_M$, satisfying $\sigma({\bf m}) = {\bf m}$, and let the interferometer $U$ satisfying
\begin{equation}
    P_\sigma U = U \Lambda 
    ,
    \label{sym_in}
\end{equation}
where $\Lambda = (\lambda_1,...,\lambda_2)$ is a diagonal matrix containing the eigenvalues of of the permutation operator $P_\sigma$, defined by its action on a column vector as follows
\begin{equation}
    P_\sigma 
    \begin{pmatrix}
    x_1 \\
    \vdots \\
    x_M
    \end{pmatrix} 
    =
    \begin{pmatrix}
    x_{\sigma^{-1}(1)} \\
    \vdots \\
    x_{\sigma^{-1}(M)} 
    \end{pmatrix} .
\end{equation}
Any diagonal matrix contributing only global or external phases can be ignored. Under these conditions, according to~\cite{Dittel1,Dittel2}, the output configurations $\ket{{\bf n}}_{\bf b}$ satisfying $\lambda_1^{n_1} \cdots \lambda_M^{n_M} \neq 1$ are suppressed.

In this chapter, we employ an alternative approach based on recurrence relations satisfied by the matrix permanent, which are derived from a generating-function method. For a given output configuration with only a few photons occupying some of the output modes, the multivariate polynomial structure of the permanent simplifies and factorizes into a product of much simpler polynomials, whose roots can be readily determined. Within this framework, we derive new families of suppression laws for Fock-state interference in unitary multiports that are not captured by the permutation symmetry principle discussed above. This chapter is organized as follows: in Sec.~\ref{sec:Gen_Func_Fock} we present the generating-function formalism, in Sec.~\ref{Sec:Families_SuppressionLaws} we derive new families of suppression laws using this approach, and finally, in Sec.~\ref{sec:SupLaws_PartialDist} we investigate how the suppression laws obtained are affected by the partial distinguishability of the photons.

\section{Generating function and recurrence relations for  quantum  amplitudes}
\label{sec:Gen_Func_Fock}

We first recall that the $N$-photon  quantum amplitude between  two Fock states given by Eq.~(\ref{amplitude_permanent}) also has an  interesting   statistical interpretation in terms of a \textit{contingency table}~\cite{BookContTable}. For each photon, two labels are assigned: the input port number $k$, from which the photon originates, and the output port number $l$, where it is detected, the  same as  in Eq.(\ref{permanent}). The corresponding contingency table is then defined as the $M\times M$  matrix $S$, whose entries  give a partition of  $N$ photons by the two labels $k$ and $l$. It turns out that  the Fock state  amplitude  ${}_b\langle  \nb  |\mb\rangle_a$ is proportional to the   average over the contingency tables $S$ with fixed margins, $m_k= \sum_{l=1}^M S_{kl}$ and $n_l= \sum_{k=1}^M S_{kl}$~\cite{ValeryAssymptotic2013},
\begin{equation}
    {}_b\langle  \nb  |\mb\rangle_a = \frac{N!}{ \sqrt{\mb!\nb!}} \sum_{\{S \}} P(S |\mb,\nb) \prod_{k=1}^M \prod_{l=1}^M U_{kl}^{S_{kl}},
    \label{FY_amplitude}
\end{equation}
where  $ P(S |\mb,\nb) $ is   the    Fisher-Yates distribution for  two independent sets~\cite{BookContTable},
\begin{equation}
    P(S |\mb,\nb) = \frac{\binom{N}{S }}{\binom{N}{ \mb}\binom{N}{ \nb}}=\frac{1}{N!}\prod_{k=1}^M\prod_{l=1}^M\frac{ m_k!n_l!}{S_{kl}!}.
    \label{FY}
\end{equation}
The  multinomials $\binom{N}{ \mb}$, $\binom{N}{ \nb}$ and $\binom{N}{S }$  give, respectively,  the number of choices of $N$  photons for the input configuration, the output configuration, and for  a table    with given margins. To count the total number of  large-size contingency tables with fixed margins     is a hard computational problem   \cite{BookContTable}, in agreement with the hardness of the quantum amplitude given by the permanent \cite{BSAA}. Indeed the summation over the tables with fixed margins in Eq.~(\ref{FY_amplitude}) can be recast in terms of a summation over all possible $S_{kl} \geq 0$ constrained by the corresponding delta function on each margin,
\begin{equation}
    \sum_{ \{ S \} } = \sum_{\{ S_{kl} \geq 0 \}} \prod_{k=1}^M \prod_{l=1}^M \delta_{\sum_l S_{kl},m_k} \, \delta_{\sum_k S_{kl},n_l}
    \label{sum_ContingencyTable}
\end{equation}

Below we will prove that the averaging over the contingency tables with   fixed  margins, given in Eq.~(\ref{FY_amplitude}), can  be written in the form of partial derivatives of a generating function. Introducing the   formal variables, $\alpha_1, \, ... \, , \alpha_M$, the output margin constraints in Eq.~(\ref{sum_ContingencyTable}), $\sum^M_{k=1} S_{kl}=n_l$, can be recast in terms of the following partial derivatives
\begin{equation}
    \frac{1}{n_l!} \frac{\partial^{n_l}}{\partial \alpha_l^{n_l}} \left. \left( \prod^M_{k=1} \alpha^{S_{kl}}_l \right) \right|_{\alpha_l=0} = \delta_{\sum_k S_{kl},n_l}.
    \label{output_margin}
\end{equation}
Thus, writing the amplitude in Eq.~(\ref{FY_amplitude}) explicitly as the summation in Eq.~(\ref{sum_ContingencyTable}), with the output constraint written in terms of the derivatives, and simplifying the factorial terms of the Fisher–Yates distribution, we obtain
\begin{align}
    {}_b\langle  \nb  |\mb\rangle_a = & ~ \frac{\sqrt{\mb!}}{\sqrt{\nb!}} \sum_{\{ S_{kl} \geq 0 \}} \prod_{k=1}^M \prod_{l=1}^M \delta_{\sum_l S_{kl},m_k} \, \frac{\partial^{n_l}}{\partial \alpha_l^{n_l}} \left. \left( \frac{(U_{kl}\alpha_l)}{S_{kl}!}^{S_{kl}} \right) \right|_{\alpha_l=0} \nonumber\\
    %= & ~ \frac{1}{\sqrt{\nb!}} \left( \prod^M_{l=1} \frac{\partial^{n_l}}{\partial \alpha_l^{n_l}} \right) \prod^M_{k=1} \left. \left( \sum_{\{ S_{kl} \geq 0 \}} \delta_{\sum_l S_{kl},m_k} \prod_{l=1}^M \frac{\sqrt{m_k!}}{S_{kl}!} \left(U_{kl}\alpha_l\right)^{S_{kl}} \right) \right|_{{\bm \alpha}=0} \nonumber\\
    = & ~ \frac{1}{\sqrt{\nb!}} \left( \prod^M_{l=1} \frac{\partial^{n_l}}{\partial \alpha_l^{n_l}} \right)  \frac{1}{\sqrt{\mb!}} \prod^M_{k=1} 
    \Bigg( \sum_{\sum_l S_{kl}=m_k} \frac{m_k!}{\prod_l S_{kl}!} \left(U_{kl}\alpha_l \right)^{S_{kl}} \Bigg)
    \Bigg|_{{\bm \alpha} =0}  .
\end{align}
where in the last parentheses, we identify:
\begin{equation}
     \sum_{\sum_l S_{kl}=m_k} \frac{m_k!}{\prod_l S_{kl}!} \left(U_{kl}\alpha_l \right)^{S_{kl}} = \left(\sum^M_{l=1} U_{kl} \alpha_l \right)^{m_k}
     .
\end{equation}

The product over $k$ in the previous equation corresponds to a product of multinomial expansions,  each enforcing the respective input margin  $\sum^M_{l=1} S_{kl}=m_k$, and results in the generating function.  Finally, from the previous equation, the transition amplitude can be recast in a compact form,
\begin{equation}
    _b\langle { \bf n } | { \bf m } \rangle_a = \frac{1}{\sqrt{\nb!}} \left. \prod^M_{l=1} \frac{\partial^{n_l}}{\partial  \alpha_l^{n_l} } G_{{\bf m}}( {\bm \alpha} ) \right|_{{\bm \alpha} =0} 
    \label{GenFunc_amplitude}
\end{equation}
where the generating function is identified as
\begin{equation}
    G_{{\bf m}}( {\bm \alpha} ) = \frac{1}{\sqrt{\mb!}} \prod^M_{k=1} \left( \sum^M_{l=1} U_{kl}\alpha_l \right)^{m_k}.
    \label{GenFunc_Fock}
\end{equation}

The  function defined in the previous equation admits recurrence relations  between  the generating functions with different total numbers of photons $N$. For instance,  taking the  partial derivative over $\alpha_l$ we obtain
\begin{equation}
    \frac{\partial}{\partial \alpha_l} G_{{\bf m}} ( {\bm \alpha} ) = \sum^M_{k=1} \sqrt{m_k} U_{kl} G_{{\bf m} - {\bf 1}_k} ( {\bm \alpha} ) ,
    \label{recurrence_GenFunc}
\end{equation}
where $ {\bf m} - {\bf 1}_k \equiv (m_1,\ldots,m_k-1,\ldots,m_M)$ is the input configuration with one photon subtracted from the $k$-th input mode.   This recurrence relation is possible due to the absence of interactions between the photons, reflected in the  multi-linearity of the quantum amplitude of $N$ photons in Eq. (\ref{amplitude_permanent}) in the  one-photon amplitudes  $U_{kl}$:  the partial derivative over $ \alpha_l$ corresponds to  detection of a photon in the output mode $l$ in Eq.(\ref{GenFunc_amplitude}) and the sum over the  input modes $k$  is the  expansion over the generating functions  having one photon less in this mode, i.e. $G_{{\bf m} - {\bf 1}_k} ({\bm \alpha})$, with the coefficients being the one-photon amplitudes  $U_{kl}$.

Finally, we can also write the corresponding recurrence relation for the amplitude, rather than for the generating function by inserting Eq.~(\ref{recurrence_GenFunc}) in Eq.~(\ref{GenFunc_amplitude}), recovering the recurrence relation presented in Ref.~\cite{Quesada_GenFunc}. Indeed, on the phase space, this method can also be viewed through the simple inner product of two coherent states on a unitary operator, resulting in a direct generating function for the transition amplitudes~\cite{Quesada_GenFunc}. Alternatively, the transition probabilities of Fock states (rather than amplitudes) can be expressed through a generating function obtained from the inner product with a thermal state~\cite{Kim_GenFunc,Jabbour_GenFunc}.

\section{Families of suppression laws}
\label{Sec:Families_SuppressionLaws}

Let us now  focus on a specific output configuration, given by ${\bf n}=(n_1, {\bf n}_S)$ where we selected the mode $n_1 \geq 1$ with an arbitrary number of photons and  ${\bf n}_S=(n_2,...,n_M)$ has few photons. Following this method, for the output configuration ${\bf n}=(n_1, {\bf n}_S)$, we detect the photons in each mode in ${\bf n}_S$  using the  recurrence relation  of  Eq. (\ref{recurrence_GenFunc})  repeatedly   $n_l$ times for each one of these output modes. Following this procedure, we obtain  the amplitude $_b\langle {\bf n} | {\bf m} \rangle_a$ as a linear combination of amplitudes in the form
\begin{eqnarray}
    _b\langle n_1, {\bm 0}_S | {\bf m}'  \rangle_a &=&
    \frac{1}{\sqrt{n_1!}}  \left. \frac{\partial^{n_1}}{\partial \alpha_1^{n_1}}  G_{{\bf m}'} (\alpha_1,0,...,0) \right|_{\alpha_1 = 0} = \sqrt{\frac{n_1!}{{\bf m}'!}} \hspace{0.5mm} U^{m'_k}_{k1} ,
    \label{trivial_amp}
\end{eqnarray}
where ${\bf m}' $ is the input configuration with fewer photons,  subtracted according to Eq.~(\ref{recurrence_GenFunc}). Finally, factoring ${\bf m}!$ and the smallest order of $U_{kl}$, that is $m_k-|{\bf n}_S|$, we obtain the amplitude in the  form 
\begin{equation}
    _b\langle n_1, {\bf n}_S | {\bf m}  \rangle_a = \sqrt{ \frac{n_1!}{{\bf n}_S! \mb! }} \left( \prod^M_{k=1} U^{m_k- |{\bf n}_S|}_{k1} \right) f^{\bf n}_{\bf m}(U),
    \label{Form}
\end{equation}
where  $f^{\bf n}_{\bf m}(U) $ is referred to as the \textit{suppression function}, whose roots produce the non-trivial suppression laws. Here, non-trivial refers to suppression laws that do not result merely from some $U_{k1}=0$ on the previous equation. The suppression function is obtained by collecting the matrix elements that appear in Eqs.(\ref{recurrence_GenFunc}),(\ref{trivial_amp}) and the terms remaining in the factorization. This function is a multivariate polynomial in the parameters of the interferometers, namely transmissivities $\tau_j$ and phase $\varphi$ when convenient. Additionally, we can consider an arbitrary number of photons in the selected mode $n_1$ without increasing the calculation complexity, because the total photon number, $N=m_1+...+m_M$, is fixed by the input and consequently determines the photon number at this output, $n_1=N-|{\bf n}_S|$, a simplification already embedded in the structure of the generating function.

The method scales unfavorably with the total number of photons in the separated output modes ${\bf n}_S$, since an increase in photon number requires applying the recurrence relation multiple times to reach Eq.~(\ref{Form}), which ultimately falls back to the computational complexity of evaluating a permanent. Even so, the approach remains advantageous for non-large photon numbers ${\bf n}_S$. At the same time, the reduction procedure itself follows from the multilinearity of the amplitude and therefore remains formally valid for arbitrary photon numbers. As a result, even though the method does not allow direct analysis for arbitrarily large numbers of photons, one can still expect  the existence  of similar suppression laws. 

In order to illustrate the existence of suppression laws beyond the permutation symmetry principle~\cite{Dittel1,Dittel2} mentioned at the beginning of this chapter, we focus our analysis to output configurations with $|{\bf n}_S|=1,2$, and consider two-mode and three-mode interferometers, namely a beamsplitter and a tritter, shown in Fig.~\ref{interferometers23}. In addition, we find that the suppression laws can be grouped into distinct classes, which we refer to as \textit{families of suppression laws}, as each one lies along a smooth curve in the parameter space.

\subsection{Beamsplitter}
\label{subsec:beamsplitter}

We first apply our method to a beamsplitter, illustrated in Fig.~\ref{interferometers23}(a), with a free transmissivity $\tau$,
\begin{equation}
    B(\tau)   =
    \begin{pmatrix}
    \sqrt{\tau} & i \sqrt{1-\tau}  \\
    i \sqrt{1-\tau} & \sqrt{\tau}
    \end{pmatrix},
    \label{beamsplitter}
\end{equation}
which can be obtained from Eq.~(\ref{beamsplitter_general}). Note that, for Fock state inputs, the transmission and reflection phases of the beamsplitter become relevant only when combining in sequence with more optical elements, as done for the tritter. We also recall that this matrix represents a balanced beamsplitter when $\tau = 1/2$, which corresponds to the \textit{symmetric beamsplitter}, 
since it satisfies the symmetry operation associated with the group $\mathbb{S}_2$. This will be discussed in more detail in Subsec.~\ref{subsec:symmetry}.

%In addition, the   beamsplitter of Eq.(\ref{beamsplitter}) with an arbitrary $\tau$  can be realized by the  composition of a balanced beamsplitter, followed by two phase shifters and another balanced beamsplitter, where  the transmissivity    $\tau$ is controlled by the  phase shifters \cite{matrix1}.

Here we have outputs on the form ${\bf n}=(n_1, n_2)$ with $n_1 \geq 1$ and simply $|{\bf n}_S|=n_2$. To begin, we examine the case $n_2=1$, which is sufficiently simple for the detailed calculations be presented in the main text as a starting point. Applying Eq.(\ref{recurrence_GenFunc}) to the output mode $l=2$, we have the recurrence relation:
\begin{equation}
    \frac{\partial}{\partial \alpha_2} G_{(m_1,m_2)} ( \alpha_1, \alpha_2 ) = \sqrt{m_1} U_{12} G_{(m_1-1,m_2)} (\alpha_1, \alpha_2) + \sqrt{m_2} U_{22} G_{(m_1,m_2-1)} (\alpha_1, \alpha_2) , 
\end{equation}
thus, evaluating at $\alpha_2=0$ and using the definition of the generating function from Eq.~(\ref{GenFunc_Fock}), we obtain that
\begin{equation}
    \left. \frac{\partial}{\partial \alpha_2} G_{(m_1,m_2)} ( \alpha_1, \alpha_2 ) \right|_{\alpha_2=0} = \big( m_1 U_{12} U^{m_1-1}_{11} U^{m_2}_{21} + m_2 U_{22} U^{m_1}_{11} U^{m_2-1}_{21} \big) \frac{\alpha^{m_1+m_2-1}_1}{\sqrt{m_1! m_2!}}.
    \label{sup_BS_Jabbour_auxiliar}
\end{equation}
Finally, replacing the previous equation in Eq.~(\ref{GenFunc_amplitude}) we can easily write the transition amplitude in the form given by Eq.~(\ref{trivial_amp}),
\begin{align}
    _b\langle n_1, 1 | m_1, m_2  \rangle_a = & ~ \frac{1}{\sqrt{n_1! m_1! m_2!}} \, \big( m_1 U_{12} U^{m_1-1}_{11} U^{m_2}_{21} + m_2 U_{22} U^{m_1}_{11} U^{m_2-1}_{21} \big) \,
    \underset{= n_1!}{\underbrace{\frac{\partial^{n_1} \alpha^{m_1+m_2-1}_1}{\partial \alpha^{n_1}_1} \Bigg|_{\alpha_1=0} }} 
    \nonumber\\
    = & ~ \sqrt{ \frac{n_1!}{m_1! m_2!}} \, U^{m_1-1}_{11} U^{m_2-1}_{21} f^{(n_1,1)}_{(m_1,m_2)}\big(B(\tau)\big),
\end{align}
with the corresponding generating function:
\begin{align}
    f^{(n_1,1)} _{(m_1,m_1)} \big(B(\tau)\big) = & ~ m_1 B_{12} B_{21} + m_2 B_{11} B_{22} \nonumber\\ 
    = & ~ (m_1+m_2) \, \tau - m_1 .
\end{align}
It follows that the family of quantum amplitudes \mbox{${}_b\langle n_1, 1 | m_1 , m_2 \rangle_a=0$} for the transmissivities
\begin{equation}
    \tau^{(1)}  = \frac{m_1}{m_1+m_2},
    \label{sup_BS_Jabbour}
\end{equation}
recovering  the  suppression law presented in Ref.  \cite{Jabbour_GenFunc}, obtained by another method. The whole  family of such suppression laws also includes  the  HOM effect \cite{HOM1987} for the symmetric beamsplitter, i.e., $m_1=m_2=1$. As an additional result, we can also consider the amplitudes corresponding to the output configuration ${\bf n} = (1, n_2)$. By applying the recurrence relation given in Eq.~(\ref{GenFunc_Fock}) to the mode $l = 1$ and performing analogous calculations, we obtain the corresponding suppression function:
\begin{equation}
    f^{(1,n_2)} _{(m_1,m_1)} \big(B(\tau)\big) = (m_1+m_2) \tau - m_2 .
\end{equation}
Therefore, these new suppression laws have the same form, differing only by interchanging the input configurations $m_1$ and $m_2$, as expected. Already from these initial results, one can understand our method as: \textit{given the defined input and output configurations, one can determine which interferometer suppresses the corresponding transition amplitudes}.

In sequence, for $n_2 = 2$ we apply Eq.~(\ref{recurrence_GenFunc}) to the output mode $l = 2$ twice. After evaluating $\alpha_2 = 0$, we obtain an equation with the same structure as that given in Eq.~(\ref{sup_BS_Jabbour_auxiliar}), but with a more complicated polynomial inside the parentheses and with exponent $\alpha_1^{m_1 + m_2 - 2}$. Finally, we arrive at the corresponding suppression function,
\begin{eqnarray}
    f^{(n_1,2)} _{(m_1,m_2)} \big(B(\tau)\big) = (m_1 + m_2 -1) (m_1+m_2) \Bigl[  \tau^2  - \frac{2m_1}{m_1+m_2} \tau  + \frac{m_1(m_1-1)}{(m_1+m_2)(m_1 + m_2 -1)} \Bigr],
    \label{fun_b2}
\end{eqnarray}
giving  another  (previously unknown)  suppression  law   $\langle n_1, 2 | m_1 , m_2 \rangle=0$ for the  transmission 
\begin{equation}
    \tau^{(2)} =\frac{m_1}{m_1+m_2} \left(1  \pm \sqrt{\frac{ m_2/m_1 }{ m_1+m_2-1}}\right).
    \label{sup_BS_two}
\end{equation}
This family of  suppression  laws   also  contains  the symmetric beamsplitter $\tau^{(2)}=1/2$ for specific inputs, as, for example,  for four input photons  ${}_b\langle 2,2|1,3\rangle_a=0$  \cite{CamposSU2,4PhotonInterference1999}. Only such cases  can be explained by the permutation symmetry principle, where the  symmetry corresponds to the  transposition of the two output ports with $n_1=n_2=2$)\cite{Dittel1,Dittel2}. 

%The  above approach allows us  to derive all possible  suppression laws for the beamsplitter,   at least in principle.  However, the computations become quite involved  as the   numbers of bosons in  the input and output ports scale up.  Nevertheless, some general conclusions  are allowed by the  fact that  the quantum amplitudes ${}_b\langle n_1,n_2 |  m_1,m_2 \rangle_a$ on  a beamsplitter  can be made  real-valued  functions of its  transmission $\tau$ by removing the overall phase.  Numerical simulations    with various distributions of bosons  revealed that  the  number of   zeros in   a quantum amplitude   is given by  the minimum number of bosons $\mathrm{min}(n_l,m_k)$  in the four ports.   Moreover, two quantum amplitudes   related to the  exchange of a single boson  have interlaced zeros: between two zeros of one of them there is one  zero of the other, as shown in Fig. \ref{FIG2}  (at the end points,  $\tau =0 $ and $\tau = 1$,   a real-valued quantum amplitude  can be  either equal to zero or to  $\pm 1$, which  explains the above  bound on the total number of zeros).  

\subsection{Tritter} 

\begin{figure}[t]
\centering
\includegraphics[width=0.8 \textwidth]{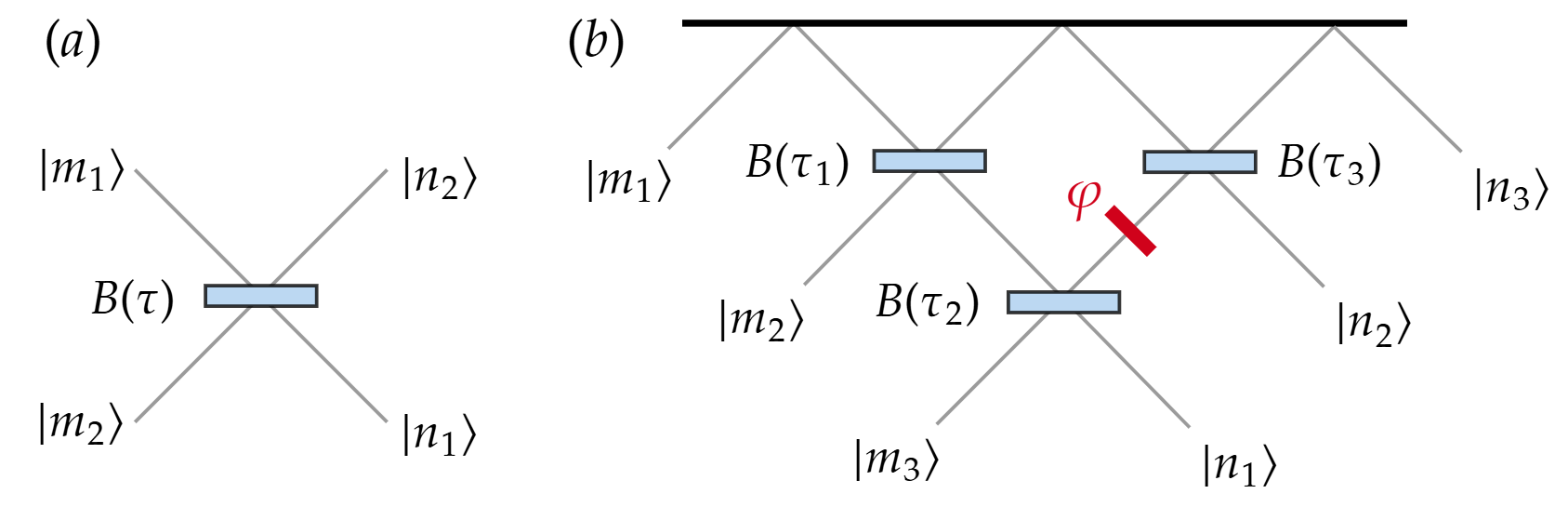}
\caption{Schematic representation of the two interferometers used to illustrate the proposed method: a) Beamsplitter, that transforms two input modes into two output modes; b) Tritter, that is a composition of three different beamsplitters $B_1$, $B_2$, $B_3$ and a control phase shifter $\varphi$. Here, each $m_k$ denotes the number of photons prepared in the input mode $k$ and $n_l$ denotes the number of photons detected in the output mode $l$.}
\label{interferometers23}
\end{figure}

In sequence, we apply our method to a tritter, which is constructed from a combination of three beamsplitters, a phase shifter, and a mirror. There are many realizations for such an interferometer, and here we considered the well-known triangular configuration~\cite{InterferometerRealization1994, InterferometerRealization1997, CamposSU3}, illustrated in Fig. \ref{interferometers23}(b). Each of the beamsplitters has a matrix  $B(\tau_k)$ as given in  Eq. (\ref{beamsplitter})  with  transmissivities $\tau_k$ and fixed reflection phase $\phi_r =\pi/2$. We also account for an additional phase plate with phase $\varphi$ inserted into one of the paths, and we consider the reflection phase $\chi$ imposed by the mirror. Thus, the full matrix of this tritter is constructed from the sequence action of the following matrices~\cite{CamposSU3}:
\begin{align}
    T(\{ \tau_k \},\varphi) = & ~ 
    \begin{pmatrix}
        \sqrt{\tau_1} & i \sqrt{1-\tau_1} & 0 \\
        i \sqrt{1-\tau_1} &  \sqrt{\tau_1} & 0 \\
        0 & 0 & 1 
    \end{pmatrix}
    \begin{pmatrix}
        1 & 0 & 0 \\
        0 & \text{e}^{i \chi} & 0 \\
        0 & 0 & 1
    \end{pmatrix}
    \begin{pmatrix}
        \sqrt{\tau_2} & 0 & i \sqrt{1-\tau_2} \\
        0 & 1 & 0 \\
        i \sqrt{1-\tau_2} & 0 & \sqrt{\tau_2}
    \end{pmatrix}
    \nonumber\\
    & ~ \times
    \begin{pmatrix}
        1 & 0 & 0 \\
        0 & 1 & 0 \\
        0 & 0 & \text{e}^{i \varphi}
    \end{pmatrix}
    \begin{pmatrix}
        1 & 0 & 0 \\
        0 & \sqrt{\tau_3} & i \sqrt{1-\tau_3} \\
        0 & i \sqrt{1-\tau_3} & \sqrt{\tau_3} 
    \end{pmatrix} .
    \label{construction}
\end{align}
From this point, we set the reflection phase of the mirror to $\chi = \pi$ for simplicity. Note that, for Fock state inputs, any phase becomes relevant only when inserted between some of these matrices, since a phase applied on either side contributes merely as a global factor that can be ignored. This matrix contains four free parameters, which makes a general analysis difficult and thus, we need to reduce the number of free parameters by fixing some of them.  To select an appropriate parametrization, we first recall that the balanced tritter is obtained by setting $\tau_1 = \tau_3 = 1/2$, $\tau_2 = 2/3$, and $\varphi = 0$~\cite{CamposSU3}, resulting in a $3$-mode Fourier matrix (also refereed to as a $3$-mode Bell multiport)~\cite{LimFourier2005, TichyFourier2010,TichyFourier2012,CrespiFourier2016}. Indeed, the balanced tritter also corresponds to the \textit{symmetric tritter}, for which the permutation symmetry principle applies~\cite{Dittel1,Dittel2}. In that case, this principle is based on the permutation symmetry operations of the symmetric group $\mathbb{S}_3$, which will be discussed in more detail in Subsec.~\ref{subsec:symmetry}. Since our goal is to reveal suppression laws that are not predicted by the permutation symmetry principle, we intentionally ``break'' this symmetry by varying two of the parameters, namely one transmissivity $\tau_k$ and the additional phase $\varphi$, while keeping the remaining two transmissivities $\tau_{j_1}$ and $\tau_{j_2}$ fixed according to the parametrization of the symmetric tritter.

For the interference in a tritter, the output configurations have the form ${\bf n} = (n_1, n_2, n_3)$ with $n_1 \geq 1$ and ${\bf n}_S= (n_2,n_3)$, where we focus on the cases $|{\bf n}_S| = 1, 2$. In contrast to the beamsplitter, the input configurations in this case admit many possible parametrization. For simplicity, we therefore focus our analysis on two particular classes of input states ${\bf m}^{(I)} = (n_1, 1, 1)$ with $n_1 \ge 1$ and ${\bf m}^{(II)} = (m, m, m)$ with $m \ge 1$. This choice allows us to examine explicitly the dependence of the suppression laws on the interferometer parameters and to facilitate comparison with the suppression laws predicted by the permutation symmetry principle.

Thus, following the discussion, we define a first class of tritter, obtained from Eq.~(\ref{construction}) by setting $\tau_2=2/3$ and $\tau_3=1/2$, which corresponds to the parametrization of the symmetric tritter, and letting $\tau_1 \in (0,1)$ and $\varphi \in [0, 2\pi]$ as the free parameters. Introducing the reflectivity $\rho_1 = 1 - \tau_1$, this tritter is represented by the following matrix
\begin{eqnarray}
    && T(\tau_1, \varphi)  =
    \frac{1}{\sqrt{6}}
    \begin{pmatrix}
    2 \sqrt{\tau_1}, & - \sqrt{\tau_1} e ^{i \varphi} - i \sqrt{3 \rho_1} ,& - \sqrt{\tau_1} e ^{i \varphi} + i \sqrt{3 \rho_1} \\
    2 \sqrt{\rho_1}, &  - \sqrt{\rho_1} e ^{i \varphi} + i \sqrt{3 \tau_1} ,  &  - \sqrt{\rho_1} e ^{i \varphi} - i \sqrt{3 \tau_1} \\
    \sqrt{ 2}, &  \sqrt{2} e^{i \varphi}, &  \sqrt{2} e^{i \varphi}
    \end{pmatrix}. 
    \label{rho1_theta}
\end{eqnarray}

\begin{figure}[t]
    \centering
\includegraphics[width=1 \columnwidth]{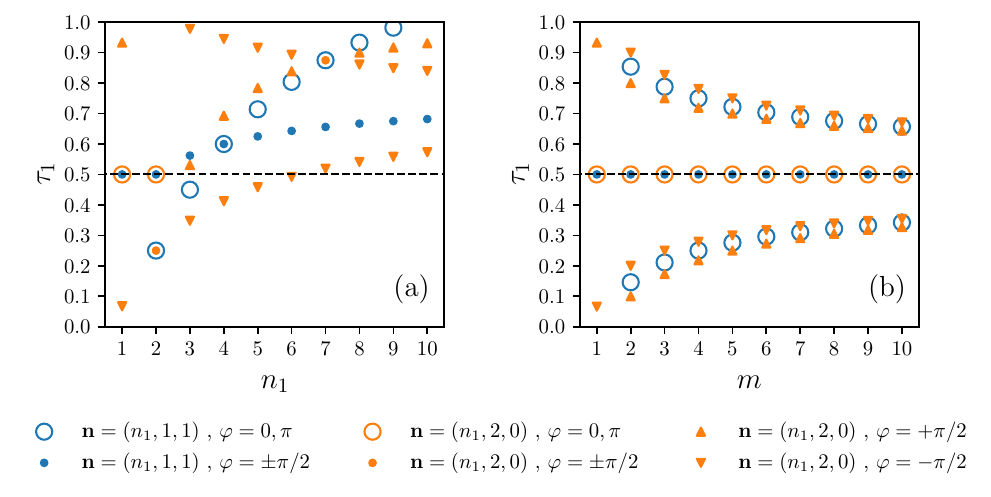}
\caption{Suppression laws for the tritter $T(\tau_1, \varphi)$ defined in Eq.~(\ref{rho1_theta}). In both graphs is shown value of the transmissivity $\tau_1$ that suppresses the corresponding amplitudes. Panel (a) presents the suppression laws for the amplitudes ${}_b\langle n_1, {\bf n}_S | n_1 , 1, 1\rangle_a=0$ while panel (b) presents the suppression laws for the amplitudes ${}_b\langle n_1, {\bf n}_S  | m , m, m \rangle_a=0$.}
\label{suppressionlaws_t1}
\end{figure}

Starting with the simplest case, $|{\bf n}_S| = 1$, with ${\bf n}_S = (1,0)$ or ${\bf n}_S = (0,1)$,  the recurrence relation given in Eq.~(\ref{recurrence_GenFunc}) is applied to the corresponding mode of each configuration, $l=2$ or $l=3$. Thus, following the procedure described above, the corresponding transition amplitudes have suppression functions given by:
\begin{equation}
    f^{(n_1,1,0)}_{m,m,m} \big(T(\tau_1, \varphi)\big) =  - f^{(n_1,0,1)}_{m,m,m} \big(T(\tau_1, \varphi)\big) =\frac{m}{3} (2 \tau_1 - 1) .
\end{equation}
The roots of these functions give, respectively, the suppression laws $_b\langle n_1,1,0  | m,m,m \rangle_a = 0$ and $_b\langle n_1,0,1 | m,m,m \rangle_a = 0$, which are imposed by setting $\tau_1=1/2$. It corresponds exactly to the symmetric tritter, and therefore no additional suppression laws appear beyond the permutation symmetry principle for this simple configuration. Now, considering the output configuration $|{\bf n}_S|=2$ with ${\bf n}_S = (1,1)$, the recurrence relation given in Eq.~(\ref{recurrence_GenFunc}) is applied once for each of the modes $l=2$ and $l=3$. Thus, for the input configurations ${\bf m}^{(I)} = (n_1, 1, 1)$ and ${\bf m}^{(II)} = (m,m,m)$, we have the  respective suppression functions:
\begin{equation}    
    f^{(n_1,1,1)}_{(n_1,1,1)} \big(T(\tau_1, \varphi)\big) = \frac{\sqrt{2}}{18} \Big[ \left( 4  \text{e}^{i 2 \varphi} + 3 (1+  \text{e}^{i 2 \varphi} ) n_1 + (3- \text{e}^{i 2 \varphi} )n^2_1 \right) \tau_1  - 3 n_1 (n_1 - 1 )   \Big] \sqrt{1-\tau_1}, 
\end{equation}
\begin{equation} 
    f^{(n_1,1,1)}_{(m,m,m)} \big(T(\tau_1, \varphi)\big) =  \frac{m}{9} \Big[ 2(2m+ \text{e}^{i 2 \varphi}-1)(\tau_1-1)\tau_1 + m -1 \Big] .
\end{equation}
Here, the suppression laws are identified from the roots of two functions, which establish a relation between the interferometer parameters $\tau_1$ and $\varphi$ as functions of the input parameters $n_1$ (for ${\bf m}^{(I)}$) or $m$ (for ${\bf m}^{(II)}$). The roots of the first function give the suppression laws $_b\langle n_1,1,1  | n_1,1,1 \rangle_a = 0$, plotted in blue at Fig.~\ref{suppressionlaws_t1}(a); whereas the roots of the second function give the suppression laws $_b\langle n_1,1,1  | m,m,m \rangle_a = 0$, plotted in blue at Fig.~\ref{suppressionlaws_t1}(b). 

Finally, for this tritter, we also consider the output configuration $|{\bf n}_S|=2$ with ${\bf n}_S = (2,0)$, where the recurrence relation in Eq.~(\ref{recurrence_GenFunc}) is applied twice for the mode $l=2$. In the same way, for the input configurations ${\bf m}^{(I)} = (n_1, 1, 1)$ and ${\bf m}^{(II)} = (m,m,m)$, we have the respective suppression functions:
\begin{align}
    f^{(n_1,2,0)}_{(n_1,1,1)} \big(T(\tau_1, \varphi)\big) = & ~ \frac{\sqrt{2}}{18} \left[ (4 \text{e}^{i 2 \varphi} + 3(\text{e}^{i 2 \varphi}-1)n_1-(3+\text{e}^{i 2 \varphi})n^2_1) \tau_1 +3n_1(n_1-1) \right] \sqrt{1-\tau_1}  \nonumber \\ 
    & + \frac{\sqrt{6}}{9} i \text{e}^{i \varphi} \left[ n_1(2-n_1) - (2+n_1-n^2_1) \tau_1 \right] \sqrt{\tau_1} , 
\end{align}
\begin{align}
    f^{(n_1,2,0)}_{(m,m,m)} \big(T(\tau_1, \varphi)\big) = & ~  \frac{m}{9} \left[ (4m-2-2 \text{e}^{i 2 \varphi})(1-\tau_1)\tau_1-m+1 \right]  \nonumber\\
    & +  \frac{2m}{27} i  \text{e}^{i \varphi} (2 \tau_1-1) \sqrt{3 \tau_1(1-\tau_1)}  .
    \label{f1_n20_mmm}
\end{align}
The roots of the first function give the suppression laws $_b\langle n_1,2,0  | n_1,1,1 \rangle_a = 0$, plotted in orange at Fig.~\ref{suppressionlaws_t1}(a); whereas the roots of the second function give the suppression laws $_b\langle n_1,2,0  | m,m,m \rangle_a = 0$, plotted in orange at Fig.~\ref{suppressionlaws_t1}(b). Some of the suppression functions have roots that can be calculated analytically, as shown in Table~\ref{table1}.

\begin{figure}[t]
    \centering
\includegraphics[width=1 \columnwidth]{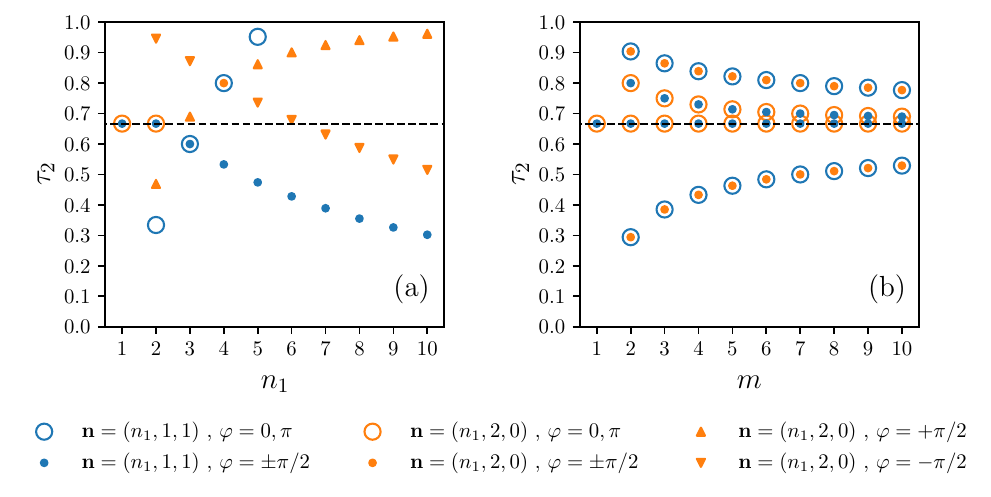}
\caption{Suppression laws for the tritter $T(\tau_2, \varphi)$ defined in Eq.~(\ref{rho2_theta}). In both graphs is shown value of the transmissivity $\tau_2$ that suppresses the corresponding amplitudes. Panel (a) presents the suppression laws for the amplitudes ${}_b\langle n_1, {\bf n}_S | n_1 , 1, 1\rangle_a=0$ while panel (b) presents the suppression laws for the amplitudes ${}_b\langle n_1, {\bf n}_S  | m , m, m \rangle_a=0$.}
\label{suppressionlaws_t2}
\end{figure}

For a more complete analysis, we also consider a second class of tritter, obtained from Eq.~(\ref{construction}) by setting $\tau_1=\tau_3=1/2$, which corresponds to the parametrization of the symmetric tritter, and letting $\tau_2 \in (0,1)$ and $\varphi \in [0, 2\pi]$ as the free parameters. It has the following matrix
\begin{equation}
    T(\tau_2, \varphi)   = \frac{1}{2}
    \begin{pmatrix}
    \sqrt{2 \tau_2}, & -i - \sqrt{\rho_2} e^{i \varphi}, & i - \sqrt{\rho_2} e^{i \varphi} \\
    \sqrt{2 \tau_2}, & i - \sqrt{\rho_2} e^{i \varphi} ,& -i - \sqrt{\rho_2} e^{i \varphi} \\
    2 \sqrt{ \rho_2},& \sqrt{2 \tau_2} e^{i \varphi}  ,& \sqrt{2 \tau_2} e^{i \varphi}
    \end{pmatrix}.
    \label{rho2_theta}
\end{equation}
where we introduce the reflectivity $\rho_2 = 1 - \tau_2$. Since the formal derivation is analogous to the previous case, differing only by the tritter, the explanation presented here will be more direct. We begin with the simplest case, $|{\bf n}_S| = 1$,  which have the possible transition amplitudes with the suppression functions:
\begin{equation}
    f^{(n_1,1,0)}_{m,m,m} \big(T(\tau_2, \varphi)\big) =   f^{(n_1,0,1)}_{m,m,m} \big(T(\tau_2, \varphi)\big) =\frac{m \sqrt{2}}{4} (3 \tau_2 - 2) \sqrt{\tau_2} \text{e}^{i \varphi} .
\end{equation}
In the same way as noticed in the previous case, it corresponds exactly to the symmetric tritter, and therefore no new suppression laws is obtained, since it falls into the permutation symmetry principle. 

Moving forward, considering the output configuration $|{\bf n}_S|=2$ with ${\bf n}_S = (1,1)$, we have the following suppression functions:
\begin{equation}
    f^{(n_1,1,1)}_{(n_1,1,1)} \big(T(\tau_2, \varphi)\big) = \frac{\sqrt{2}}{4} \Big[ \text{e}^{i 2 \varphi} \left( 2 + 3 n_1 + n^2_1 \right) \tau_2 + 3(1-\text{e}^{i 2 \varphi})n_1 - (1+\text{e}^{i 2 \varphi}) n_1^2 \Big] \sqrt{\tau_2 (1-\tau_2)},
\end{equation}
\begin{equation}
    f^{(n_1,1,1)}_{(m,m,m)} \big(T(\tau_2, \varphi)\big) = \frac{m}{8} \Big[ 3(3m-1)\text{e}^{i 2 \varphi} \tau^2_2-2 \left((6m-2)\text{e}^{i 2 \varphi}-1 \right) \tau_2 + (4m-2)\text{e}^{i 2 \varphi} - 2  \Big] \tau_2 .
\end{equation}
The roots of the first function give the suppression laws $_b\langle n_1,1,1  | n_1,1,1 \rangle_a = 0$, plotted in blue at Fig.~\ref{suppressionlaws_t1}(a); whereas the roots of the second function give the suppression laws $_b\langle n_1,1,1  | m,m,m \rangle_a = 0$, plotted in blue at Fig.~\ref{suppressionlaws_t1}(b). Finally, the last output configuration is $|{\bf n}_S|=2$ with ${\bf n}_S = (2,0)$, which give the following suppression laws:
\begin{align}
     f^{(n_1,2,0)}_{(n_1,1,1)} \big(T(\tau_2, \varphi)\big) = & ~  \frac{\sqrt{2}}{4} \left[ (2+3n_1+n^2_1) \text{e}^{i 2 \varphi} \tau_2 -\left( 3+ \text{e}^{i 2 \varphi} +  ( \text{e}^{i 2 \varphi}-1)n_1 \right) n_1 \right] \sqrt{\tau_2 (1-\tau_2)} \nonumber\\ 
     & ~ + \frac{1}{\sqrt{2}} i \text{e}^{i  \varphi} (1-n_1) \left[ n_1-(1+n_1)\tau_2 \right] \sqrt{\tau_2} ,
\end{align}
\begin{equation}
     f^{(n_1,2,0)}_{(m,m,m)} \big(T(\tau_2, \varphi)\big) =   \frac{m}{8} \left[ (9m-3)\text{e}^{i 2 \varphi} \tau^2_2  -2\left((6m-2)\text{e}^{i 2 \varphi}+1 \right) \tau_2 + 2 \left((2m-1)\text{e}^{i 2 \varphi}+1 \right) \right] \tau_2 .
     \label{f2_n20_mmm}
\end{equation}
The roots of the first function give the suppression laws $_b\langle n_1,1,1  | n_1,1,1 \rangle_a = 0$, plotted in orange at Fig.~\ref{suppressionlaws_t2}(a); whereas the roots of the second function give the suppression laws $_b\langle n_1,1,1  | m,m,m \rangle_a = 0$, plotted in orange at Fig.~\ref{suppressionlaws_t2}(b). Some of the suppression functions have roots that can be calculated analytically, as shown in Table~\ref{table1}.

\begin{table*}[h]
\caption{Suppression laws for the tritter.}
\renewcommand{\arraystretch}{1.5} % increases row height
\setlength{\tabcolsep}{8pt} % increases column spacing
\begin{tabular}{ccccc}
\toprule
 & $\theta = 0, \pi$ & $\theta = \pm \frac{\pi}{2}$ & $\theta = 0, \pi$ & $\theta = \pm \frac{\pi}{2}$ \\
\cmidrule(lr){2-5}
$_b\langle {\bf n} | {\bf m} \rangle_a$ & $\tau_1$ & $\tau_1$ & $\tau_2$ & $\tau_2$ \\
\midrule
$_b\langle n_1,1,0 | {\bf m}^{(II)} \rangle_a$ 
& $\tfrac{1}{2}$ 
& $\tfrac{1}{2}$ 
& $\tfrac{2}{3}$ 
& $\tfrac{2}{3}$ \\

$_b\langle n_1,1,1 | {\bf m}^{(I)} \rangle_a$ 
& $\tfrac{3 n_1 (n_1 - 1)}{2 (n_1 + 1)(n_1 + 2)}$ 
& $\tfrac{3 n_1}{4 (n_1 + 1)},\ n_1 \neq 1$ 
& $\tfrac{2 n_1 (n_1 - 1)}{(n_1 + 1)(n_1 + 2)}$ 
& $\tfrac{4 n_1}{(n_1 + 1)(n_1 + 2)}$ \\

$_b\langle n_1,2,0 | {\bf m}^{(I)} \rangle_a$ 
& $\tfrac{1}{2},\ n_1 = 1,2$ 
& --- 
& $\tfrac{2}{3},\ n_1 = 1,2$ 
& --- \\

$_b\langle n_1,1,1 | {\bf m}^{(II)} \rangle_a$ 
& $\tfrac{1}{2}\!\left(1 \pm \tfrac{1}{\sqrt{m}}\right)$ 
& $\tfrac{1}{2}$ 
& $\tfrac{2m - 1}{3m - 1} \pm \sqrt{\tfrac{12(4m - 1)}{6(3m - 1)}}$ 
& $\tfrac{2}{3},\ \tfrac{2m}{3m - 1}$ \\

$_b\langle n_1,2,0 | {\bf m}^{(II)} \rangle_a$ 
& $\tfrac{1}{2}$ 
& --- 
& $\tfrac{2}{3},\ \tfrac{2m}{3m - 1}$ 
& $\tfrac{2m - 1}{3m - 1} \pm \sqrt{\tfrac{12(4m - 1)}{6(3m - 1)}}$ \\
\bottomrule
\end{tabular}
\label{table1}
\end{table*}

\subsection{\label{subsec:symmetry} Suppression laws from the permutation symmetry}

Finally, we revisit the permutation symmetry principle derived in Refs.~\cite{Dittel1,Dittel2} and compare it with the results obtained from the generating function method presented here. In particular, we show that only a fraction of the suppression laws derived from our approach can be associated with this permutation symmetry principle. 

Starting with the beamsplitter, the permutation symmetries are analyzed under the group $\mathbb{S}_2 = \{ \mathbb{I}, (12) \}$. In this case, the symmetric beamsplitter corresponds to the standard balanced beamsplitter, given in Eq.(\ref{beamsplitter_balanced}). 
For the amplitudes $_b\langle n_1,1 | m_1,m_2 \rangle_a$, the only suppression laws related to the symmetry principle are those with $m_1=m_2$, resulting in $\tau=1/2$, according to Eq.~(\ref{sup_BS_Jabbour}). For the other amplitude considered $_b\langle n_1,2 | m_1,m_2 \rangle_a$, according to Eq.~(\ref{sup_BS_two}), only very specific suppression laws are related to this symmetry principle, for example ${}_b\langle 2,2|1,3\rangle_a$.

Considering the tritter, the permutation symmetries are analyzed under the group $\mathbb{S}_3=\{ \mathbb{I}, (12), (13), (23), (123), (132) \}$. The symmetric tritter, corresponding to the standard balanced tritter, is obtained either from Eq.~(\ref{rho1_theta}) with $T(\tau_1=1/2, \varphi=0)$, or from Eq.~(\ref{rho2_theta}) with $T(\tau_2=2/3, \varphi=0)$, resulting in
\begin{equation}
    T = \frac{1}{\sqrt{3}}
    \begin{pmatrix}
    1 & \e^{-i 2 \pi/3}  & \e^{i 2 \pi/3}  \\
    1 & \e^{i 2 \pi/3}  & \e^{-i 2 \pi/3}  \\
    1 & 1 & 1
    \end{pmatrix},
    \label{tritter_balanced}
\end{equation}
which is common choice of tritter on the literature, i.e., a $3$-mode Fourier matrix \cite{CollectivePhase1, DistinguishabilityAndMixedness}. The previous matrix is symmetric under the permutation operations $\sigma = (123),(321)$ in Eq.~(\ref{sym_in}). In Figs.~\ref{suppressionlaws_t1} and~\ref{suppressionlaws_t2}, the suppression laws arising from this permutation symmetry principle are those within the dashed lines, for $\varphi = 0$. This symmetry is manifested in our approach from a particular factorization of the corresponding suppression function, where one factor provides a fixed root associated with the symmetric configuration. It can be proven by setting $\varphi=0$ in Eqs.~(\ref{f1_n20_mmm}) and (\ref{f2_n20_mmm}), which results in
\begin{equation}
    f^{(n_1,2,0)}_{(m,m,m)} \big(T(\tau_1, 0)\big)  \overset{\varphi=0}{=} \frac{m}{27} \left[ 3(m-1)(2\tau_1-1)+2i \sqrt{3 (1-\tau_1) \tau_1} \right] (2 \tau_1-1) ,
    \label{123_1_n20}
\end{equation}
\begin{equation}
    f^{(n_1,2,0)}_{(m,m,m)}  \big(T(\tau_2, 0)\big)  \overset{\varphi=0}{=} - \frac{m}{8} \left[ (3m-1)\tau_2-2m  \right] (3 \tau_2 - 2) \tau_2 ,
    \label{123_2_n20}
\end{equation}
where the constant roots $\tau_1=1/2$ and $\tau_2=2/3$ correspond to the constant orange points in Figs.~\ref{suppressionlaws_t1}.(b) and~\ref{suppressionlaws_t2}.(b), and give the symmetric tritter in Eq.~(\ref{tritter_balanced}). 

Another tritter,  $\widetilde{T}$, possessing a different type of symmetry was also reported. This tritter corresponds to a real (orthogonal) matrix in a form similar to the usual balanced tritter  in Eq. (\ref{tritter_balanced}), which is obtained either from Eq.~(\ref{rho1_theta}) with $T(\tau_1=1/2, \varphi=\pi/2)$, or from Eq.~(\ref{rho2_theta}) with $T(\tau_2=2/3, \varphi=\pi/2)$,
\begin{equation}
    \widetilde{T} =  \frac{1}{\sqrt{3}}
    \begin{pmatrix}
    1 &- \frac{1+\sqrt{3}}{2} & \frac{-1+\sqrt{3}}{2} \\
    1 & \frac{-1+\sqrt{3}}{2} & -\frac{1+\sqrt{3}}{2} \\
    1 & 1 & 1
    \end{pmatrix} .
    \label{Trs2}
\end{equation}
Such matrix is  invariant under the simultaneous  permutation of rows $1$ and $2$ and  columns $2$ and $3$, which is not the  same  symmetry property described in Eq.~(\ref{sym_in}). The  suppression laws  for this tritter correspond to the blue points on the dashed lines in Figs.~\ref{suppressionlaws_t2}.(b) and Fig.~\ref{suppressionlaws_t2}.(b), resulting in $_b \langle n_1, 1, 1 | m, m, m \rangle_a =0$.

\section{Effects of the partial distinguishability on the suppression laws}
\label{sec:SupLaws_PartialDist}

In the previous discussion, we investigated suppression laws under the assumption of completely indistinguishable photons. However, as discussed at the beginning of this chapter in the context of the HOM effect, photon distinguishability affects quantum interference by degrading the destructive interference responsible for these suppression laws. 

%In Ref.~\cite{ValeryBS2}, it was conjectured that an exactly zero output probability in the presence of partial distinguishability results from an exact cancellation of transition amplitudes of a subset of completely indistinguishable photons, and that this condition is independent of the coherence between only partially indistinguishable photons. This conjecture generalizes the HOM effect to more than two photons and an arbitrary interferometer, which  is confirmed  in Refs.~\cite{TichyBS, Dittel1,Dittel2}. Let us now illustrate how the above conjecture applies to our framework.

Recall that photons are partially distinguishable due to the degrees of freedom not affected by the action of these interferometers. These degrees of freedom are encoded in the \textit{internal states} of the photons, denoted by $|\psi_k\rangle$, with $k=1,...,N$. We can also define an orthonormal basis for the internal states, $\{|e_1\rangle, \ldots, |e_D\rangle\}$, such that $|\psi_k\rangle = \sum_{s=1}^D \phi_{ks} |e_s\rangle$. It motivates us to define the creation and annihilation operators for the internal state, denoted respectively by $\hat{a}^\dagger_{\psi_k}$ and $\hat{a}_{\psi_k}$, of the photons:
\begin{equation}
    \hat{a}^\dagger_{\psi_k} \ket{0} = \ket{\psi_k}
    , \qquad
    \hat{a}_{\psi_k} \ket{\psi_k} = \ket{0},
\end{equation}
which generalizes the bosonic commutator algebra introduced in Eq.(\ref{creation_annihilation_def}),
\begin{align}
    & [\hat{a}^\dagger_{\psi_k}, \hat{a}^\dagger_{\phi_k}] = [\hat{a}_{\psi_k}, \hat{a}_{\phi_k}] = 0 ,
    \\
    & [\hat{a}_{\psi_k}, \hat{a}^\dagger_{\phi_k}] = \langle \psi_k | \phi_k \rangle .
    \label{creation_annihilation_def}
\end{align}
From that definition, we can trivially also introduce the creation and annihilation operators for the orthonormal basis of the internal state, $\hat{a}^\dagger_{s}$ and $\hat{a}_{s'}$, where that the last commutator becomes simply $[\hat{a}_{s}, \hat{a}^\dagger_{s'}] = \delta_{s,s'}$, in such a way that:
\begin{equation}
    \hat{a}^\dagger_{\psi_k} = \sum^M_{s=1} \phi_{ks} \hat{a}^\dagger_{k,s},
    \label{CreationOperator_InternalState}
\end{equation}

Thus, we generalize the input state given by Eq.~(\ref{Fock_input}), as follows:
\begin{equation}
     | \mb \rangle_{\bf a} = \prod_{k=1}^M \frac{(\hat{a}^\dagger_{\psi_k})^{m_k}}{\sqrt{m_k!}}|0\rangle ,
     \label{Fock_input_internalstate}
\end{equation}
In order to account for the effects of partial distinguishability on the photon-counting probabilities, the POVM $\hat{\Pi}_{\nb}$ used previously in Eq.~(\ref{born_rule}) also need to be generalized. Thus, we introduce a second index $s$ creation and annihilation operators, which correspond to the internal state basis $| e_s \rangle$, as discussed below. Finally, since the detectors are not able to resolve the internal states of the photons, we sum over the internal-state basis, resulting in the following POVM:
~\cite{ValeryBS1, ValeryBS2}
\begin{equation}
    \hat{\Pi}_{\bf n} =  \frac{1}{\bf n!} \sum_{s_1,...,s_N} \prod^N_{i=1} \hat{b}^\dagger_{l_i , s_i} | 0 \rangle \langle 0 |  \prod^N_{i=1} \hat{b}_{l_i , s_i},
    \label{POVM_Counting_PartialDist}
\end{equation}
where the summations on the second index $\{ s_i \}$ are performed over the internal state basis for each photon $ | e_{s_i} \rangle $ with $s_i=1, \, ... \, , j$. Note that this operator is written in the first quantization formalism, where the state is described in terms of the modes occupied by each photon; rather than in the second quantization formalism, where one specifies the number of photons in each mode. For a detailed discussion, see for instance Refs.~\cite{ValeryBS1, ValeryBS2}.

Consider now a simple case of input configuration in which one mode is occupied by a single photon that is partially distinguishable from the remaining $N-1$ photons. Thus, it is sufficient to consider only two orthogonal internal states, denoted by $|\phi\rangle \equiv |e_1\rangle$ and $|\phi_\perp\rangle \equiv |e_2\rangle$. Considering that the partially distinguishable photon is injected in the input mode ${k_N}$, the internal states of this photon can be written as $|\psi_N\rangle = \sqrt{1-\epsilon}\,|\phi\rangle + \sqrt{\epsilon}\,|\phi_\perp\rangle$; while the remaining photons have internal states $\ket{\psi_1}=\ket{\psi_2}=\,...\,=\ket{\psi_{N-1}} = |\phi\rangle$. Thus, from Eqs.~(\ref{Fock_input_internalstate}) and (\ref{CreationOperator_InternalState}), we define this state of $N$ as
\begin{equation}
    | \mb \rangle_{\bf a} = \frac{1}{\sqrt{\bf m!}} \left( \prod^{N-1}_{i=1} \hat{a}^\dagger_{k_i , \phi} \right) \left( \sqrt{1-\epsilon} \, \hat{a}^\dagger_{k_N, \phi} + \sqrt{\epsilon} \, \hat{a}^\dagger_{k_N, \phi_\perp} \right) | 0 \rangle ,
\end{equation}
where $k_i \neq k_N$ for any $i =1,2, \, ... \, , N-1$. Introducing $ {\bf n} - {\bf 1}_l = (n_1,\ldots,n_l-1,\ldots,n_M)$, from Eq.~(\ref{born_rule}), we obtain the following expression for the probability
\begin{equation}
    P_\epsilon({\bf n}) = \frac{1}{{\bf m}!{\bf n}!}  \sum_{s_1,...,s_N} \left|
    \langle 0 | \prod^N_{i=1} \hat{b}_{l_i , s_i} \prod^{N-1}_{i=1} \hat{a}^\dagger_{k_i , \phi} \left( \sqrt{1-\epsilon} \, \hat{a}^\dagger_{k_N,  \phi} + \sqrt{\epsilon} \, \hat{a}^\dagger_{k_N, \phi_\perp} \right) | 0 \rangle
    \right|^2 ,
\end{equation}
and since the two internal states are orthogonal, $\langle \phi | \phi_\perp \rangle = 0$, we have
\begin{align}
    P_\epsilon({\bf n})
    = & ~ \frac{1-\epsilon}{{\bf m}!{\bf n}!}  \left|
    \langle 0 | \prod^N_{i=1} \hat{b}_{l_i , \phi} \prod^{N}_{i=1} \hat{a}^\dagger_{k_i , \phi} | 0 \rangle
    \right|^2 +  \frac{\epsilon}{{\bf m}!{\bf n}!} \sum_{\bm{j}}  \left|
    \langle 0 | \prod^N_{i=1} \hat{b}_{l_i , s_i} \prod^{N-1}_{i=1} \hat{a}^\dagger_{k_i , \phi} \hat{a}_{k_N , \phi_\perp} | 0 \rangle
    \right|^2  \nonumber\\
    = & ~ (1-\epsilon) \left| _b\langle {\bf n} | {\bf m} \rangle_a \right|^2 +  \epsilon \sum^M_{l=1} |U_{{k_N} l}|^2 \left| _b\langle {\bf n}- {\bm 1}_l | {\bf m}- {\bm 1}_{k_N}  \rangle_a \right|^2 .
\end{align}

On the previous equation, The sum over $l$  has  $M$  non-negative   terms, each one being a product of two probabilities:   the probability  of the transition of one  distinguishable photon from the input mode ${k_N}$ to one   output mode $l$, i.e. $|U_{{k_N} l}|^2$, multiplied by the probability of  detecting  the remaining $N-1$ indistinguishable photons in  the reduced output, i.e. $ \left| _b\langle {\bf n}- {\bm 1}_l | {\bf m}- {\bm 1}_k  \rangle_a \right|^2$. Therefore, for zero  probability $P_\epsilon({\bf n})$,   all probabilities of detecting $N-1$ photons in the outputs  $ \nb-\bm{1}_l$  should be zero. Observe that the terms $_b\langle {\bf n}- {\bm 1}_l | {\bf m}- {\bm 1}_k  \rangle_a$ are generally non-zero in the above probability, thus we need to  employ the  recurrence relations also  for ${\bf n}- {\bm 1}_l$, which leads to some additional conditions.   Below we illustrate  how the  latter conditions lead to breaking of the suppression laws by considering two examples. 

As a first example, let us analyze how the partial distinguishability affects the suppression laws on a beamsplitter. Assuming that the partially distinguishable photon is injected in the input mode ${k_N}=1$, we arrive at the following probability:
\begin{align}
    P_\epsilon(n_1,1) = & ~ \epsilon \Big( |U_{11}|^2 | _b\langle n_1-1,1|m_1-1,m_2 \rangle_a |^2 + |U_{12}|^2 | _b\langle n_1,0|m_1-1,m_2 \rangle_a |^2 \Big)
    \nonumber\\
    & ~ + (1-\epsilon) | _b\langle n_1,1|m_1,m_2, \rangle_a |^2
\end{align}
where, according to Eq.~(\ref{sup_BS_Jabbour}), the first term is zero for $\tau=(m_1-1)/(m_1+m_2-1) $, while the last term is zero for $\tau = m_1/(m_1+m_2)$. The middle term is zero only in the trivial case when $\tau=0,1$, which we are interested. Therefore, the suppression law is indeed broken if we have a partially distinguishable photon,  because the three terms cannot be simultaneously zero for $\tau \neq 0,1$.

In sequence, we  consider the interference on the tritter $T(\tau_1, \varphi)$ given in Eq.~(\ref{rho1_theta}) with phase $\varphi=\pi/2$. If the partially distinguishable photon is injected in the input mode ${k_N}=1$, we have
\begin{align}
    P_\epsilon(n_1,1,1) = & \, \epsilon \Big( 
    |U_{11}|^2 | _b\langle n_1-1,1,1|n_1,1,0 \rangle_a |^2 + |U_{12}|^2  | _b\langle n_1,0,1|n_1,1,0 \rangle_a |^2 + \nonumber\\ 
    & \, +|U_{13}|^2  | _b\langle n_1,1,0|n_1,1,0 \rangle_a |^2 \Big)  + (1-\epsilon) | _b\langle n_1,1,1|n_1,1,1 \rangle_a |^2
\end{align}
where the first term is zero for $\tau_1=3n_1/4(n_1+1)$, according to Table \ref{table1}. By using the same procedure described in the previous section, the other three amplitudes are zero when, respectively, the following equations are satisfied:
\begin{eqnarray}
&& (n_1+1) \sqrt{\tau_1 (1-\tau_1)} + \sqrt{3}(n_1+1) \tau_1 - \sqrt{3} n_1 = 0,
\nonumber\\
&& (n_1+1) \sqrt{\tau_1 (1-\tau_1)} - \sqrt{3}(n_1+1) \tau_1 + \sqrt{3} n_1 = 0,
\nonumber\\
&& 4(n_1+1) \tau_1 \sqrt{1-\tau_1} - 3 \sqrt{1-\tau_1} = 0,
\end{eqnarray}
where the  last equation  leads to $\tau_1=1$ or $\tau_1=3/4(n_1+1)$,  but these are not solutions of the first two equations. Therefore, the probability $ P_\epsilon(n_1,1,1)$ cannot be zero when we have  a partially distinguishable photon in the input.

Therefore, these simple examples illustrate how partial distinguishability degrades the suppression laws derived by our method, cancelling the destructive interference that emerges from multiphoton interference.

%%%%%%%%%%%%%%%%%%%%%%%%%%%%%%%%%%%%%%%%%%%%%%%%%%%%%%%%%%%%%%%%%%%%%%%
%%%%%%%%%%%%%%%%%%%%%%%%%%%%%%%%%%%%%%%%%%%%%%%%%%%%%%%%%%%%%%%%%%%%%%%
%%%%%%%%%%%%%%%%%%%%%%%%%%%%%%%%%%%%%%%%%%%%%%%%%%%%%%%%%%%%%%%%%%%%%%%

\begin{center}
\myclearpage
\par
\end{center}

\chapter{Interference of single-mode squeezed states with partially distinguishable photons}
\label{chapter:GBS}

\begin{tcolorbox}[colback=gray!5,colframe=black,title={This chapter is based on the following publication:}]
\underline{M.~E.~O.~Bezerra} and V.~S.~Shchesnovich,  
``A generating-function approach to the interference of squeezed states with partial distinguishability'',   
arXiv Preprint (2026).  
\href{https://arxiv.org/abs/2602.00071}{DOI: 10.1088/1367-2630/acfa1e}.
\end{tcolorbox}

\section{Background}

In this chapter, we discuss the interference of single-mode squeezed states and connect it to \textit{Gaussian Boson Sampling} (GBS)~\cite{GBS,GBSdetailed}, an alternative to the traditional Boson Sampling (BS). The GBS relies on multiphoton interference generated by single-mode squeezed-vacuum inputs propagating through a multiport, followed by photon-number-resolving (PNR) detection at the output. A vacuum single-mode squeezed state $\ket{r_k}$, simply referred to as a squeezed state, is defined as the action of the single-mode squeezing operator $\hat{S}_k(r)$, given in Eq.~(\ref{sm_operator_def}), on the vacuum state, $\ket{r_k} = \hat{S}_k(r)\ket{0}$. The squeezing operator can be written in normally-ordered form as follows:~\cite{BookBarnett}:
\begin{align}
    \hat{S}_k (r) = & \,
    \exp \left[  \frac{1}{2} \tanh \, r_k \, \e^{i \theta} (\hat{a}^\dagger_k)^2 \right] 
    \,
    \exp \left[ \ln \left( \cosh \, r_k \right) \left( \hat{a}^\dagger_k \hat{a}_k + \frac{1}{2} \right) \right]
    \nonumber\\
    & \, \times
    \exp \left[-  \frac{1}{2} \tanh \, r_k \, \e^{-i \theta} \hat{a}^2_k \, \right] ,
\end{align}
and  acting on the vacuum results in
\begin{equation}
    | r_k \rangle 
    =
    \sqrt{1-|S_k|^2} \, \text{exp} \left[ \frac{S_k}{2} ( \hat{a}^\dagger_k )^2 \right] \ket{0}
        , \qquad
    S_k = \text{tanh} \, r_k \, \text{e}^{i \theta_k}.
    \label{smss_def}
\end{equation}
Therefore, the full input state is given by
$\ket{\mathbf{r}} = \bigotimes_{k=1}^{M} \ket{r_k},$
where each mode $k$ is prepared in a single-mode squeezed vacuum state with squeezing parameter $r_k$. In general, it is assumed that only $N$ modes are squeezed, while the remaining $M - N$ modes are left in the vacuum state, which corresponds to setting $r_k = 0$ for $k \in N+1,...,M$. This model has many advantages over the original BS. Indeed, the original BS requires $N$ single-photon sources, each typically generated via heralded parametric down-conversion with a very low success probability per pump pulse, making the overall preparation a highly nontrivial task. In contrast, squeezed states $\ket{r_k}$ can be readily produced with existing quantum-optical technology, by second-order \cite{GenerationSMSS1,GenerationSMSS2} and third-order nonlinear optical processes \cite{GenerationSMSS3}. This feature makes GBS considerably easier to scale up the number of modes and the average photon number in the experiment, as already demonstrated in several outstanding experimental realizations~\cite{Pan2020,Pan2023,Pan2025,Madsen2022,Paesani2019}.

In the formalism of GBS, the photon-counting probabilities $P(\mathbf{n})$ can be expressed using the phase-space methods introduced in Sec.~\ref{sec:Phase_Space}. In particular, the Glauber-Sudarshan $\mathcal{P}$-function associated with a Fock state $\ket{\mathbf{n}}$ is given by a highly singular distribution~\cite{BookBarnett}:
\begin{equation}
    \mathcal{P}_{\bf n} ({\bm \alpha}) = \prod^M_{l=1} \frac{\e^{|\alpha_l|^2}}{n_l!}\left(\frac{\partial^2}{\partial \alpha_l \partial \alpha^*_l}\right)^{n_l} \delta^{(2)}(\alpha_l)
    \label{P_function_fock}.
\end{equation}
Following this, denoting by $\mathcal{Q}_{\mathbf{r}}(\bm{\alpha})$ the Husimi
$\mathcal{Q}$-function of the input state
$\ket{\mathbf{r}}$,
the photon-counting probability can be written, using the phase-space integral given in Eq.~(\ref{trace_PhaseSpace}), as
\begin{align}
    P({\bf n}) 
    = \int d^2{\bm \alpha}  \, \mathcal{Q}_{{\bf r}} ({\bm \alpha})  \, \mathcal{P}_{\bf n} ({\bm \alpha})
    = \left. \prod^M_{l=1} \frac{1}{n_l!}\left(\frac{\partial^2}{\partial \alpha_l \partial \alpha^*_l}\right)^{n_l} F_{\bf r}({\bm \alpha})   \right|_{{\bm \alpha} = 0}
    ,
    \label{probability_PhaseSpace}
\end{align}
where, by introducing the complete phase-space coordinates
$\mathcal{A} \equiv (\alpha_1, \ldots, \alpha_M, \alpha_1^*, \ldots, \alpha_M^*)$,
we define the generating function~\cite{GBS,GBSdetailed}:
\begin{align}
     F_{\bf r}({\bm \alpha}) 
     \equiv \mathcal{Q}_{\bf r} ({\bm \alpha}) \, \e^{{\bm \alpha}^\dagger {\bm \alpha}} 
     = \frac{1}{\det[ \sqrt{\sigma(-1)]}} \, \exp \left[ 
     \frac{1}{2} \, \mathcal{A}^{t}
     \begin{pmatrix}
         B & 0_M \\
         0_M & \overline{B}
     \end{pmatrix}
     \mathcal{A} \, \right]
     ,
\end{align}
where $\sigma(-1)$ is the antinormally-ordered covariance matrix, as defined in Eq.~(\ref{CovMatrix_ordering}), and $B$ is the following $M \times M$ symmetric matrix:
\begin{equation}
    B = U  D_{\bf r}  U^t
    , \qquad
    D_{\bf r} = 
    \text{diag}(
    \tanh{r_1} \, \e^{i \theta_1}, \cdots, \tanh{r_M} \, \e^{i \theta_M})
    .
\end{equation}
Following the derivation presented in Refs.~\cite{GBS,GBSdetailed}, the resulting
probabilities in Eq.~(\ref{probability_PhaseSpace}) can be expressed in terms of the \textit{hafnian} of a $2N \times 2N$ matrix $B_{\mathbf{n}}$, obtained by repeating the rows
and columns of the symmetric matrix $B$ according to the output occupation
configuration $\mathbf{n}$,
\begin{equation}
    P({\bf n}) =
    \frac{ \text{haf} \left(B_{ \mb,\nb} \right)}{\sqrt{{\bf n}! \, \det[ \sigma(-1)]}},
    \label{amplitude_hafnian}
\end{equation}
which is given by~\cite{BookMinc,GBSdetailed}
\begin{equation}
    \mathrm{haf}(B)=\frac{1}{N!\,2^N}
    \sum_{\sigma\in \mathbb{S}_{2N}}
    \prod_{j=1}^{N} B_{\sigma(2j-1),\,\sigma(2j)},
    \label{eq:hafnian_def}
\end{equation}
Analogously to the permanent, the computation of the hafnian scales exponentially with the matrix size, rendering the problem intractable for classical computers at large values of $N$ and $M$ and providing the basis for the quantum advantage of
GBS, which has already been demonstrated experimentally~\cite{Pan2020,Pan2023,Pan2025,Madsen2022,Paesani2019}. In addition, GBS has applications beyond quantum-supremacy tasks, including problems in graph theory~\cite{GBSGraph2018,GBSGraph2018_2} and molecular vibronic simulations~\cite{GBSMolecule2015,GBSMolecule2020}, due to the role of the hafnian in counting perfect matchings.

As in any photonic device, the performance of GBS is also affected by noise, which
degrades the quantum signatures arising from multiphoton interference
~\cite{PatronSimulabilityGBS}. Similarly to standard boson sampling (BS), partial
distinguishability~\cite{RenemaSimulabilityGBS2020,Shi2020,QuesadaLoopHafnian} and
losses~\cite{PatronSimulabilityGBS,Shi2020,BressaniGBS2024,QuesadaLoopHafnian} constitute
some of the most significant sources of noise. In the case of standard BS, the effects
of partial distinguishability are already well understood, as they are characterized
by the \emph{overlap matrix}, defined in terms of the inner products of the internal states of the photons~\cite{distGramMatrix,ValeryBS2,TichyBS}. However, in contrast to Fock-state interference, the effects of partial distinguishability in squeezed-state interference have not yet been fully explored. Previous works have addressed distinguishability using virtual-mode \cite{Shi2022,GBSBinnedValidation} and coarse-grained \cite{QuesadaLoopHafnian} models, which are well suited for validation and simulability investigations. At the same time, such approaches do not capture the full structure of partial distinguishability effects in squeezed-state interference. While this problem can also be formulated in a general way using a first-quantization framework~\cite{RenemaSimulabilityGBS2020,ValeryGBS}, this description is not the most convenient when dealing with Gaussian states. From a complementary physical perspective, partial distinguishability has also been shown to arise from the intrinsic spectral mixedness of photons generated from the same source, being quantified by the Schmidt number of the spectral correlations~\cite{Christ_2011}.

In this chapter, we investigate the interference of squeezed states using the phase-space formalism, providing a natural and clear framework to treat partial distinguishability, quantified by the widely used overlap matrix of the internal states. This chapter is organized as follows. In Sec.~\ref{sec:genfunc_gaussian}, we revisit key results from Ref.~\cite{Valery2017} and extend them to the interference of single-mode squeezed states, deriving a general expression for the output probabilities of squeezed-state interference with partially distinguishable photons in arbitrary linear interferometers. We then recover the indistinguishable-photon limit and, finally, derive the corresponding expressions for threshold detectors. Following this, we present two applications. In Sec.~\ref{sec:application1}, we reduce our results to the simple model where the photons have homogeneous overlap, providing a clear physical interpretation of the noise effects emerging from partial distinguishability. Finally, in Sec.~\ref{sec:application2}, we investigate partial distinguishability from a more fundamental perspective,
including zero-probability events and the phase effects emerging from the internal state overlap.

%--------------------------------------------------------------------------------------------------------------------------------------------------------------------------%

\section{Probability events and Generating function}
\label{sec:genfunc_gaussian}

%In this section, we revisit the method introduced by Shchesnovich in Ref.~\cite{Valery2017} and provide a detailed discussion to clarify how partial distinguishability is incorporated into the phase-space formalism, in order to apply on the GBS problem in the next section. These results are essential for understanding how partial distinguishability affects the interference of Gaussian states in the quantum phase space and, moreover, provide an alternative method to that adopted in other works~\cite{RenemaSimulabilityGBS2020,ValeryGBS,Shi2020,QuesadaLoopHafnian}. 

To begin, we recall an alternative derivation of the photon-counting probabilities,
based on recasting the PNR operator defined in Eq.~(\ref{born_rule}) as a Poisson-like
operator~\cite{BookMandel,BookVogel}. Writing the density matrix of the state in the Glauber-Sudarshan
$\mathcal{P}$ representation according to Eq.~(\ref{Glauber_def}),
the photon-counting probability can be written as
\begin{align}
    P({\bf n}) = & ~ \Tr \left[
    \left( \int \frac{d^2{\bm \alpha}}{\pi^M} \, \mathcal{P} ({\bm \alpha}) \, | {\bm \alpha} \rangle \langle {\bm \alpha} | \right) | {\bf n} \rangle \langle {\bf n} | \right] \nonumber\\
    = & ~ \int d^2{\bm \alpha} \, \mathcal{P} ({\bm \alpha}) \prod^M_{l=1} \frac{|\alpha_l|^2}{n_l!} \e^{-|\alpha_l|^2} 
    ,
\end{align}
Identifying the second line as a normally ordered expectation value,as given in Eq.~(\ref{expectation_NormalOrdering}), we obtain
\begin{equation}
    P({\bf n}) = 
    \langle \, \mathcal{N} \big\{ \prod^M_{l=1} \frac{\hat{N}^{n_l}_l}{n_l!} \text{e}^{-\hat{N}_l} \big\} \, \rangle ,
    \label{probability_Poisson}
\end{equation}
where $\mathcal{N} \{ ... \}$ reads for the normal ordering operator. Here we introduce the detector efficiencies as $0 \leq \eta_l \leq 1$ and define the photon-number operator as 
\begin{equation}
    \hat{N}_l = \eta_l \sum^D_{s=1} \hat{b}^\dagger_{l,s} \hat{b}_{l,s} 
    .
    \label{General_PhotonNumber_Output}
\end{equation} 
Note that the efficiencies $\eta_l$ are equivalent to the loss parameters
introduced in Eq.~(\ref{loss_def}), since they quantify the probability that a
photon incident on the detector is actually registered as a detection event.
In the same way as for the PNR operator defined in Eq.~(\ref{POVM_Counting_PartialDist}),
in Eq.~(\ref{General_PhotonNumber_Output}) the creation and annihilation operators carry two
indices: the first labels the spatial output mode $l$ in which the photon is
detected, while the second index $s$ denotes the internal-state basis, which has
dimension $D$. We then perform the summation over the internal-state basis, since
the detectors are insensitive to the internal degrees of freedom of the photons. Thus, introducing the efficiency vector ${\bm \eta} = (\eta_1, ... ,\eta_M)$, the photon-counting probability can be recast in terms of a generating function as
\begin{equation}
    P({\bf n}) = \left. \prod^M_{l=1} \frac{\eta^{n_l}_l}{n_l!} \left( - \frac{\partial}{\partial \eta_l} \right)^{n_l} G({\bm \eta}) \right|_{{\bm \eta} = {\bm 1}} ,
    \label{probability_gen_func}
\end{equation}
where we assume full efficient detectors, evaluating the derivatives at ${\bm 1} = (1,\ldots,1)$, and the generating function is defined as
\begin{equation}
    G({\bm \eta}) \equiv \left\langle \mathcal{N} \left\{ \prod^M_{l=1} \text{e}^{-\hat{N}_l} \right\} \right\rangle  . 
    \label{vacuum_prob}
\end{equation}
Note that, here the probability is expressed in terms of derivatives with respect to the detector efficiencies ${\bm \eta}$, in contrast to GBS formulation, in Eq.~(\ref{probability_PhaseSpace}), where the formal variables of the generating correspond to the phase-space coordinates ${\bm \alpha}$ and ${\bm \alpha^*}$. In the following, we show that this method is useful and naturally incorporates the overlap matrix of the internal states into the structure of the generating function.

We now compute the generating function defined in Eq.~(\ref{vacuum_prob}). To begin,
we express the photon-number operator $\hat{N}_l$, given in
Eq.~(\ref{General_PhotonNumber_Output}), in the input-mode basis. The mode evolution
is straightforward, since the multiport implements a linear transformation between
the input and output modes. A direct application
of Eq.~(\ref{multiport_evolution}) gives:
\begin{equation}
    \hat{N}_l = \eta_l \sum^D_{s=1} \sum^M_{k_1=1} \sum^M_{k_2=1} U_{l,k_1} U^*_{l,k_2} \, \hat{a}^\dagger_{k_1,s} \hat{a}_{k_2,s} .
    \label{General_PhotonNumber_Output_1}
\end{equation}

For the internal states, we consider inputs defined in terms of the generalized
creation operator $\hat{a}^\dagger_{\psi_k}$, introduced in
Eq.~(\ref{creation_annihilation_def}). As discussed in
Sec.~\ref{sec:SupLaws_PartialDist}, this operator creates an internal state in each
input mode $k$ according to
\begin{equation}
    \hat{a}^\dagger_{\psi_k}|0\rangle \equiv | \psi_k \rangle = \sum_{s=1}^D \phi_{ks} |e_s\rangle ,
    \label{ketExpansion_InternalState}
\end{equation}
where $\mathcal{B} = \{\ket{e_1}, \ldots, \ket{e_D}\}$ denotes an orthonormal basis of
the internal-state Hilbert space.  From Eq.~(\ref{ketExpansion_InternalState}), we define the overlap matrix as the corresponding Gram matrix of the internal state of the photons, which provides a general
quantitative measure of partial distinguishability.
~\cite{distGramMatrix,ValeryBS2, TichyBS},
 \begin{equation}
	 V_{kj} \equiv \langle \psi_k | \psi_j \rangle = \sum^D_{s=1} \phi^*_{ks} \phi_{js} 
	\label{Gram_matrix}
 \end{equation}

In general, the set of internal states does not form a complete basis; consequently,
the canonical basis vectors cannot be expanded directly in terms of these states. For that, we need to complement the internal state of each input mode $k$ to a new basis, let us say $\widetilde{B}^{(k)} = \{ |\psi_k \rangle, | h^{(k)}_2 \rangle, \, ... \, , | h^{(k)}_D \rangle \}$, in such a way that we can establish a unitary relation between the two basis $B$ and $\widetilde{B}^{(k)}$. Each element of the latter set can be written as $|h^{(k)}_j\rangle = \sum^D_{s=1} \chi^{(k)}_{js} | e_s \rangle$, where $\chi^{(k)}_{1s} \equiv \phi_{ks}$ according to Eq.~(\ref{ketExpansion_InternalState}). This previous expansion can be equivalently written as $\hat{c}^\dagger_{k,j} = \sum^D_{s=1} \chi^{(k)}_{js} \hat{a}^\dagger_{k,s}$, where we introduce the creation operators for this new basis $\hat{c}^\dagger_{k,j} | 0 \rangle = | h^{(k)}_j \rangle$, with $\hat{c}^\dagger_{k1} \equiv \hat{a}^\dagger_{\psi_k}$. The coefficients $\chi^{(k)}_{js}$ of this expansion can be explicitly determined  from a orthonormalization process (e.g., Gram-Schmidt) following the procedure:
\begin{align}
    & | h^{(k)}_1 \rangle = | \psi_k \rangle \nonumber\\
    & | h^{(k)}_2 \rangle = | e_1 \rangle - \langle e_1 | h^{(k)}_1 \rangle | h^{(k)}_1 \rangle \nonumber\\
    & | h^{(k)}_2 \rangle = | e_2 \rangle - \langle e_2 | h^{(k)}_1 \rangle | h^{(k)}_1 \rangle - \langle e_2 | h^{(k)}_2 \rangle | h^{(k)}_2 \rangle \nonumber\\
    & ~ \vdots \nonumber\\
    & | h^{(k)}_D \rangle = | e_{D-1} \rangle - \sum^{D-1}_{j=1} \langle e_{D-1} | h^{(k)}_j \rangle | h^{(k)}_j \rangle
\end{align}
where in each line we perform the normalization $| h^{(k)}_j \rangle \mapsto | h^{(k)}_j \rangle / \langle h^{(k)}_j | h^{(k)}_j \rangle$. As a result, we can relate the creation operators defined in terms of each of the basis from the following relation:
\begin{equation}
    \begin{pmatrix}
        \hat{a}^\dagger_{\psi_k} \\
        \hat{c}^\dagger_{k,2} \\
        \vdots\\
        \hat{c}^\dagger_{k,D}
    \end{pmatrix}
    =
    \begin{pmatrix}
        \phi_{k1} & \phi_{k2} & \cdots & \phi_{kD} \\  
        \chi_{21} & \chi_{22} & \cdots & \chi_{kD} \\
        \vdots & \vdots & & \vdots \\
        \chi_{D1} & \chi_{D2} & \cdots & \chi_{DD}
    \end{pmatrix}
    \begin{pmatrix}
        \hat{a}^\dagger_{k,1} \\
        \hat{a}^\dagger_{k,2} \\
        \vdots\\
        \hat{a}^\dagger_{k,D}
    \end{pmatrix} .
\end{equation}
Finally, by construction, such a transformation is unitary, and then we can easily invert the relation between the two vectors and write
\begin{equation}
    \hat{a}^\dagger_{k,s} = \phi^*_{ks} \hat{a}^\dagger_{\psi_k} + \hat{d}^\dagger_k
    , \qquad
    \hat{d}^\dagger_k = \sum^D_{s=2} \chi^*_{ks} \hat{c}^\dagger_{k,s}.
\end{equation}
Note that, since there is only vacuum on the modes produced by $\hat{d}^\dagger_k$, i.e., $\langle \hat{d}^\dagger_k \hat{d}_k \rangle = 0$, we can ignore the contributions arising from this operator. Thus, replacing the previous equation in Eq.~(\ref{General_PhotonNumber_Output_1}) and summing over the output modes we have:
\begin{align}
    \sum^M_{l=1} \hat{N}_l = & ~ \sum^M_{l=1} \eta_l \sum^D_{s=1} \sum^M_{k_1=1} \sum^M_{k_2=1} U_{l,k_1} U^*_{l,k_2} \, \phi^*_{k1,s} \phi_{ks} \hat{A}^\dagger_{\psi_{k_1}} \hat{A}_{\psi_{k_2}} \nonumber\\
    =  & ~ \sum^M_{k_1=1} \sum^M_{k_2=1} \hat{A}^\dagger_{\psi_{k_1}} \left[ U \Lambda U^\dagger  \right]_{k_1,k_2} V_{k_1,k_2}  \hat{A}_{\psi_{k_2}} \nonumber\\
    = & ~ \hat{\bf a}^\dagger_{\bm \psi} \left(U \Lambda U^\dagger \circ V \right) \hat{\bf a}_{\bm \psi} \nonumber\\[0.8em]
    = & ~ \hat{\bf a}^\dagger_{\bm \psi} \left(I_M - H \right) \hat{\bf a}_{\bm \psi}
    \label{General_PhotonNumber_Output_2}
\end{align}
where we introduce the diagonal matrix of the detector efficiencies $\Lambda_{il} = (1-\eta_l) \delta_{il}$, the generalized creation operator vector $\hat{\bf a}^\dagger_{\bm \psi} = (\hat{a}^\dagger_{\psi_1}, \, ... \, , \hat{a}^\dagger_{\psi_M})$, and ``$\circ$'' denotes the Hadamard (or element-wise) product with the overlap matrix $V$ defined in Eq.~(\ref{Gram_matrix}). For convenience, we also define the following Hermitian matrix 
\begin{equation}
    H = U \Lambda U^\dagger \circ V .
    \label{def_H}
\end{equation}

In sequence, the expectation value in Eq.~(\ref{vacuum_prob}) can be easily evaluated by writing the operators in the anti-normal ordering, since it results in a simple integral containing the Husimi function. Thus, substituting Eq.~(\ref{General_PhotonNumber_Output_2}) at the exponent of the generating function given in Eq.~(\ref{vacuum_prob}), and generalizing the relation between normal and anti-normal ordering presented in Eq.~(\ref{normal_antinormal_relation}),
\begin{equation}
    \langle \, \mathcal{N} \big\{ \exp \left[\hat{\bf a}^\dagger (H-I_M)\hat{\bf a} \right] \big\} \, \rangle
    = \frac{1}{\det(H)}
    \langle \, \mathcal{A} \big\{ \exp \left[\hat{\bf a}^\dagger (I_M-H^{-1})\hat{\bf a} \right] \big\} \, \rangle ,
    \label{normal_antinormal}
\end{equation}
resulting in the generating function: 
\begin{equation}
    G({\bm \eta}) =  \frac{1}{\det(H)}
    \langle \, \mathcal{A} \big\{ \exp \left[-\hat{\bf a}^\dagger_{\bm \psi} (H^{-1}-I_M) \hat{\bf a}_{\bm \psi} \right] \big\} \, \rangle .
    \label{vacuum_prob_aux}
\end{equation}
We can formally extend the definitions of displacement operator and coherent state, given respectively by Eqs.~(\ref{displacement_op_def}) and (\ref{coherent_def}), to the corresponding creation and annihilation operators possessing internal states, as follows:
\begin{equation}
    \hat{D}_{\psi_k} (\alpha_k) = \exp \left( \alpha_k \hat{a}^\dagger_{\psi_k} - \alpha^*_k \hat{a}_{\psi_k} \right)
    , \qquad
    \hat{D}_{\psi_k} (\alpha_k) | \alpha_{\psi_k} \rangle = \alpha_k | \alpha_{\psi_k} \rangle .
    \label{displacement_coherent_generalized}
\end{equation}
Therefore, according to Eq.~(\ref{expectation_AntiNormalOrdering}), integral in the generating function $G({\bm \eta})$ can be calculated in terms of the Husimi function of the input state considered, resulting in the following general expression for the generating function:
\begin{equation}
    G({\bm \eta}) = \int d^2{\bm \alpha} \, Q({\bm \alpha}) \, \frac{\text{exp}\left[- {\bm \alpha}^\dagger \left(H^{-1} - I_M \right) {\bm \alpha} \right]}{\det(H)} .
    \label{gen_func_definition}
\end{equation}
Then, the matrix $H$ defined in Eq.~(\ref{def_H}) encodes the effects of internal-state
overlaps, state evolution, and detector efficiencies through the matrices $V$, $U$,
and $\Lambda$, respectively; while the Husimi function $Q(\bm{\alpha})$ fully encodes all the phase-space
representation of the input state.

%----------------------------------------------------------------------------------------------------------------------------------------------------------------------------------------------------------------------%

\section{Interference of single-mode squeezed states in $M$-mode multiports}
\label{sec:genfunc_smss}

\begin{figure}[t]
    \centering
\includegraphics[width=0.5 \columnwidth]{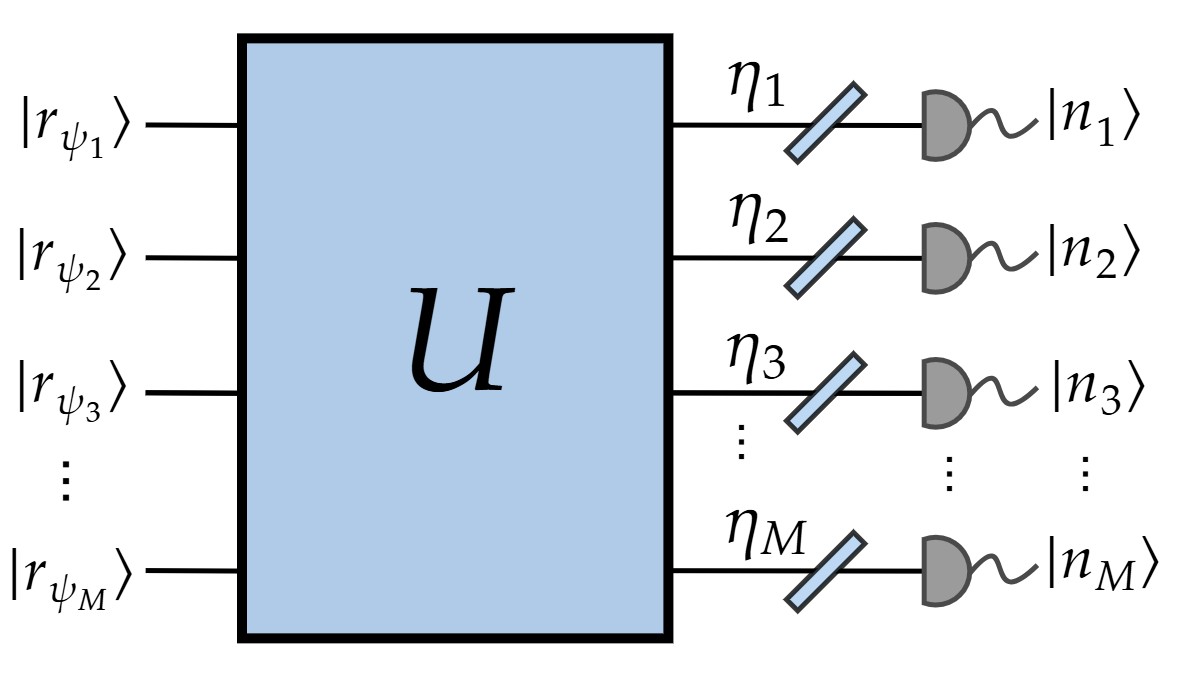}
\caption{Schematic representation of the $M$-mode GBS setup considered in this work. Each input mode $k=1,\ldots,M$ of a linear optical interferometer (multiport) described by a unitary matrix $U$ is injected with a single-mode squeezed vacuum state $|r_{\psi_k}\rangle$, whose photons occupy an internal state $|\psi_k\rangle$.  The output modes are measured by PNR detectors with efficiencies $\eta_1,\ldots,\eta_M$.}
\label{scheme_smss}
\end{figure}

We consider an $M$-mode linear optical interferometer, or multiport, described by a unitary matrix $U$. Each input mode $k = 1, \ldots, M$ is injected with a single-mode squeezed vacuum state $|r_{\psi_k}\rangle$, which we refer to simply as a squeezed state, as illustrated in Fig.~\ref{scheme_smss}. We assume that the photons
generated by the source in input mode $k$ occupy an internal state
$|\psi_k\rangle$. The overall $M$-mode input state is therefore given by
$|{\bf r}_{\bm \psi}\rangle = \bigotimes_{k=1}^M |r_{\psi_k}\rangle$, where the state in each mode is given by
\begin{equation}
    | r_{\psi_k} \rangle = \left(1-|S_k|^2\right)^{\frac{1}{4}} \, \text{exp} \left[ \frac{S_k}{2} ( \hat{a}^\dagger_{k, \psi_k} )^2 \right] |0 \rangle ,
    \label{smss_input}
\end{equation}
where $S_k = \tanh r_k\, e^{i\theta_k}$, with $r_k$ denoting the squeezing parameter and $\theta_k$ the squeezing phase. As commonly assumed in the GBS, we consider that the first $N$ modes of the interferometer are injected with
squeezed light, while the remaining $M-N$ modes are prepared in the vacuum state, by setting $r_k = 0$ for $k = N+1,\ldots,M$. The operator $\hat{a}^\dagger_{k,\psi_k}$ denotes the creation operator
associated with input spatial mode $k$, where the first index labels the spatial optical mode of the interferometer and the second index labels the corresponding internal state of the photon. 

The corresponding Husimi function is calculated by the inner product presented in Eq.~(\ref{Husimi_def}), where we use the definition of the coherent state $\ket{\alpha_{\psi_k}}$ given by Eq.~(\ref{displacement_coherent_generalized}), which results in
\begin{equation}
    Q_{\bf r}({\bm \alpha}) = \prod^M_{k=1} \frac{1}{\pi} |\langle \alpha_{\psi_k}  | r_{\psi_k} \rangle|^2 ,
    \label{hussimi_smss_def}
\end{equation}
with the following inner product 
\begin{equation}
    \langle \alpha_k | S_k \rangle = \sqrt{1-|S_k|^2} \, \text{exp} \left[-|\alpha_k|^2 - \frac{1}{2} \left( S_k \alpha^2_k  + c.c. \right) \right] .
    \label{overlap_coherent_smss}
\end{equation}
In sequence, it is convenient to explicitly introduce the complete vector of phase-space coordinates $\mathcal{A} = (\alpha_1, \ldots, \alpha_M, \alpha_1^*, \ldots, \alpha_M^*)$, such that for any $M \times M$ Hermitian matrix $X$, the following relation holds:
\begin{equation}
    {\bm \alpha}^\dagger X {\bm \alpha} =
    \frac{1}{2} \mathcal{A}^\dagger
    \begin{pmatrix}
        X & 0_M \\
        0_M & X^*
    \end{pmatrix}
    \mathcal{A}.
    \label{identity_phasespace_coordinates}
\end{equation}
Thus, defining the global normalization coefficient $c_{\bf r} = \prod^M_{k=1} \sqrt{1-|S_k|^2}$, the diagonal matrix $D_{\bf r} = \text{diag}(S_1,...,S_M)$ that encodes the squeezing parameters of each mode, and using Eqs.~(\ref{hussimi_smss_def}) and (\ref{overlap_coherent_smss}), we obtain the Husimi function for $M$ squeezed states at the input~\footnote{An alternative derivation consists in considering directly the function
$\mathcal{W}_{\mathbf{r}}(\bm{\alpha},-1)$, defined in
Eq.~(\ref{GaussianState_general_form}), which can be obtained by applying the symplectic matrix of the squeezing operator $\hat{S}_k$, given in Eq.~(\ref{sm_squeezing_matrix}), to the covariance matrix of an $M$-mode
vacuum state.}:
\begin{eqnarray}
    Q_{\bf r}({\bm \alpha}) = \frac{c_{\bf r}}{\pi^M}~\text{exp} \left[- \frac{1}{2} \mathcal{A}^\dagger
    \begin{pmatrix}
        I_M & D^*_{\bf r} \\
        D_{\bf r} & I_M
    \end{pmatrix}
    \mathcal{A} \, \right] .
    \label{hussimi_smss}
\end{eqnarray}

The antinormally-ordered covariance matrix $\boldsymbol{\sigma}(-1)$ follows directly from Eq.~(\ref{hussimi_smss}). By applying the ordering-conversion relation given in Eq.~(\ref{CovMatrix_ordering}), we then obtain the usual symmetrically-ordered covariance matrix for $M$ single-mode squeezed states as:

\begin{equation}
    \sigma =    
    \frac{1}{2}
    \begin{pmatrix}
        I_M & D^*_{\bf r} \\
        D_{\bf r} & I_M
    \end{pmatrix}
    + \frac{1}{2} I_{2M}.
\end{equation}

In order to calculate the corresponding generating function, we first rewrite ${\bm \alpha}^\dagger \left(H^{-1} - I_M \right) {\bm \alpha}$ in terms of the coordinates $\mathcal{A}$, according to Eq.~(\ref{identity_phasespace_coordinates}). Then, replacing the previous equation in Eq.~(\ref{gen_func_definition}), we arrive at the following integral
\begin{align}
    G({\bm \eta}) = & ~ \frac{c_{\bf r}}{\text{det}(H)} \int \frac{d^2{\bm \alpha}}{\pi^M} ~ \text{exp}\left[ - \frac{1}{2} \mathcal{A}^\dagger 
    \begin{pmatrix}
        H^{-1} & D^*_{\bf r} \\
        D_{\bf r} & (H^*)^{-1}
    \end{pmatrix}
    \mathcal{A} \right] \nonumber\\
    = & ~ \frac{c_{\bf r}}{\text{det}(H)} 
    \, \text{det}
    \begin{pmatrix}
        H^{-1} & D^*_{\bf r} \\
        D_{\bf r} & (H^*)^{-1}
    \end{pmatrix}^{-1/2} ,
    \label{gen_func_smss_aux1}
\end{align}
where, using the fact that $H^*=H^t$, the determinant can be simplified as follows:
\begin{align}
    \text{det}
    \begin{pmatrix}
        H^{-1} & D^*_{\bf r} \\
        D_{\bf r} & (H^*)^{-1}
    \end{pmatrix} 
    = & ~ \det(H)^{-1} \det \big((H^*)^{-1} - D_{\bf r} H D^*_{\bf r}\big) \nonumber\\
    = & ~ \det(H)^{-2}  \det \big(I_M - H^*D_{\bf r} H D^*_{\bf r} \big) \nonumber\\[0.5em]
    = & ~ \det(H)^{-2} \det \Big[I_M - ( \sqrt{D^*_{\bf r}} H^*D^{1/2}_{\bf r}) (\sqrt{D_{\bf r}} H \sqrt{D^*_{\bf r}}) \Big] \nonumber\\[0.5em]
    = & ~ \det(H)^{-2} \det \big(I_M - H^*_{\bf r}  H_{\bf r} \big) \nonumber\\[0.5em]
    = & ~ \det(H)^{-2} \det 
    \begin{pmatrix}
        I_M & H_{\bf r} \\
        H^*_{\bf r} & I_M
    \end{pmatrix}
    \label{gen_func_smss_aux2}
\end{align}
where from the second to the third line we have permuted the factor $\sqrt{D_{\bf r}}$ according to Sylvester’s determinant identity, and in the fourth line, we introduced the following Hermitian matrix
\begin{equation}
    H_{\bf r} = \sqrt{D_{\bf r}} \, H \, \sqrt{D^*_{\bf r}} .
    \label{H_r}
\end{equation}
Finally, combining the Eqs.~(\ref{gen_func_smss_aux1}) and ~(\ref{gen_func_smss_aux2}), we arrive in the final expression of the generating function
\begin{equation}
    G_{\bf r}({\bm \eta}) 
    = c_{\bf r} \, \text{det}
    \left( I_{2M} - \mathcal{M}_{\bf r} \right)^{-1/2},
    \label{gen_func_smss} 
\end{equation}
where we have defined the Hermitian matrix:
\begin{equation}
    \mathcal{M}_{\bf r}=
    \begin{pmatrix}
    0_M & H_{\bf r}\\
    H^*_{\bf r} & 0_M
    \end{pmatrix}.
    \label{H_r}
\end{equation}

\begin{tcolorbox}[colback=gray!5,colframe=black,title={Code Availability}]
The code for this work is available at \href{https://github.com/Matheus-Eiji/smss_partial_dist}{GitHub Repository}.
\end{tcolorbox}

%----------------------------------------------------------------------------------------------------------------------------------------------%

\vspace{10mm}
\subsection{\label{subsec:hafnian} Output probability for indistinguishable photons and the relation with the hafnian}

For consistency, we reduce our general result to the limit of completely indistinguishable photons in order to establish a connection with previous results. In this limit, the output probability is expressed in terms of a matrix hafnian, which lies at the core of the computational hardness of GBS~\cite{GBS,GBSdetailed}. By making use of the Hafnian Master Theorem recently proven in Ref.~\cite{HafnianMasterTheorem}, we recover the standard hafnian expression for the output probabilities, also in agreement with the generating function derived in Ref.~\cite{QuesadaMoments}. In this way, we assume that all photons occupy the same internal state, so that the overlap matrix satisfies $V_{ij}=1$ for all $i,j$. Consequently, the Hermitian matrix in Eq.~(\ref{H_r}) becomes
\begin{equation}
    \mathcal{M}^0_{\bf r}=
    \begin{pmatrix}
    0_M & H^0_{\bf r}\\
    (H^0_{\bf r})^* & 0_M
    \end{pmatrix}
    ,
    \quad
    H^0_{\bf r} \equiv \big. H_{\bf r} \big|_{V_{ij}=1}
    ,
    \label{H_r_ind}
\end{equation}
which follows that $H^0_{\bf r} = \sqrt{D_{\bf r}} U \Lambda U^\dagger \sqrt{D^*_{\bf r}}$.

After properly decomposing $H_{\bf r}^{0}$ in terms of the matrices $\Lambda$, $U$, and $D_{\bf r}$, and reordering the matrices inside the determinant using Sylvester's identity, we can rewrite the determinant in Eq.~(\ref{gen_func_smss}) as
\begin{align}
    \text{det}
    (I_{2M} - \mathcal{M}^0_{\bf r}) 
    & =
    \text{det}
    \left[ I_{2M} -
    \begin{pmatrix}
        0_M &  \Lambda \\
        \Lambda & 0_M
    \end{pmatrix}
    \begin{pmatrix}
        B & 0_M \\
        0_M & \overline{B}
    \end{pmatrix} 
    \right] 
    \label{gen_func_ind_aux}
\end{align}
where in the last line we introduced the symmetric matrix $B = U^t D_{\bf r} U$, which recovers the generating function reported in Ref.~\cite{QuesadaMoments} with a different parametrization and zero displacement. Following the Hafnian Master Theorem~\cite{HafnianMasterTheorem}, we recognize in
Eq.~(\ref{gen_func_ind_aux}) the generating-function structure corresponding to the hafnian of the matrix $B_{\mathbf n}\oplus\overline{B}_{\mathbf n}$,
\begin{equation}
    \text{det}
    (I_{2M} - \mathcal{M}^0_{\bf r})^{-1/2}  =  \sum_{\bf n} \big| \text{haf}(B_{\bf n}) \big|^2 \prod^M_{k=1} \frac{(1-\eta_k)}{n_k!}^{n_k} ,
    \label{gen_func_ind}
\end{equation}
where the $|{\bf n}|\times |{\bf n}|$ matrix $B_{\bf n}$ is obtained from the $M\times M$ matrix $B$ by repeating each $l$-th row and column $n_l$ times. Additionally, within the coarse-grained model of distinguishability, the internal modes can be partitioned into interfering and noninterfering subspaces, such that the resulting output probabilities are expressed in terms of a blocked loop hafnian~\cite{QuesadaLoopHafnian}.

Finally, by inverting Eq.~(\ref{gen_func_ind}) and using the general expression
for the probability in Eq.~(\ref{probability_gen_func}), we recover the standard GBS probability for indistinguishable photons with fully efficient detectors~\cite{GBS,GBSdetailed},
\begin{align}
    P^0_{\bf r} ({\bf n})
    &= c_{\bf r} \left. \prod^M_{k=1} \frac{\eta^{n_k}_k}{n_k!} \left( - \frac{\partial}{\partial \eta_k} \right)^{n_k} \text{det}
    (I_{2M} - \mathcal{M}^0_{\bf r})^{-1/2} \right|_{{\bm \eta} = {\bm 1}} 
    = \, \frac{c_{\bf r}}{\prod^M_{k=1}n_k!} \, \big| \text{haf}(B_{\bf n}) \big|^2 .
    \label{probability_hafnian}
\end{align}

%----------------------------------------------------------------------------------------------------------------------------------------------%

\vspace{5mm}
\subsection{\label{subsec:ClickProb} Threshold detectors}

As an alternative to the photon-number-resolving detection considered above, we also consider threshold detectors, which gives a binary outcome: either no photons are detected (no-click) or one or more photons are detected (click). Such a scheme was first proposed in Ref.~\cite{QuesadaGBSClick} and has since been adopted as the detection strategy in the main experimental demonstrations of Gaussian boson sampling~\cite{Pan2020,Pan2023}. For perfectly indistinguishable photons, the corresponding output probabilities are expressed in terms of the matrix function known as the torontonian, which is believed to be classically hard to compute in the low-collision regime ~\cite{QuesadaGBSClick}. In the following, we show that, using our method, we can express the probability of such click events in terms of the generating function defined in Eq.~(\ref{gen_func_smss}).

To begin, denoting the vacuum POVM element in the output mode $k$ as
\begin{equation}
    \hat{\Pi}_{0_k}=|0_k\rangle\langle 0_k|
    ,
    \label{probability_no_click}
\end{equation}
it follows directly from
Eqs.~(\ref{probability_gen_func}) and~(\ref{vacuum_prob}) that the probability of detecting no photon in the corresponding output mode is given by
\begin{equation}
    P_{\bf r}(0_k) \equiv \langle \hat{\Pi}_{0_k}  \rangle = \left. G_{\bf r}({\bm \eta}) \right|_{\eta_k=1}
    .
    \label{}
\end{equation}
In this way, a click event in the output mode $k$, corresponding to the detection of one or more photons, is defined as the complement of the vacuum event and is described by the POVM element
\begin{equation}
    \hat{\Pi}_{l_k} = \hat{I}_k - \hat{\Pi}_{0_k} ,
    \label{POVM_click}
\end{equation}
where $\hat{I}_k$ is the identity operator in the corresponding mode $k$. Note that, from Eq.~(\ref{gen_func_smss}), we can readily identify how the
identity operator $\hat{I}_k$ acts on each mode via the property
$\left. G_{\bf r}({\bm \eta}) \right|_{{\bm \eta} = {\bm 0}} = 1$.
Therefore, it follows that the probability of a click event is given by
\begin{equation}
    P_{\bf r}(l_k) \equiv \langle \hat{\Pi}_{l_k}  \rangle = \left. G_{\bf r}({\bm \eta}) \right|_{\eta_k=0} -  \left. G_{\bf r}({\bm \eta}) \right|_{\eta_k=1}.
    \label{probability_click}
\end{equation}

We now consider the event in which $N$ clicks are registered in the set of output modes $L\equiv\{k_1,\ldots,k_N\}$, where each $k_j\in\{1,\ldots,M\}$ labels a distinct mode in which a click is detected, while in the remaining output modes no photons are detected. We denote such a configuration by ${\bf l}=(l_{k_1},\ldots,l_{k_N})$.  In this way, it is convenient to introduce the reduced generating function $G_{\bf r}({\bm \eta}')$, defined by setting $\eta'_k=\eta_k$ if $k \in L$ and
$\eta'_k=1$ if $k \notin L$, where the latter accounts for the vacuum detections according to Eq.~(\ref{probability_no_click}). Therefore, the probability of detecting the corresponding click-event configuration ${\bf l}$ follows directly from Eqs.~(\ref{POVM_click}) and~(\ref{probability_click}) as
\begin{align}
    P_{\bf r}({\bf l}) & \equiv \langle \prod^M_{k=1} \hat{\Pi}_{l_k} \rangle 
    \nonumber\\
    &= \sum_{S  \subseteq L} (-1)^{|S|} \, \langle  \prod_{k \in S} \hat{\Pi}_{0_k} \prod_{k \notin S} \hat{I}_k  \rangle 
    \nonumber\\
    &= \sum_{S  \subseteq L} (-1)^{|S|} \, G_{\bf r}({\bm \eta}') \Big|_{\substack{\eta'_k = 1, k \in S \\ \eta'_k = 0, k  \notin S}}
    \label{probability_click}
    ,
\end{align}
where the sum over $S\subseteq L$ runs over all subsets of $L$, that is, over all possible choices of modes selected from $L$. Therefore, the equation~(\ref{probability_click}) extends the result of Ref.~\cite{QuesadaGBSClick} beyond the fully indistinguishable case, providing an explicit expression for click probabilities in the presence of partial distinguishability.

%----------------------------------------------------------------------------------------------------------------------------------------------------------------------------------------------------------------------%

\section{\label{sec:application1} Internal states with homogeneous overlap}

In this section, we specialize our general framework to the homogeneous distinguishability model, which provides a useful way to investigate partial distinguishability as an effective noise source in the boson sampling schemes~\cite{BSAlg2018, Shi2022, GBSBinnedValidation}. This enables us to
investigate how our formalism operates within this model and cleanly isolate the contribution of distinguishability to the output statistics. Thus, the partial distinguishability is modeled by decomposing the internal state of each photon into a common (indistinguishable) mode $|\phi_0\rangle$ and an orthogonal mode $|\phi_k^\perp\rangle$, 
\begin{equation}
    | \psi_k \rangle = \sqrt{1-\epsilon} \, | \phi_0 \rangle + \sqrt{\epsilon} \, | \phi^\perp_k \rangle ,
    \label{model_homogeneous_dist}
\end{equation}
where $0 \leq \epsilon \leq 1$ quantifies the degree of distinguishability, with $ \langle \phi_0 | \phi^\perp_j \rangle = 0$ and $ \langle \phi^\perp_k | \phi^\perp_j \rangle = \delta_{kj}$. In this decomposition, the component $|\phi_0\rangle$ results in the genuine multiphoton interference, whereas the orthogonal components $|\phi_k^\perp\rangle$ contribute without interference and
are therefore interpreted as the classical contribution. Thus, the overlap matrix V, defined in Eq.~(\ref{def_H}), reduces to
\begin{equation}
    V_{ij} = (1-\epsilon) + \epsilon \, \delta_{ij} ,
    \label{v_m_model}
\end{equation}
where $I_M$ denotes the usual $M \times M$ identity matrix. Inserting Eq.~(\ref{v_m_model}) into Eqs.~(\ref{def_H}) and (\ref{H_r}), we obtain the matrix $\mathcal{M}_{\bf r}$ as a convex sum of the two extremal cases, as follows:
\begin{equation}
    \mathcal{M}_{\bf r} = (1-\epsilon) \, \mathcal{M}^0_{\bf r} + \epsilon \, \mathcal{M}^\perp_{\bf r},
    \label{H_r_homogeneous}
\end{equation}
where the first term corresponds to the indistinguishable contribution introduced in Eq.~(\ref{H_r_ind}), while the second term defines the distinguishable (orthogonal) contribution:
\begin{equation}
    \mathcal{M}^\perp_{\bf r}=
    \begin{pmatrix}
    0_M & H^\perp_{\bf r}\\
    H^\perp_{\bf r} & 0_M
    \end{pmatrix}
    ,
    \quad
    H^\perp_{\bf r} \equiv \big. H_{\bf r} \big|_{V_{ij}=\delta_{ij}}
    ,
    \label{H_r_dist}
\end{equation}
which follows that we have the real matrix $H^\perp_{\bf r} = \text{diag}(H^0_{\bf r})$. Thus, substituting Eq.~(\ref{H_r_homogeneous}) into Eq.~(\ref{gen_func_smss}), we obtain the corresponding expression for the generating function
\begin{equation}
    G_{\bf r}({\bm \eta}, \epsilon) = c_{\bf r} \,
    \text{det}\left[I_{2M} - (1-\epsilon) \, \mathcal{M}^0_{\bf r} - \epsilon \, \mathcal{M}^\perp_{\bf r} \right]^{-1/2} 
    .
    \label{gen_func_homog}
\end{equation}

Note that, when the generating function factorizes as
$G_{\bf r}({\bm \eta},\epsilon)=g_1({\bm \eta})\,g_2({\bm \eta})$, it follows that the probability can be expressed as a sum of two contributions. Using the compact notation $\partial_{\eta_l}\equiv\partial/\partial\eta_l$, this property is proven by applying the generalized Leibniz rule to the factorized generating function, 
\begin{equation}
    (-\partial_{\eta_l})^{n_l}
    \, g_1({\bm \eta})\, g_2({\bm \eta})
    = \sum_{m_l=0}^{n_l} \binom{n_l}{m_l}
    \Big[(-\partial_{\eta_l})^{m_l} g_1({\bm \eta})\Big]
    \Big[(-\partial_{\eta_l})^{n_l-m_l} g_2({\bm \eta})\Big] 
    \label{Leibniz_rule}
    ,
\end{equation}
and then substituting in Eq.~(\ref{probability_gen_func}).
Motivated by this observation, we present a convenient factorization of the generating function in Eq.~(\ref{gen_func_homog}), separating the orthogonal component defined in Eq.~(\ref{H_r_dist}), resulting in
\begin{equation}
    G_{\bf r}({\bm \eta}, \epsilon) 
    = c_{\bf r} \, 
    \text{det}(I_{2M} - \epsilon \, \mathcal{M}^\perp_{\bf r})^{-1/2} 
    \,
    \text{det}\left[I_{2M} - (1-\epsilon) \, \widetilde{\mathcal{M}}^0_{\bf r} \right]^{-1/2}
    ,
    \label{gen_func_factor}
\end{equation}
where we define the noise-modified version of the matrix  $\mathcal{M}^0_{\bf r}$ defined in Eq.~(\ref{H_r_ind}), as follows
\begin{equation} 
    \widetilde{\mathcal{M}}^0_{\bf r}  = 
    (I_{2M} - \epsilon \, \mathcal{M}^\perp_{\bf r})^{-1}
    \mathcal{M}^0_{\bf r}
    .
    \label{tilde_matrix_M}
\end{equation}
Thus, substituting  Eq.~(\ref{gen_func_factor}) into Eq.~(\ref{probability_gen_func}) and applying Eq.~(\ref{Leibniz_rule}), the output probability can be written as a sum of a classical contribution $P^{\perp}_{\mathbf r}(\mathbf m)$ and a noisy quantum contribution $\widetilde{P}^{0}_{\mathbf r}(\mathbf n-\mathbf m)$,
\begin{equation}
    P_{\bf r}({\bf n}) = c_{\bf r}  \sum^{n_1}_{m_1=0} ... \sum^{n_M}_{m_M=0} \epsilon^{|{\bf m }|} \, 
    P^\perp_{\bf r}({\bf m }) \, 
    \widetilde{P}^0_{\bf r}({\bf n}-{\bf m }) 
    .
    \label{prob_dist_indist}
\end{equation}

We now focus on the explicit evaluation of the classical contribution in Eq.~(\ref{prob_dist_indist}). From Eq.~(\ref{H_r_dist}), we have
\begin{align}
    \text{det}
    (I_{2M} - \epsilon \, \mathcal{M}^\perp_{\bf r}) 
    = \prod^M_{k=1} \left[ 1 - \epsilon^2 \tanh^2 r_k \left( \sum^M_{l=1} |U_{kl}|^2 (1-\eta_l) \right)^2 \right]
    .
    \label{det_matrix_N}
\end{align}
Recalling that squeezing is applied only to the first $N$ modes, so that $\tanh r_k=\tanh r$ for $1\le k\le N$ and $\tanh r_k=0$ for $N<k\le M$, and additionally assuming a uniform multiport $|U_{kl}|^2=1/M$, the classical probability
in Eq.~(\ref{prob_dist_indist}) follows directly from
Eqs.~(\ref{det_matrix_N}) and (\ref{probability_gen_func}) as
\begin{align}
    P^\perp_{\bf r} ({\bf m})
    &=  \left. \prod^M_{k=1} \frac{\eta^{m_k}_k}{m_k!} \left( - \frac{\partial}{\partial \eta_k} \right)^{m_k} \text{det}
    (I_{2M} - \mathcal{M}^\perp_{\bf r})^{-1/2} \right|_{{\bm \eta} = {\bm 1}} 
    \nonumber\\
    & = 
    \begin{cases}
    \frac{2m!}{m!} \left(\frac{N}{2} \right)_m \left( \frac{\tanh \, r}{M}\right)^{2m} & , |{\bf m}| = 2 m \\
    0 & , |{\bf m}| = 2 m+1.
    \end{cases}
    \label{dist_probability}
\end{align}
where $(N/2)_m=\prod_{r=0}^{m-1}(N/2+r)$ denotes the rising factorial. The same final expression for $P^\perp_{\bf r} ({\bf m})$ in Eq.~(\ref{dist_probability}) was also reported in Ref.~\cite{ValeryGBS}, obtained by a first-quantization method. The remaining term $\widetilde{P}^{0}_{\bf r}({\bf n})$ can be formally obtained by substituting the generating-function factor containing $\widetilde{\mathcal{M}}^0_{\bf r}$ into Eq.~(\ref{probability_gen_func}). In the indistinguishable limit ($\epsilon=0$), it reduces to the ideal probability $P^{0}_{\bf r}({\bf n})$ defined in Eq.~(\ref{probability_hafnian}).

An important observation is that the distinguishable component in Eq.~(\ref{gen_func_factor}) gives
the classical probabilities $P^\perp_{\bf r}({\bf n})$, which are easily computed, as shown in Eq.~(\ref{dist_probability}). Thus, the remaining term $\widetilde{P}^0_{\bf r}({\bf n})$ contains all the genuine
multiphoton-interference effects and therefore encodes the quantum contribution responsible for computational hardness. Such a decomposition of the probability into a classical (distinguishable)
part and a quantum (interfering) part was also identified in FBS~\cite{ValeryPRA2019}.

\subsection{Distinguishability contribution as an average over displaced states}

In the large noise regime ($\epsilon \approx 1$), the distinguishability arising from orthogonal internal modes suppresses the multiphoton interference, collapsing the probability into Eq.~(\ref{dist_probability}). We therefore focus now on the low-noise regime ($\epsilon \approx 0$), to elucidate how the partial distinguishability affects the ideal squeezed state interference. In this regime, we identify an additional interpretation in which the contribution from the orthogonal modes can be understood as an average over displaced states.

We start by expressing the input creation operators in the exponent of Eq.~(\ref{smss_input}) explicitly in the internal-state basis defined in Eq.~(\ref{model_homogeneous_dist}), resulting in
\begin{align}
    | {\bf r} (\epsilon) \rangle &= \sqrt{c_{\bf r}} \, \prod_{k=1}^M \exp 
    \left[ \frac{S_k}{2}
    \left( \sqrt{1-\epsilon} \, \hat{a}^\dagger_{k, \phi_0}
    + \sqrt{\epsilon} \, \hat{a}^\dagger_{k, \phi^\perp_k} \right)^2
    \right]  |0 \rangle 
    \nonumber\\
    &= \sqrt{c_{\bf r}} \, \prod_{k=1}^M \exp \left[
    \frac{S_k}{2}(\hat{a}^\dagger_{k, \phi_0})^2  + 
    S_k \sqrt{\epsilon} \, \hat{a}^\dagger_{k, \phi_0} \hat{a}^\dagger_{k, \phi^\perp_k} + \mathcal{O}(\epsilon) 
    \right] |0 \rangle
    .
    \label{smss_homog}
\end{align}
Note that, in this approximation, we neglect second-order contributions from the orthogonal modes $(\hat{a}^\dagger_{k,\phi_k^\perp})^2$, which effectively produce the trivial probabilities given by
Eq.~(\ref{dist_probability}).

In sequence, we decompose the photon-counting operator by separating the detection of the common mode $|\phi_0\rangle$, and the detection of the orthogonal modes $|\phi_k^\perp\rangle$, in Eq.~(\ref{photon_number_def}), resulting:
\begin{equation}
    \hat{N}_l = \eta_l \, \hat{b}^\dagger_{l,\phi_0} \hat{b}_{l,\phi_0} + \eta_l \sum^M_{k=1} \hat{b}^\dagger_{l,\phi^\perp_k} \hat{b}_{l,\phi^\perp_k} .
    \label{detec_oper_homog}
\end{equation}
Using Eqs.~(\ref{vacuum_prob}) and (\ref{gen_func_definition}), the generating function can be rewritten in a form where the detection acts explicitly only on the common mode, while the all the contributions from the orthogonal components are absorbed into an effective Husimi function that retains all the noise effects,
\begin{equation}
    G_{\bf r}({\bm \eta}, \epsilon) = \int d^2 {\bm \alpha}_{\phi_0} \, Q_{\text{eff}}({\bm \alpha}_{\phi_0}) \frac{\text{exp}\left[- {\bm \alpha}^\dagger_{\phi_0} \left(H^{-1}_0 - I_M \right) {\bm \alpha}_{\phi_0} \right]}{\text{det}(H_0)} ,
    \label{gen_func_aux2}
\end{equation}
where from Eq.~(\ref{def_H}) we define $H_0 \equiv U \Lambda U^\dagger$. In Eq.~\ref{gen_func_aux2} we define the phase-space coordinates that parameterize the Husimi function statistics of the common mode as ${\bm \alpha}_{\phi_0} = (\alpha_{1,\phi_0}, ..., \alpha_{M,\phi_0})$. The detailed derivation is presented in Appendix.~\ref{detailed_aver_coherent}

Under the approximation in Eq.~(\ref{smss_homog}), the effective Husimi function $Q_{\mathrm{eff}}({\bm \alpha}_{\phi_0})$ in Eq.~(\ref{gen_func_aux2}) takes the form
\begin{equation}
    Q_{\text{eff}}({\bm \alpha}_{\phi_0}) \approx \frac{c_{\bf r}}{\pi^M} \, \langle {\bm \alpha}_{\phi_0} |
    \prod_{k=1}^M \exp \left[
    \frac{S_k}{2}(\hat{a}^\dagger_{k, \phi_0})^2 \right] \, \hat{\rho}_\epsilon  \, \prod_{k=1}^M \exp \left[ \frac{S^*_k}{2}\hat{a}_{k, \phi_0}^2 \right]
    | {\bm \alpha}_{\phi_0} \rangle
    \label{Husimi_effective}
    ,
\end{equation}
where $\hat{\rho}_\epsilon$ collects all noise contributions in the model. In the following, we introduce the phase-space coordinates of the orthogonal modes as ${\bm \alpha}_{\phi_\perp}=(\alpha_{1,\phi^\perp_1},\ldots,\alpha_{M,\phi^\perp_M})$. Therfore,  the state $\hat{\rho}_\epsilon$ in Eq.~(\ref{Husimi_effective}) is given as a mixture of displaced states, in which the orthogonal modes enter as displacement parameters, weighted by a Gaussian distribution with diagonal covariance matrix $\Sigma=\epsilon D_{\bf r}\,\mathrm{diag}(H_0)\,\overline{D}_{\bf r}$,
\begin{equation}
    \hat{\rho}_\epsilon =  \int \frac{d^2 {\bm \beta}}{\pi^M} \, \frac{\text{exp}\left(- {\bm \beta}^\dagger \Sigma^{-1} {\bm \beta} \right)}{\text{det}(\Sigma)} \, \text{exp}\left( {\bm \beta}^t {\bf \hat{a}}^\dagger_{\phi_0}\right) | 0 \rangle \langle 0 | \,\text{exp}\left( {\bm \beta}^\dagger {\bf \hat{a}}_{\phi_0}\right)
    .
    \label{rho_effective}
\end{equation}
where we introduce the effective displacement
${\bm \beta}=\sqrt{\epsilon}\,\overline{D}_{\bf r}\,{\bm \alpha}_{\phi_\perp}$, the creation-operator vector on the common mode ${\bf \hat{a}}^\dagger_{\phi_0}=(\hat{a}^\dagger_{1,\phi_0},\ldots,\hat{a}^\dagger_{M,\phi_0})$,
and define ${\bf \hat{a}}_{\phi_0}$ analogously. In the noiseless limit, Eq.~(\ref{rho_effective})  becomes the vacuum,
\begin{equation}
    \big. \hat{\rho}_\epsilon \big|_{\epsilon=0} =|0\rangle \langle 0|
    ,
\end{equation}
and therefore the effective Husimi function in Eq.~(\ref{Husimi_effective}) reduces to that of (vacuum) squeezed-states, as expected. Detailed derivations of Eqs.~(\ref{gen_func_aux2}), (\ref{Husimi_effective}), and
(\ref{rho_effective}) are given in Appendix~\ref{detailed_aver_coherent}.

Our approach also recovers a known result for the effect of losses on squeezed states, first derived in Ref.~\cite{PatronSimulabilityGBS}, where a homogeneous loss maps a pure squeezed state to a mixed Gaussian state, equivalently to a
squeezed thermal state. In our framework, this corresponds to the case where photons occupying orthogonal modes are not detected. Thus, this result is recovered by setting ${\bm \eta}={\bm 0}$ in Eq.~(\ref{rho_effective}), which is equivalent to taking $\eta_l=0$ for all $l$ in the detection operator for the orthogonal modes in Eq.~(\ref{detec_oper_homog}). In this way, Eq.~(\ref{rho_effective}) reduces to a thermal state in the
$P$ representation, with the following parametrization:
\begin{equation}
    \left. \Sigma_{kk} \right|_{\eta_k=0} = \epsilon  \tanh^2 r_k = \frac{\aver{n_k}+1}{\aver{n_k}},
    \label{aver_thermal_photons}
\end{equation}
where $\aver{n_k}$ denotes the average number of thermal photons. In addition, $\aver{n_k}$ can be written as $\aver{n_k} = \sinh^2 \widetilde{r}_k$, with $\tanh \widetilde{r}_k = \sqrt{\epsilon}\,\tanh r_k$.

%%%%%%%%%%%%%%%%%%%%%%%%%%%%%%%%%%%%%%%%%%%%%%%%%%%%%%%%%%%%%%%%%%%%%%%%%%%%%%%%%%%%%%%%%%%%%%%%%%%%%%%%%%%%%%%%%%%%%%%%%%%%%%%%%%%%%%%%%%%%%%%%%%%%%%%%

\section{\label{sec:application2} Gaussian model for the internal states and zero-probability events}

In this section, we address a more fundamental aspect of partial distinguishability. In two-photon interference, described by the HOM effect~\cite{HOM1987}, the distinguishability is fully captured by the squared modulus of the mutual overlap, $|V_{12}|^{2}$ \cite{HOM1987,Loudon1989,Mandel1991}, often referred to as the visibility. By contrast, for interferences of three or more photons in Fock states, partial distinguishability depends not only on the moduli of pairwise overlaps but also on phases encoded in the internal states. In particular, three-photon interference is sensitive to a collective three-particle phase \cite{CollectivePhase1} given by $\arg\!\left(V_{12}V_{23}V_{31}\right)$, which is not determined by the pairwise moduli $|V_{ij}|^{2}$ alone. Likewise, four-photon interference exhibits an analogous four-particle phase \cite{CollectivePhase3} and, more generally, collective phases associated with genuine multiphoton overlaps \cite{CollectivePhase1}. To the best of our knowledge, such phase-dependent effects have not yet been investigated in the interference of single-mode squeezed states.

\begin{figure}[t]
    \centering
\includegraphics[width=1 \columnwidth]{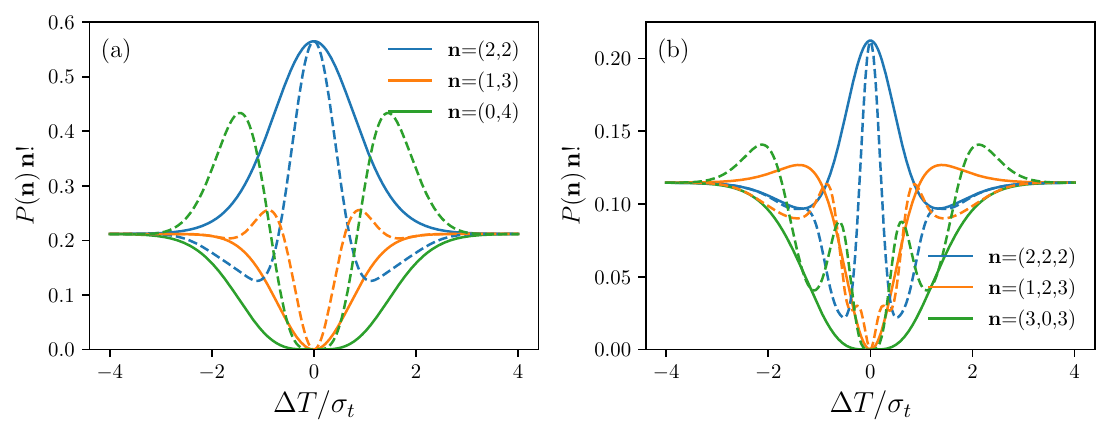}
\caption{Probabilities at the output for the interference of single-mode squeezed states with squeezing parameters $r_k=1$ in a: (a) ballanced beamsplitter and, (b) ballanced tritter. The probabilities are shown in function of the relative photon delay $\Delta T / \sigma_t$ for different central frequencies $\Omega_0=0$ (solid) and $\Omega_0 = \pi/4$ (dashed).}
\label{probabilities_smss}
\end{figure}

For this purpose, we adopt a realistic Gaussian model for the internal state of the photons, which has been widely employed in previous studies~\cite{ValeryBS1,ValeryBS2,CollectivePhase1,CollectivePhase2,CollectivePhase3} because it provides a clear physical interpretation and an experimentally motivated description of the internal state overlap. In this way, the internal state of the photon prepared in the input mode $k$ is defined in a temporal-mode basis as
\begin{equation}
    | \psi_k \rangle = \int d t \, \psi_k(t-T_k) | t \rangle
    ,
    \label{smss_gauss}
\end{equation}
where we define the temporal wave-packet as the following Gaussian distribution
\begin{equation}
    \psi_k(t) = \frac{1}{\left(\pi \sigma^2_t \right)^{\frac{1}{4}}} \exp \left[ - \left(\frac{t}{2 \sigma_t} \right)^2 + i \Omega_0 t \right] 
    ,
\end{equation}
parameterized by the relative time delay $T_k$, central frequency $\Omega_0$, and pulse width $\sigma_t$, resulting in the overlap
\begin{equation}
    V_{kj} = \exp \left[ - \left(\frac{T_k-T_j}{2 \sigma_t} \right)^2 + i \, \Omega_0 (T_k-T_j) \right] 
    .
    \label{smss_gauss_product}
\end{equation}
Here, the central frequency $\Omega_0$ denotes the angular frequency of the photon wave packet, around which the spectral distribution is peaked.

In the following, we first derive the zero-output probabilities for perfectly indistinguishable photons and sources with identical squeezing parameters,
and then investigate how these events are modified by partial
distinguishability.

To begin, we consider interference at a balanced beamsplitter with a reflection phase of $\pi/2$, which implements the unitary evolution via the matrix
\begin{equation}
    B = \frac{1}{\sqrt{2}}
    \begin{pmatrix}
        1 & i \\
        i & 1
    \end{pmatrix} 
    .
    \label{matrix_beamsplitter}
\end{equation}
For perfectly indistinguishable photons, all modes share the same internal state, $|\psi_k \rangle=|\phi_0 \rangle$ in Eq.~(\ref{smss_input}). Applying the state evolution
performed by Eq.~(\ref{matrix_beamsplitter}) to the creation operators in the exponent of Eq.~(\ref{smss_input}), we obtain the following output state
\begin{align}
    | r_1, r_2 \rangle_{(out)} 
    &= \sqrt{c_{\bf r}} \, \text{exp} \left[ \frac{S}{2} \sum^2_{k=1}\left( \sum^2_{l=1} B_{kl} \hat{b}^\dagger_{l, \phi_0} \right)^2 \right] |0 \rangle
    \nonumber\\
    &= \sqrt{c_{\bf r}} \exp \left( i \, S \, \hat{b}^\dagger_{1,\phi_0} \hat{b}^\dagger_{2,\phi_0}  \right) |0 \rangle 
    ,
\end{align}
where we assumed $S_1=S_2=S$. Therefore, the output state $| r_1, r_2 \rangle_{(out)}$ is effectively a two-mode squeezed state, implying that only outcomes with equal photon numbers in the two output modes are allowed, resulting in $P_{\bf r}(n_1, n_2)=0$ for any $n_1\neq n_2$.

To investigate the effects of partial distinguishability on these probabilities, Fig.~\ref{probabilities_smss}.(a) shows the selected output probabilities $P_{\bf r}(n_1,n_2)$ as a function of the relative time delay $\Delta T=T_1-T_2$, for two different central frequencies $\Omega_0$. As in the original HOM effect, the destructive interference is maximal at $\Delta T=0$ and continuously degraded as $|\Delta T|$ increases. Additionally, Fig.~\ref{probabilities_smss}.(a) reveals a dependence on the complex phase of $V_{12}$, arising by the difference choices of the central frequency $\Omega_0$ in the dashed and solid curves. In contrast to the two-photon Fock-state case, for squeezed-state inputs, this phase dependence arises already at the two-mode level because the input is a coherent superposition of photon-pair components rather than a state with a definite photon number.

In sequence, we consider interference at a balanced tritter, which implements the unitary evolution via the matrix
\begin{equation}
    T = \frac{1}{\sqrt{3}}
    \begin{pmatrix}
        1 & 1 & 1 \\
        1 & \text{e}^{i\frac{2\pi}{3}} & \text{e}^{-i\frac{2\pi}{3}} \\
        1 & \text{e}^{-i\frac{2\pi}{3}} & \text{e}^{i\frac{2\pi}{3}}
    \end{pmatrix} .
    \label{matrix_tritter}
\end{equation}
Analogously to the two-mode case, substituting the transformation in Eq.~(\ref{matrix_tritter}) into Eq.~(\ref{smss_input}) for perfectly indistinguishable photons gives the output state
\begin{align}
    | r_1, r_2, r_3 \rangle_{(out)} 
    &=c^{\frac{1}{2}}_{\bf r} \, \text{exp} \left[ \frac{S}{2} \sum^3_{k=1}\left( \sum^3_{l=1} T_{kl} \hat{b}^\dagger_{l, \phi_0} \right)^2 \right] |0 \rangle
    \nonumber\\
    &= c^{\frac{1}{2}}_{\bf r}  \exp \left[ S \left( (\hat{b}^\dagger_{1,\phi_0})^2 + 2 \, \hat{b}^\dagger_{2,\phi_0} \hat{b}^\dagger_{3,\phi_0} \right) \right] |0 \rangle ,
\end{align}
where $S_1=S_2=S_3$. Therefore, the output state $| r_1, r_2, r_3 \rangle_{(out)} $ factorizes into an effective single-mode squeezed state in mode $l=1$ and a two-mode squeezed state shared by modes $l=2,3$. It follows that $P_{\bf r}(n_1,n_2,n_3)=0$ whenever $n_1$ is odd or $n_2\neq n_3$. 

Finally, to investigate the effects of partial distinguishability on these
probabilities, Fig.~\ref{probabilities_smss}(b) shows selected output probabilities $P_{\mathbf r}(n_1,n_2,n_3)$ for equal temporal delays between neighbouring inputs, so that $\Delta T \equiv T_1-T_2 = T_2-T_3$ and $T_1-T_3 = 2\Delta T$. As expected, interference is maximal at $\Delta T=0$ and gradually degrades
as $|\Delta T|$ increases. The probabilities also exhibit a pronounced dependence on the overlap phases, controlled by the choices of $\Omega_0$.

%%%%%%%%%%%%%%%%%%%%%%%%%%%%%%%%%%%%%%%%%%%%%%%%%%%%%%%%%%%%%%%%%%%%%%%%%%%%%%%%%%%%%%%%%%%%%%%%%%%%%%%%%%%%%%%%%%%%%%%%%%%%%%%%%%%%%%%%%%%%%%%%%%%%%%%%%%%%%%%%%%%%%%%%%%%%%%%%%%%%%%%%%%%%%%%%%%%%%%%%%%%%%%%%%%%%%%%%

\begin{center}
\myclearpage
\par
\end{center}

\chapter{Quantum multiparameter estimation in optical interferometry}
\label{chapter:parameter_estimation}

\begin{tcolorbox}[colback=gray!5,colframe=black,title={Publication on which this Chapter is based}]
\underline{M.~E.~O.~Bezerra}, F.~Albarelli, R.~Demkowicz-Dobrzanski,  
``Simultaneous optical phase and loss estimation revisited: measurement and probe incompatibility'',   
J. Phys. A: Math. Theor. {\bf 58} 265303 (2025).  
\href{https://iopscience.iop.org/article/10.1088/1751-8121/ade516}{DOI:10.1088/1751-8121/ade516} \, 
(\href{https://arxiv.org/abs/2504.02893}{arXiv:2504.02893})
\end{tcolorbox}

\section{Background}

The optical \textit{phase estimation} problem stands out as one of the most important taskss in quantum metrology, on which all practical applications of quantum-enhanced interferometry are based~\cite{Caves1981, Giovannetti2004, Bollinger1996, Anisimov2010}. In optical phase estimation, a coherent state achieves the standard quantum limit (SQL) scaling, whereas interfering with a single-mode squeezed state enhances the precision, achieving sub-SQL scaling~\cite{Caves1981}. By SQL, one refers to an uncertainty that decreases with the photon number as 
$\Delta^2 \varphi \propto 1/N$. 
In contrast, sub-SQL scaling denotes an uncertainty that decreases even faster with $N$, 
therefore enhancing the achievable precision.
However, their performance is highly susceptible to photon losses, which rapidly degrade that advantage.
Phase estimation in the presence of loss was first investigated in~\cite{Demkowicz-Dobrzanski2009, Dorner2009}, where it was numerically shown that the optimal states exhibit a non-trivial 
function of the loss. 

\begin{figure}[t]
    \centering
    \includegraphics[width=0.65\columnwidth]{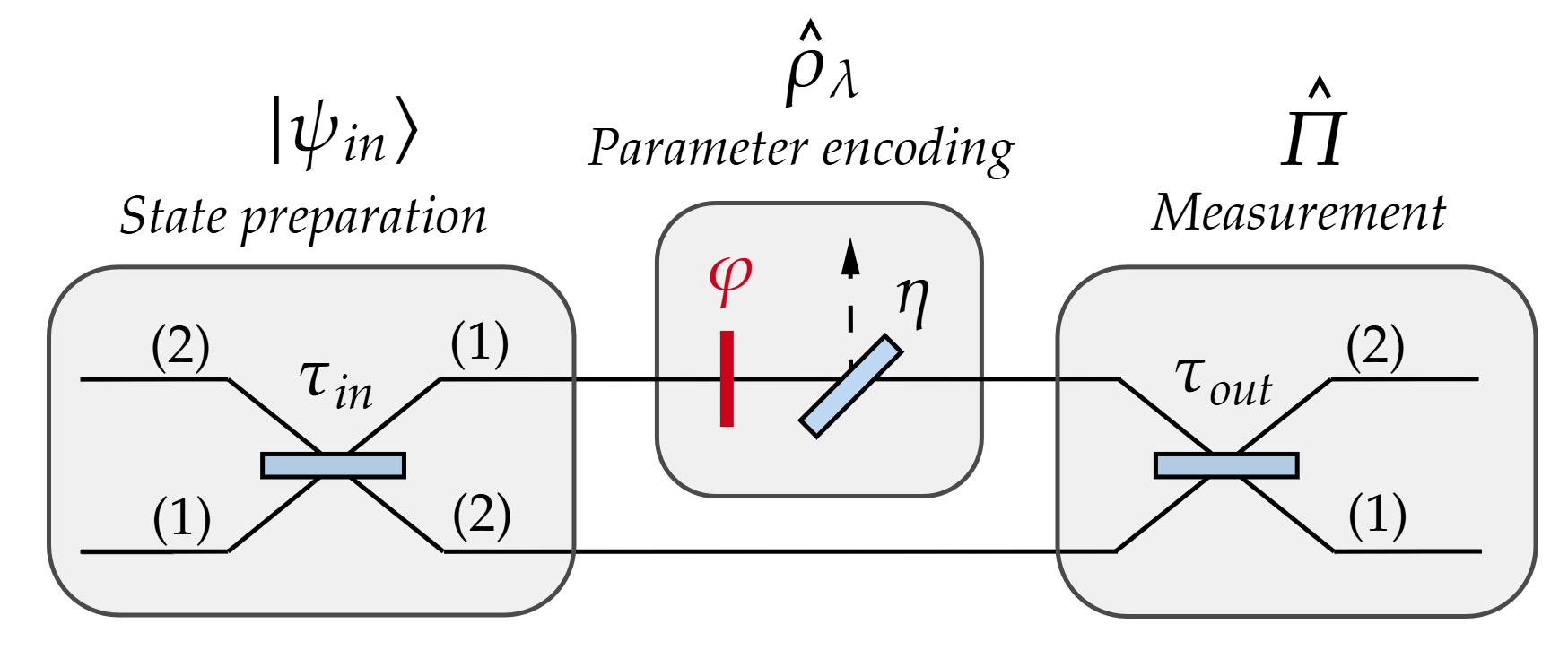}
    \caption{Illustration of the scheme considered. At the input, a two-mode pure probe state
    $\ket{\psi_{\text{in}}}$ is prepared, comprising a probe mode (labelled by $1$) in which the parameters of interest are are encoded, and an ancillary mode (labelled by $2$), which serves as a reference. The state then evolves through a quantum channel $\Lambda_{\bm{\lambda}}$ that encodes both the phase $\varphi$ and the loss parameter $\eta$, resulting in the output state $\hat{\rho}_{\bm{\lambda}}$. Finally, at the output, a POVM measurement is performed.
    }
    \label{phase_loss_GeneralScheme}
\end{figure}

Recalling the definition of the \textit{quantum Fisher information} (QFI) given in
Eq.~(\ref{qfi_def}), it provides a fundamental precision bound for a given input state and does not depend on the specific measurement chosen to estimate the
parameter of interest. However, different quantum states exhibit distinct sensitivities to variations in the parameter. States with larger QFI values encode
the parameter more efficiently and therefore allow for a lower minimum uncertainty the estimation of this parameter. In this way, one can determine the ultimate bound on the maximal QFI for phase estimation in the presence of loss, which is given by
\begin{equation}
    F^{\text{(max)}}_\varphi \equiv \max_{\hat{\rho}_{\mathrm{in}}} F_\lambda (\hat{\rho}_\lambda) = \frac{4 \eta N}{1-\eta} , 
    \label{upper_phase}
\end{equation}
where $F^{\text{(max)}}_\varphi$ represents the maximal achievable QFI in the scenario
illustrated in Fig.~\ref{phase_loss_GeneralScheme}. In general, this bound can be
saturated only in the asymptotic limit of large photon numbers,
$N \rightarrow \infty$~\cite{Koodynski2010,Knysh2011,Escher2011,Demkowicz2012}. Besides being a fundamental source of noise in any optical implementation, \textit{loss estimation} has also
received considerable attention due to its applications such as absorption imaging and
spectroscopy~\cite{Taylor2016,Shi2020}. In this case, the ultimate bound on the QFI is
given by~\cite{Monras2007,Adesso2009,Nair2018}
\begin{equation}
    F^{\text{(max)}}_\eta \equiv \max_{\hat{\rho}_{\mathrm{in}}} F_\eta (\hat{\rho}_\lambda)  = \frac{N}{\eta(1 -\eta)},
    \label{upper_loss}
\end{equation}
which is achieved by a Fock state~\cite{Adesso2009} for any finite $N$, or, when
interpreting $N$ as the mean photon number, by any pure state diagonal in the Fock
basis~\cite{Nair2018}.

Therefore, for any given state $\hat{\rho}_\lambda$, one can compute the corresponding
QFI, and the associated Cramér--Rao bound follows from Eq.~(\ref{CRB}) as
\begin{equation}
    \Delta^2 \lambda \geq \frac{1}{F(\hat{\rho}_\lambda)} \geq \frac{1}{F_\lambda^{\text{(max)}}}.
    \label{explanation_optimal_qfi}
\end{equation}
In the single-parameter estimation scenario, the objective is to identify the states
$\hat{\rho}_\lambda$ that achieve the optimal QFI, satisfying
$F_\lambda(\hat{\rho}_\lambda) = F^{\text{(max)}}$, which we refer to here as the
\textit{ultimate quantum limit}.

Moving forward, the fundamental task in \textit{multiparameter quantum metrology} is to estimate with the best precision a set of parameters ${\bm \lambda} = \{ \lambda_1,...,\lambda_d \}$ encoded in a probe state $\rho_{\bm \lambda}$, that has been evolved through a quantum channel $\rho_{\bm \lambda} = \Lambda_{\bm \lambda}(\rho)$. 
In the multiparameter estimation scenario, the uncertainty is quantified by the covariance matrix $\Sigma$, given by
\begin{equation}
    \Sigma_{i j} = \langle \left( \hat{\lambda}_i -  \langle \hat{\lambda}_i \rangle  \right) \left( \hat{\lambda}_j -  \langle \hat{\lambda}_j \rangle \right) \rangle,
\end{equation}
which is a generalization of the uncertainty defined in Eq.~(\ref{def_variance_lambda}). 
Note that, in quantum metrology, when referring to the estimation of a parameter $\lambda$, 
the terms “uncertainty” and “variance” are often used interchangeably when discussing scaling laws or bounds involving $\Delta^2 \lambda$. 
The diagonal elements are the variances $\Delta^2 \lambda_i = \Sigma_{i i}$. For any unbiased estimator, the \textit{quantum Cramér-Rao bound} (quantum CRB) gives a fundamental lower bound to the covariance matrix, which in the multiparameter case reads: 
\begin{equation}
    \Sigma \geq F(\rho_{\bm \lambda})^{-1},
    \label{CRB_matrix}
\end{equation}
where $ F(\rho_{\bm \lambda})$ is the quantum Fisher information (QFI) matrix, defined similarly to Eq.~(\ref{qfi_def}),
\begin{equation}
    F_{i j} (\rho_{\bm \lambda}) = \frac{1}{2} \Tr \big( \rho_{\bm \lambda} \{ L_{\lambda_i} , L_{\lambda_j} \}  \big),
    \qquad
    \frac{\partial \rho_{\bm \lambda}}{\partial \lambda_i} = \frac{1}{2} \{ \rho_{\bm \lambda} , L_{\lambda_i} \} ,
    \label{qfi}
\end{equation}
where $L_{\lambda_i}$ is the symmetric logarithmic derivative (SLD) related to the parameter $\lambda_i$.
The diagonal elements of Eq.~\eqref{qfi} give the QFI for the corresponding parameter $\lambda_i$, which determines the quantum precision limit for estimating this parameter independently of the others. The off-diagonal elements of the QFI matrix indicate correlations between the estimation of different parameters $\lambda_i$ and $\lambda_j$. We can recast the matricial quantum CRB given by Eq.~\eqref{CRB_matrix} into a scalar quantum CRB as follows:
\begin{equation}
    \Tr \left(W \Sigma \right) \geq C^S(\rho_{\bm \lambda}) = \Tr \left( W F(\rho_{\bm \lambda})^{-1} \right) ,
\end{equation}
where the weight matrix $W$ is a real, positive, $d \times d $ matrix.

The quantum CRB depends only on the state $\rho_{\bm \lambda}$ and, in general, cannot be saturated for all parameters by considering one single measurement strategy \cite{BookHolevo, Demkowicz-Dobrzanski2020}. 
In other words, even if we identify the states $\hat{\rho}^{(\mathrm{opt})}$ 
that simultaneously reach the ultimate quantum limits for both parameters, 
it may still occur that the optimal measurements for each parameter are different. 
In this case, no single measurement $\hat{X}$ can simultaneously achieve the corresponding optimal variances for both estimations, 
as expressed in Eq.~(\ref{error_prop}) for example.
The optimal measurement strategy for a parameter is determined by its corresponding SLD and then, when the SLDs do not commute, we expect that the bound cannot be attained. A necessary and sufficient (assuming access to multiple copies of the state) condition for saturating the QCR bound is that the commutators of SLDs vanish when traced with the state \cite{Ragy2016,Suzuki2018}:
\begin{equation}
    I_{\lambda_i \lambda_j} (\rho_{\bm \lambda}) = \frac{1}{2} \Tr \big( \rho_{\bm \lambda} \left[ L_{\lambda_i} , L_{\lambda_j} \right]  \big).
    \label{incomp}
\end{equation}
Otherwise, if this expectation value is non-zero, it indicates measurement incompatibility, meaning the parameters cannot be estimated simultaneously with the precision given by the quantum CRB. 

Let us now restrict our attention to the simultaneous estimation of phase and loss, as illustrated in Fig.~\ref{phase_loss_GeneralScheme}. If one treats the problem of phase and loss independently, then optimal protocols that saturate fundamental bounds are well known, as mentioned in the previous discussion. However, the phase and loss ultimate quantum limits in Eqs.~\eqref{upper_phase} and~\eqref{upper_loss}, which are achieved by very different states. In the multiparameter setting, a key problem is to determine whether the above asymptotic bounds, which are saturable in single parameter scenarios, can be  saturated when phase and loss are being sensed simultaneously. In other words, in Eq.~\ref{explanation_optimal_qfi}, we want to find the optimal states $\hat{\rho}^{\text{(opt)}}$ that  gives simultaneously $F_\varphi(\hat{\rho}^{\text{(opt)}}_{\bm \lambda}) = F^{\text{(max)}}_\varphi$ and $F_\eta(\hat{\rho}^{\text{(opt)}}_{\bm \lambda}) = F^{\text{(max)}}_\eta$.

Therefore, the main objective of this Chapter is to investigate the metrological incompatibilities in the simultaneous estimation of phase and loss encoded in one arm of a two-mode interferometer, as illustrated in Fig.~\ref{phase_loss_GeneralScheme}.
This chapter is organized as follows:
In Sec.~\ref{sec:probe_comp}, we search for the probe states that simultaneously gives the optimal QFI for phase and loss, 
defined respectively in Eqs.~(\ref{upper_phase}) and~(\ref{upper_loss}). 
In Sec.~\ref{sec:measurem_incomp}, once the optimal states are identified, 
we investigate which measurement schemes are capable of saturating the lower bounds imposed by these optimal QFIs 
and if the corresponding optimal measurements coincide.

%--------------------------------------------------------------------------------------------------------------------------------------------------------------------------%

\section{Probe compatibility in the simultaneous estimation of phase and loss}
\label{sec:probe_comp}

In multiparameter quantum metrology, a central problem is to determine whether a single quantum state exists that performs optimally when estimating all the parameters of interest simultaneously, here the phase $\varphi$ and loss $\eta$. In order to investigate this problem quantitatively, we introduce the \textit{probe incompatibility quantifier} that shows the amount of probe incompatibility for a given input probe state $\ket{\psi}$. This quantifier is defined as the sum of the diagonal elements of the QFI matrix in Eq.~(\ref{qfi}), each rescaled by the respective optimal single parameter QFI, as follows
\footnote{In addition, Eq.~(\ref{probe_incomp}), can be generalized to the case of $d$ parameters estimation. Denoting such parameters as $\lambda_1 , \, ... \, , \lambda_d$, the probe incompatibility quantifier reads~\cite{Albarelli2022},
\begin{equation}
    \mathcal{F} (\rho_{\bm \lambda}) = \frac{1}{d} \sum_{j=1}^d \frac{F_{j j} (\rho_{\bm \lambda})}{F_{\lambda_j}^{(\text{max})}} \leq 1 .
\end{equation}
}:
\begin{equation}
    \mathcal{F} (\rho_{\bm \lambda}) = \frac{1}{2} \left( \frac{F_{\varphi \varphi} (\rho_{\bm \lambda})}{F_\varphi^{(\text{max})}} + \frac{F_{\eta \eta} (\rho_{\bm \lambda})}{F_\eta^{(\text{max})}} \right)
    , \qquad
    \rho_{\bm \lambda} = \Lambda_{\bm \lambda} \left( \ket{\psi} \bra{\psi} \right) .
    \label{probe_incomp}
\end{equation}
On the previous equation, $F_{\lambda_j}^{(\text{max})}$ are the QFI of the single parameter $\lambda_j$ optimized over input probe states, which results in Eq.~(\ref{upper_phase}) for the phase, and Eq.~(\ref{upper_loss}) for the loss. This quantifier is bounded as $0 \leq \mathcal{F} (\rho_{\bm \lambda}) \leq 1$, and the value $1$ indicates no probe incompatibility in the model, meaning that each diagonal element of the QFI matrix attains the optimal value of the single-parameter estimation QFI.

Even if there is no probe incompatibility, there could also be additional issues due to correlations between the estimators of the parameters. Mathematically, this is related to the fact that the quantum CRB is given by the inverse of the QFI matrix, the correlations are related to the off-diagonal elements, and they imply the inequality $\Tr\left( F(\rho_{\bm \lambda})^{-1} \right) \geq \sum_j 1/F(\rho_{\bm \lambda})_{jj}$, as shown in Eq.~(\ref{CRB_matrix}). While we will not directly consider the impact of such correlations in optimizing the probe state, it is always possible to check the full QFI matrix afterwards to see if the off-diagonal elements are sufficiently small. 

To summarize, the main objective of this section is to find the states $\rho_{\bm \lambda}$ that reaches the upper bound of Eq.~(\ref{probe_incomp}) for phase and loss estimation, at least asymptotically, i.e.,
\begin{equation}
    \mathcal{F} (\rho_{\bm \lambda}) \rightarrow 1 .
\end{equation}
Then, for such states, check if the off-diagonal elements of the QFI matrix are zero. Indeed, we call each one of the bounds given by Eqs.~(\ref{upper_phase}) and (\ref{upper_loss}) as the ultimate quantum precision limits, since they provide the most fundamental CBR for each parameter. In fact, in order to have that upper bound equal to 1 we have these to necessary conditions:
\begin{equation}
    \frac{\langle \hat{n} \rangle}{N} \rightarrow 1, \quad \frac{\langle \hat{n} \rangle}{\Delta^2 n} \rightarrow 0.
    \label{Conditions_SimultaneousEstimation}
\end{equation}
The first condition means that we should use states of light that on average contain the maximal amount of photons in the sensing arm that are allowed in the problem, while the second condition requires the photon number statistics in the sensing arm to be super-Poissonian. The proof in intricate and is fully presented at Appendix~\ref{append:sufficient_cond}

\subsection{Optimization of arbitrary single-mode and two-mode states}

We begin our analysis by considering the probe state $|\psi_{\text{in}}\rangle$, represented in Fig.~\ref{phase_loss_GeneralScheme}, corresponding to the states obtained from numerical optimization of two different cases: a general single-mode state $|\psi^{(1)}_N\rangle$ and a general two-mode states  $|\psi^{(2)}_N\rangle$. Recalling the notation of a Fock state given in Eq.~(\ref{Fock_input}), the first one is a single-mode state containing up to $N$ photons, given by
\begin{equation}
    | \psi^{(1)}_N \rangle = \sum^N_{n=0} c^{(1)}_n | n \rangle_{a_1} 
    , \qquad
    \sum^N_{n=0} | c^{(1)}_n |^2 = 1
    \label{psiN_sm}
\end{equation}
while the second one is a two-mode state containing $N$ photons,
\begin{equation}
    | \psi^{(2)}_N \rangle = \sum^N_{n=0} c^{(2)}_n | n \rangle_{a_1} | N - n \rangle_{a_2} 
    , \qquad 
    \sum^N_{n=0} | c^{(2)}_n |^2 = 1 .
    \label{psiN_tm}
\end{equation}
Note that, for these two states, we set the input beam splitter transmissivity to
$\tau_{\mathrm{in}} = 1$, since the states are fully characterized by the
coefficients $c^{(1)}_n$ and $c^{(2)}_n$.

The objective is then, perform the optimization over these two input states $\ket{\psi^{(1)}_N}$ and $\ket{\psi^{(2)}_N}$ in order to maximize the probe incompatibility quantifier defined in Eq.~(\ref{probe_incomp}). As a result, we obtain a quantity that depends only on the channel $\Lambda_{\boldsymbol{\lambda}}$ and captures the intrinsic probe incompatibility of the metrological model. We perform this maximization using an \textit{iterative see-saw (ISS) algorithm} to maximize $\mathcal{F}(\rho_{\boldsymbol{\lambda}})$ over the probe states. This algorithm is a direct generalization of the ISS method employed in the single-parameter estimation~\cite{Macieszczak2014,Macieszczak2013,Kurdzialek2025}, which, however, has not been applied to multiparameter estimation problems until now. 

We also mention that the state $|\psi^{(2)}_N\rangle$ was  considered in the first work on the trade-off between the simultaneous estimation of phase and loss by Crowley \textit{et al.}~\cite{Crowley2014}. 
However, the numerical optimization performed in that study relied on gradient-based methods, which is significantly less efficient and did not provide a complete characterization of the probe compatibility of this state. In contrast, our ISS method allows us to explore much larger photon numbers $N$ and to gain a deeper understanding of this compatibility.

To start, we recall some details about the ISS algorithm in the single parameter scenario. In this case, the optimization simply consists in the maximization of the QFI over the input states, $\max_{| \psi \rangle} F_{\lambda_i} (\rho_{\bm \lambda})$ with $\rho_{\bm \lambda} = \Lambda_{\bm \lambda}(| \psi \rangle \langle \psi | )$. We introduce the pre-QFI function for the parameter $\lambda_i$, defined as follows:
\begin{equation}
    f_{\lambda_i} (| \psi \rangle, A) = 2 \hspace{1mm} \Tr \left(\frac{\partial \rho_{\bm \lambda}}{\partial \lambda_i} A \right) - \Tr \left( \rho_{\bm \lambda} A^2 \right) ,
    \label{pre_qfi}
\end{equation}
where $A$ is a Hermitian operator and the derivative over the output state can be conveniently written in terms of the derivatives over the Kraus operators $\hat{K}_m$,
\begin{equation}
    \frac{\partial \rho_{\bm \lambda}}{\partial \lambda_i} = 
    \sum_m
    \left(
    \frac{\partial \hat{K}_m}{\partial \lambda_i} \, \ket{\psi} \bra{\psi} \, \hat{K}^\dagger_m 
    +
    \hat{K}_m \, \ket{\psi} \bra{\psi} \, \frac{\partial \hat{K}^\dagger_m}{\partial \lambda_i} 
    \right).
    \label{der_rho_kraus}
\end{equation}

A key point is that, maximizing the previous function over the operator $A$ results in the QFI of this corresponding parameter, where the optimal $A$ is the corresponding SLD, $L_{\lambda_i}$ given in Eq.~\eqref{qfi}.
Then, the ISS algorithm consists in the double maximization problem:  
\begin{equation}
    \max_{| \psi \rangle} \, F_{\lambda_i} (\rho_{\bm \lambda}) = \max_{| \psi \rangle } \, \max_A \, f_{\lambda_i} (| \psi \rangle , A)
\end{equation}

Indeed, we can write Eq.~(\ref{pre_qfi}) in terms of the dual map of the quantum channel $\Lambda^*$ by using the the standard properties of the trace and Eq.~(\ref{der_rho_kraus}), as follows:
\begin{align}
    f_{\lambda_i} (| \psi \rangle, A) = & \,
    \sum_m 
    \tr
    \left(
    2 \, \frac{\partial \hat{K}_m}{\partial \lambda_i} \, \ket{\psi} \bra{\psi} \, \hat{K}^\dagger_m \, A
    +
    2 \, \hat{K}_m \, \ket{\psi} \bra{\psi} \, \frac{\partial \hat{K}^\dagger_m}{\partial \lambda_i} \, A
    -
    \hat{K}_m \, \ket{\psi} \bra{\psi} \hat{K}^\dagger_m \, A^2
    \right) \nonumber\\
    = & \,
    \sum_m 
    \tr
    \left(
    2 \, \ket{\psi} \bra{\psi} \, \hat{K}^\dagger_m \, A \, \frac{\partial \hat{K}_m}{\partial \lambda_i}
    +
    2 \, \ket{\psi} \bra{\psi} \, \frac{\partial \hat{K}^\dagger_m}{\partial \lambda_i} \, A \, \hat{K}_m 
    -
    \ket{\psi} \bra{\psi} \hat{K}^\dagger_m \, A^2 \, \hat{K}_m
    \right) \nonumber\\
    = & \,
    \sum_m 
    \tr
    \left[ \ket{\psi} \bra{\psi} \left(
    2 \, \hat{K}^\dagger_m \, A \, \frac{\partial \hat{K}_m}{\partial \lambda_i}
    +
    2 \, \frac{\partial \hat{K}^\dagger_m}{\partial \lambda_i} \, A \, \hat{K}_m 
    -
    \hat{K}^\dagger_m \, A^2 \, \hat{K}_m
    \right) \right] \nonumber\\
    = & \, \tr
    \left[ \ket{\psi} \bra{\psi} \left(  2 \frac{ \partial \Lambda^*}{\partial \lambda_i} (A) - \Lambda^* (A^2) \right) \right]
    \label{preQFI_dualmap}
\end{align}
where $\Lambda^*(\cdot)$ is the dual map of the quantum channel,
\begin{equation}
    \Lambda^*(\cdot)=\sum_m \hat{K}_m^\dagger \cdot \hat{K}_m .
\end{equation}

Therefore, the step of maximizing over $| \psi \rangle$ becomes the problem of maximizing the following function:
\begin{equation}
    f_{\lambda_i} (| \psi \rangle, A) = \Tr \Big( | \psi \rangle 
    \langle \psi | M_{\lambda_i} \Big)
    , \qquad
    M_{\lambda_i} = 2 \frac{ \partial \Lambda^*}{\partial \lambda_i} (A) - \Lambda^* (A^2) 
    \label{qfi_algorithm}
\end{equation}
resulting in the optimal $| \psi \rangle$ being the eigenvector with the largest eigenvalue of the matrix $M_{\lambda_i}$. Thus, we plug this optimal state once again in Eq.~(\ref{pre_qfi}) and repeat the iteration till the algorithm converges. Indeed, this algorithm is convex when each step is optimized independently and, as a result, converges to a global maximum~\cite{Macieszczak2013}.

A generalization of this method for the current case of two parameter estimation can be obtained by introducing the following multiparameter pre-QFI ~\footnote{A generalization of such a function to the estimation of $d$ parameter is given as follows:
\begin{equation}
    f_{\bm \lambda} (| \psi \rangle,\{ A_{\lambda_j} \} ) = \sum_{j=1}^d \frac{f_{\lambda_j} (| \psi \rangle, A_{\lambda_j} )}{\omega_{\lambda_j}}, 
\end{equation}
where each $f_{\lambda_i} (| \psi \rangle, A_{\lambda_i})$ is defined as in Eq.~\eqref{pre_qfi} and $\omega_{\lambda_j} > 0 $ are generic positive weights.}:
\begin{equation}
    f_{\bm \lambda} (| \psi \rangle, A_\varphi, A_\eta ) = \frac{f_\varphi (| \psi \rangle, A_\varphi )}{\omega_\varphi}
    +
    \frac{f_\eta (| \psi \rangle, A_\eta )}{\omega_\eta}
    \label{pre_QFI_algorithm}
\end{equation}
where each $f_{\lambda_i} (| \psi \rangle, A_{\lambda_i})$ is defined as in Eq.~\eqref{pre_qfi} and the weights are chosen as $\omega_{\lambda_j}= F_{\lambda_j}^{(\text{max})} $ in order to compute the probe incompatibility quantifier~\eqref{probe_incomp}.
Then, the iterative algorithm for simultaneous estimation works in the following two steps optimization
\begin{itemize}
    \item[(a)] Given a random input state $| \psi^{[0]} \rangle$, the first step is the maximization of $f_{\bm \lambda} (| \psi \rangle,  A_\varphi, A_\eta  )$ over the Hermitian operators $A_\varphi$ and $A_\eta$, obtaining the corresponding SLDs $L_\varphi$ and $L_\eta$. 
    \item[(b)] For the second step, we plug the previous result in each term of Eq.~\eqref{qfi_algorithm}, rewrite the entire trace in terms of the dual map following Eq.~(\ref{preQFI_dualmap}), and perform the maximization:
    \begin{equation}
        \max_{| \psi \rangle} \left( \frac{f_\varphi (| \psi \rangle, A_\varphi )}{\omega_\varphi}
        +
        \frac{f_\eta (| \psi \rangle, A_\eta )}{\omega_\eta} \right) = \max_{| \psi \rangle} \Tr \big( | \psi \rangle \langle \psi | M \big) 
        , \qquad
        M =  \frac{M_{\varphi}}{\omega_\varphi} + \frac{M_{\eta}}{\omega_\eta},
        \label{normalized_m_matrix}
    \end{equation}
    which results in the optimal state $| \psi \rangle$ being the eigenvector with the largest eigenvalue of the matrix $M$, denoted by $\ket{\psi^{[1]}}$. Then, we plug this state in Eq.~(\ref{pre_QFI_algorithm}) and repeat the steps $(a)$ and $(b)$ once again, obtaining another state $\ket{\psi^{[2]}}$, and so on.
    We repeat this iteration until the algorithm converges (e.g., when the last five results do not differ by more than $0.1 \%$). 
\end{itemize}

Once the optimization algorithm is formulated, we then  investigate how exactly these states are transformed under the quantum channel evolution $\Lambda_{\boldsymbol{\lambda}}$, which encodes both the phase $\varphi$ and the loss parameter $\eta$ on the first mode $\ket{n}_{a_1}$, as illustrated in Fig.~\ref{phase_loss_GeneralScheme}.
As presented in the general Kraus evolution defined in Eq.~(\ref{general_evolution_Kraus}), we assume that the output states are mixed, with the input states $\ket{\psi^{(1)}_N}$ and $\ket{\psi^{(2)}_N}$ evolving respectively into $\hat{\rho}^{(1)}_N$ and $\hat{\rho}^{(2)}_N$, as follows
\begin{equation}
    \ket{\psi^{(M)}_N} \bra{\psi^{(M)}_N}
    \mapsto
    \hat{\rho}^{(M)}_N = \sum^N_{m=0} \hat{K}_m \, \ket{\psi^{(M)}_N} \bra{\psi^{(M)}_N} \, \hat{K}^\dagger_m  .
\end{equation}
In the previous equation, the Kraus operators $\hat{K}_m$ account for both effects: the encoding of the phase $\varphi$, by the unitary transformation $\hat{U}_{\varphi}$ defined in Eq.~(\ref{phase_encoding}); and the encoding of the loss $\eta$, by the binomial process described in Eq.~(\ref{FockState_wLoss}).
Indeed, that sum over the index $m$ represents the number of photons lost in the total number of $N$, which captures the nature of the mentioned  binomial process.
We can recast such Kraus evolution in a matrix representation, by writing these states and operators in the corresponding Fock basis, which is an important step to implement the ISS algorithm.
Therefore, the phase is encoded according to the following diagonal matrix:
\begin{equation}
    [U_\varphi]_{kj} = \e^{i \varphi k} \, \delta_{kj} 
    , \qquad
    0 \leq k \leq N ,
    \label{auxiliar_matrix_phase}
\end{equation}
while the loss is encoded in each subspace with $m$ photons lost, according to this another diagonal matrix:
\begin{equation}
    [B_m]_{kj} = \sqrt{B^k_m} \, \delta_{kj} 
    , \qquad
    m \leq k \leq N,
    \label{auxiliar_matrix_loss}
\end{equation}
where $B^k_m$ are the binomial coefficients defined in Eq.~(\ref{binomial_loss}). In the following, we investigate the evolution of each state separately, since each one has its own particular features.
\begin{itemize}
    \item We start analyzing the  \textit{single-mode state} $| \psi^{(1)}_N \rangle$, defined in Eq.~(\ref{psiN_sm}). This state vector lives in the space with at most $N$ photons $\mathcal{H}^{(1)}$, which has a canonical Fock basis denoted by $B^{(1)} = \{  |n\rangle_{a_1} , \, n \in [0,N]  \}$.
    Considering the evolution of a Fock state in the presence of loss given by Eq.~(\ref{FockState_wLoss}), together with the phase shift, we have
    \begin{equation}
        | \psi^{(1)}_N \rangle \mapsto | \widetilde{\psi}^{(1)}_N \rangle = \sum^N_{n=0} c^{(1)}_n \left( \sum^n_{m=0} \sqrt{B^{n}_m} \, e^{i (n-m) \varphi} \, \ket{n-m}_{a_1} \ket{m}_e \right) ,
    \end{equation}
    and tracing out the virtual mode we obtain the corresponding density operator at the output
    \begin{equation}
        \hat{\rho}^{(1)}_N = \tr_e \big( \ket{\widetilde{\psi}^{(1)}_N} \bra{\widetilde{\psi}^{(1)}_N} \big) = \sum^N_{m=0} | \psi^{(1)}_m \rangle \langle \psi^{(1)}_m | .
        \label{rho1_output}
    \end{equation}
    Therefore, the previous equation describes a mixture of different pure states, $|\psi^{(1)}_m\rangle$, each one representing the loss of $m$ photons,
    \begin{equation}
        | \psi^{(1)}_m \rangle = \sum^N_{n = m} c^{(1)}_n \sqrt{B^{n}_m} \text{e}^{i n \varphi} | n - m  \rangle_{a_1} .
    \end{equation}
    Note that, since the input space $\mathcal{H}^{(1)}$ does not have a definite photon number, it also includes all these new states with photons lost, i.e., $|\psi^{(1)}_m\rangle \in \mathcal{H}^{(1)}$.
    In sequence, we can write the input single-mode state in the Fock state basis $B^{(1)}$ as the following column vector
    \begin{equation}
        | \psi^{(1)}_N \rangle = 
        \begin{pmatrix}
            c^{(1)}_0 & \cdots & c^{(1)}_N
        \end{pmatrix}^t .
    \end{equation}
    By using the matrices $U_\varphi $ and $B_m$, defined respectively in Eq.~(\ref{auxiliar_matrix_phase}) and ~(\ref{auxiliar_matrix_loss}), we can easily write the density matrix $\rho^{(1)}_N $ in the corresponding operator basis as follows:
    \begin{equation}
        \rho^{(1)}_N = \sum^N_{m=0} 
        \mathcal{K}^{(1)}_m 
        \begin{pmatrix}
            c^{(1)}_0 \\ \vdots \\ c^{(1)}_N
        \end{pmatrix}
        \begin{pmatrix}
            c^{(1)}_0 & \cdots & c^{(1)}_N
        \end{pmatrix}^*
        \left[\mathcal{K}^{(1)}_m \right]^\dagger,
        \label{kraus_sm_alg}
    \end{equation}
    where we introduce the corresponding Kraus operator acting on the single-mode space, $\mathcal{K}^{(1)}_m$, as follows
    \begin{equation}
        \mathcal{K}^{(1)}_m = 
        \begin{pmatrix}
        0_{(N - m+1) \times m} & B_m \\
        0_{m \times m} & 0_{m \times (N - m+1)}
        \end{pmatrix} 
        U_\varphi 
        .
    \end{equation}
    Finally, from Eq.~(\ref{kraus_sm_alg}), the action of the quantum channel on any state in $\mathcal{H}^{(1)}$ can be recast in terms of the following Kraus matrix operation:
    \begin{equation}
        \Lambda(\cdot) = 
        \sum^N_{m=0} 
        \mathcal{K}^{(1)}_m 
        \cdot
        \left[\mathcal{K}^{(1)}_m \right]^\dagger.
    \end{equation}
    
    \item In the sequence, we investigate the \textit{two-mode state} $| \psi^{(2)}_N \rangle$, defined in Eq.~(\ref{psiN_tm}). This state vector lives in the two-mode space with total number of photons $N$ , denoted by $\mathcal{H}^{(2)}_N$, which has a canonical Fock basis denoted by $B^{(2)}_N = \{  |n\rangle_{a_1} |N-n\rangle_{a_2},  \, n \in [0,N]  \}$. Proceeding with analogous calculations to those performed in the previous case, the corresponding density operator at the output is given by.
    \begin{equation}
        \hat{\rho}^{(2)}_N = \sum^N_{m=0} | \psi^{(2)}_m \rangle \langle \psi^{(2)}_m | ,
        \label{rho2_output}
    \end{equation}
    where we define the corresponding two-mode states with $m$ photons lost:
    \begin{equation}
         | \psi^{(2)}_m \rangle  = \sum^N_{n = m} c^{(2)}_n \sqrt{B^{n}_m} \text{e}^{i n \varphi} | n - m  \rangle_{a_1} | N - n \rangle_{a_2} .
    \end{equation}
    Note that the initial space $\mathcal{H}^{(2)}_N$ contains only states with exactly $N$ photons, i.e., only the states $\ket{\psi^{(2)}_0}$ on the previous equation. Therefore, for each value of $m$ there exists a corresponding space $\mathcal{H}^{(2)}_m$ that contains the states with $N-m$ photons. 
    As a result, the output density operator acts on a larger Hilbert space given by the direct sum $\mathcal{H}^{(2)} = \mathcal{H}^{(2)}_0 \oplus \cdots \oplus \mathcal{H}^{(2)}_N$, which means that the summation in Eq.~(\ref{rho2_output}) can be replaced by a direct-sum, where each block $m$ corresponds to the operator acting on the the subspace $\mathcal{H}^{(2)}_m$.
    Next, denoting the basis of each of these subspaces as $B^{(2)}_m = \{ |n\rangle_{a_1}|N-n\rangle_{a_2} \, | \, n \in [m, N] \}$, we can first express the input state in terms of the input basis $B^{(2)}_N$ as the following column vector:
    \begin{equation}
        | \psi^{(2)}_N \rangle = 
        \begin{pmatrix}
            c^{(2)}_0 & \cdots & c^{(2)}_N
        \end{pmatrix}^t .
    \end{equation}
    Analogously to the previous case, we can write the density matrix $\rho^{(2)}_N $ in the corresponding operator basis, $B^{(2)}_0 \oplus \, ... \, \oplus B^{(2)}_N$, as follows:
    \begin{equation}
        \rho^{(2)}_N = \bigoplus^N_{m=0} 
        \mathcal{K}^{(2)}_m 
        \begin{pmatrix}
            c^{(2)}_0 \\ \vdots \\ c^{(2)}_N
        \end{pmatrix}
        \begin{pmatrix}
            c^{(2)}_0 & \cdots & c^{(2)}_N
        \end{pmatrix}^*
        \left[\mathcal{K}^{(2)}_m \right]^\dagger
        \label{kraus_tm_alg}
    \end{equation}
    where we introduce the Kraus operator acting on the two-mode space, $\mathcal{K}^{(2)}_m$, as 
    \begin{equation}
        \mathcal{K}^{(2)}_m = 
        \begin{pmatrix}
        0_{(N - m+1) \times m} & B_m \\
        \end{pmatrix} 
        U_\varphi 
    \end{equation}
    Note that, each Kraus matrix projects the input state into the state with $m$ photons lost, i.e, $\mathcal{K}^{(2)}_m: \mathcal{H}^{(2)}_N \rightarrow \mathcal{H}^{(2)}_m$.     Finally, from Eq.~(\ref{kraus_tm_alg}), the action of the quantum channel on any state in $\mathcal{H}^{(2)}_N$ can be recast in terms of the following Kraus matrix operation:
    \begin{equation}
        \Lambda(\cdot) = 
        \bigoplus^N_{m=0} 
        \mathcal{K}^{(2)}_m 
        \cdot
        \left[\mathcal{K}^{(2)}_m \right]^\dagger.
    \end{equation}
\end{itemize}

Finally, all the parameters involved in the ISS optimization are easily identified with the matrices presented in the discussion above, namely: the vector form of the input states $ | \psi^{(M)}_N \rangle$, the Kraus matrices $\mathcal{K}^{(M)}_m$, and the derivatives of the Kraus matrices, which can be trivially calculated from the matrices $U_\varphi$ and $B_n$. In this way, the optimal state is obtained in the final iteration, as given in Eq.~(\ref{normalized_m_matrix}), where $\ket{\psi^{[i]}}$ provides the coefficients 
$\{ c^{(1)}_n \}$ and $\{ c^{(2)}_n \}$ of the optimal input state.

\begin{tcolorbox}[colback=gray!5,colframe=black,title={Code Availability}]
The implementation of this algorithm was developed in Python and is publicly available at the following \href{https://github.com/Matheus-Eiji/incomp_phase_loss}{GitHub Repository}.
\end{tcolorbox}

\begin{figure}[t]
    \includegraphics[width=0.99\columnwidth]{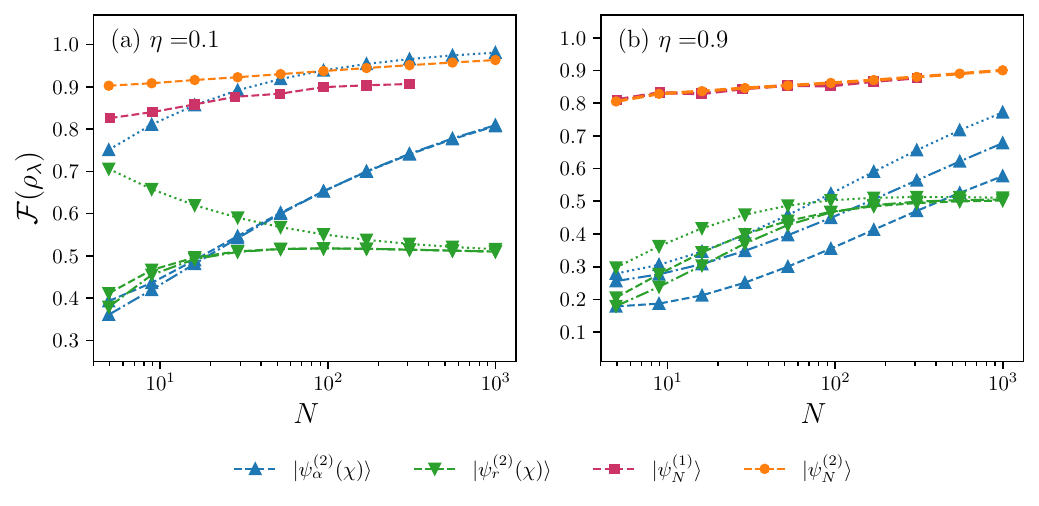}
    \caption{\label{gaussian_vs_optimization}
    Panels (a) and (b) show the probe incompatibility quantifier $\mathcal{F} (\rho_{\bm \lambda})$ in function of the photon number $N$ for different losses, respectively $\eta=0.1$ and $\eta=0.9$. On the plots we have the following states: the optimized states $| \psi^{(1)}_N \rangle$, $| \psi^{(2)}_N \rangle$; the Gaussian states with strong displacement $| \psi^{(2)}_\alpha (\chi) \rangle$ and strong squeezing $| \psi^{(2)}_r (\chi) \rangle$. For the Gaussian states we have $\chi=0$ (dashed), $\chi=\pi/4$ (dash-dotted) and $\chi=\pi/2$ (dotted).}
\end{figure}

In Fig. \ref{gaussian_vs_optimization} we show the normalized QFI for both states, $\mathcal{F}(\rho^{(1)}_N)$ and $\mathcal{F}(\rho^{(2)}_N)$, resulting from the ISS optimization.
Additionally, the off-diagonal elements of the QFI matrix vanish for the optimal two-mode state and become negligible as $N$ increases for the optimal single-mode state.
We remark that the work by Crowley \textit{et al.}~\cite{Crowley2014} employed a gradient-based method 
to optimize two-mode states $|\psi^{(2)}_N\rangle$, 
minimizing the combined variances for photon numbers up to $N = 200$.
In contrast, our iterative see-saw method is able to achieve this optimization problem for photon numbers up to $N=1000$, as shown in Fig.~\ref{gaussian_vs_optimization}.

\begin{figure}[t]
    \centering
    \includegraphics[width=0.9\columnwidth]{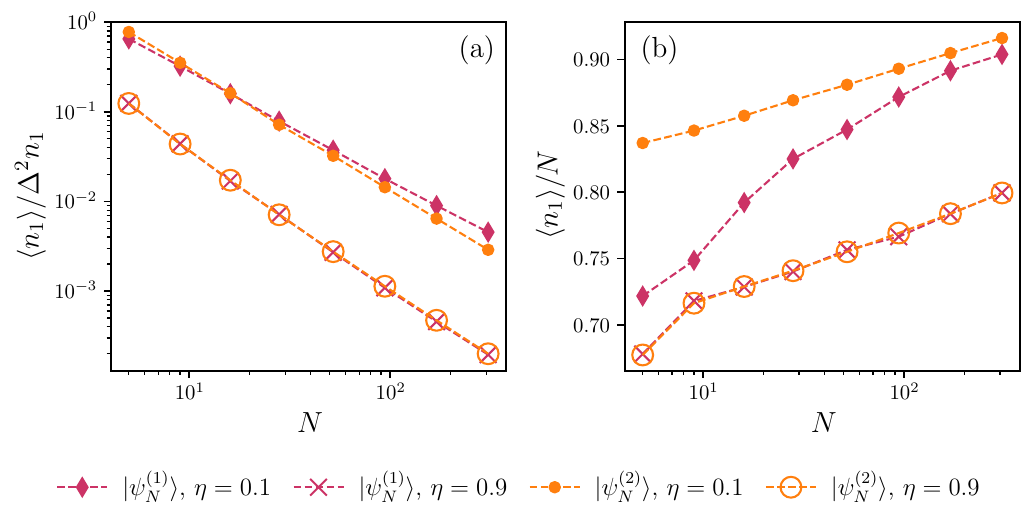}
    \caption{\label{variances_optimization}
        Panel (a) shows the photon number variance $\langle n_1 \rangle / \Delta^2 n_1$; panel (b) the photon number average $\langle n_1 \rangle / N$ of the optimal states $| \psi^{(1)}_N \rangle$ and $| \psi^{(2)}_N \rangle$ obtained from the ISS optimization method.}
\end{figure}

We can further understand how these states overcome probe incompatibility by analyzing the expectation values given in Fig.~\ref{variances_optimization} and comparing with the conditions presented in  Eq.~(\ref{Conditions_SimultaneousEstimation}).
The intuition is that the photon number variance must be sufficiently large to enable precise  phase estimation, as discussed in the beginning of this chapter. Indeed, from Fig.~\ref{variances_optimization}(a), we conclude that both states exhibit a large photon number variance, as $\langle \Delta^2 n_1 \rangle / N$ follows a super-Poissonian scaling, which benefits phase estimation. 
Furthermore, as a necessary condition for probe compatibility, 
the states must also have a mean photon number in the sensing arm approaching $N$. As shown in Fig.~\ref{variances_optimization}(b), this condition is indeed satisfied, where the average photon number approaches that of a Fock state, with $\langle \hat{n}_1 \rangle$ asymptotically reaching $N$.

In addition, the optimal single-mode state for loss estimation is a Fock state~\cite{Adesso2009}, which inherently has zero photon number variance. We, therefore, may expect to have some non-trivial trade-offs here. Interestingly, these trade-offs seem to affect only the single-mode scenario. Intuitively, the presence of the reference mode allows one to have the best of both worlds: large variance of photon number in the sensing mode, but at the same time precise information on the number of photons entering this mode, thanks to photon-number entanglement between sensing and reference modes.

\begin{figure}[t]
    \centering
    \includegraphics[width=0.9\columnwidth]{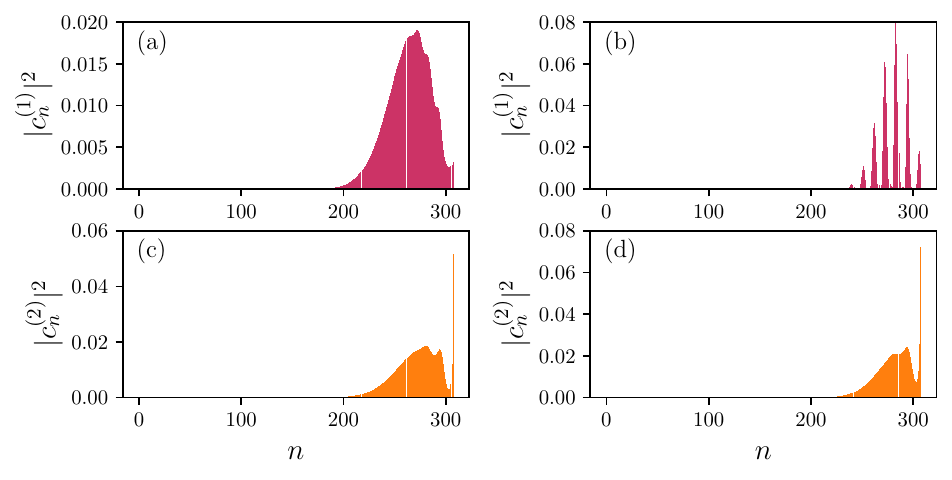}
    \caption{\label{histogram_optimalstate}
        The top panels show the photon-number distribution of the optimal single-mode state $| \psi^{(1)}_{307} \rangle$: (a) for phase estimation only, and (b) for simultaneous phase and loss estimation.
        The bottom panels show the photon-number distribution of the optimal two-mode state $| \psi^{(2)}_{307} \rangle$: (c) for phase estimation only, and (d) for simultaneous phase and loss estimation.
        In all the graphs we fix $\eta=0.1$.}
\end{figure}

This observation is reflected in the fact that the optimal states in the single-mode and two-mode scenarios exhibit very different photon number distributions in the probe mode.
This is highlighted in Fig.~\ref{histogram_optimalstate}, where we show the histogram of the photon number distribution of optimal states for: only phase estimation (at the left) and simultaneous estimation of phase and loss (at the right).
In the single-mode case, the optimal state for simultaneous estimation exhibits a teeth-like structure, which is necessary for accurately estimating loss, as shown in Fig.~\ref{histogram_optimalstate}(b). The intuition is that the separation of the teeth allows one to better determine the original number of photons based on the number of photons detected.
As a result, despite the larger photon-number variance, we obtain better loss sensitivity.
In fact, this structure is not required in the two-mode case, due to the presence of the additional reference beam.

\subsection{Single-mode and two-mode Gaussian states}
\label{probe_incomp_gauss} 

In sequence, we consider the probe state  $|\psi_{\text{in}}\rangle $, represented in Fig.~\ref{phase_loss_GeneralScheme}, corresponding to classes of single-mode and two-mode Gaussian states. At the input, we denote the selected $M$-mode Gaussian state by $|\psi^{(M)}_G\rangle$ , which evolves into the mixed state  $\hat{\rho}^{(M)}_G$  at the output. On this entire section we are considering that the covariance matrices and displacement vectors are written in terms of the complete phase-space coordinates, $\mathcal{A} = (\alpha_1, \alpha_2, \alpha_1^*, \alpha_2^*)$. 

Let a two-mode Gaussian state characterized at the input by the covariance matrix $\sigma_{\text{in}}$ and the displacement vector ${\bf d}_{\text{in}}$. Recall that, as presented in Eq.~(\ref{Gaussian_evolution_general}), the evolution of any Gaussian state is fully determined by the evolution of its  covariance matrix and displacement vector. Thus, we are interested in first investigating how these Gaussian parameters are transformed under the evolution $\Lambda_{\boldsymbol{\lambda}}$, which encodes both the phase $\varphi$ and the loss parameter $\eta$, as illustrated in Fig.~\ref{phase_loss_GeneralScheme}. The phase encoding corresponds to a simple unitary evolution defined in Eq.~(\ref{multiport_evolution_PhaseSpace}). In contrast, the loss encoding is obtained by substituting Eq.~(\ref{loss_def}) on the displacement vector and covariance matrix, given respectively by Eqs.~(\ref{displacement_def}) and ~(\ref{covariance_def}). As a result, we have:
\begin{align}
    & d_i \mapsto \sqrt{\eta_i} d_i , \nonumber\\
    & [\sigma_N]_{ij} \mapsto \sqrt{\eta_i \eta_j} [\sigma_N]_{ij} + (1-\eta_i) \left( 2 \aver{\hat{e}^\dagger_i \hat{e}_i} + 1 \right) \delta_{ij}, \nonumber\\
    & [\sigma_S]_{ij} \mapsto \sqrt{\eta_i \eta_j} [\sigma_N]_{ij},
\end{align}
where $\eta_i$ denotes the loss rate in the corresponding mode, and $\aver{\hat{e}^\dagger_i \hat{e}_i} = 0$ denotes the average number of photons in the virtual mode. Here we are considering losses on the first arm only, then $\eta_1=\eta$ and $\eta_2=1$. Finally, including also the phase effect of the phase $\varphi$ into the previous three equations, we obtain a two-mode Gaussian state with output covariance matrix $\sigma_{\boldsymbol{\lambda}}$ and output displacement vector ${\bf d}_{\boldsymbol{\lambda}}$, explicitly given by:
\begin{align}
    {\bf d}_{\bm \lambda} = A \, {\bf d}_{\text{in}}
    , \qquad
    \sigma_{\bm \lambda} = A \, \sigma_{\text{in}}  A^* + B
    \label{CovDisp_PhaseLoss}
\end{align}
where we introduce the following matrices:
\begin{align}
    A = 
    \begin{pmatrix}
       \sqrt{\eta} \, \e^{i \varphi} & 0 & 0 & 0 \\
       0 & 1 & 0 & 0 \\
       0 & 0 & \sqrt{\eta} \, \e^{-i \varphi}  & 0 \\
       0 & 0 & 0 & 1
    \end{pmatrix}
    , \qquad
    B =  
    \begin{pmatrix}
        1-\eta & 0 & 0 & 0 \\
        0 & 0 & 0 & 0 \\
        0 & 0 & 1-\eta & 0 \\
        0 & 0 & 0 & 0
    \end{pmatrix} .
\end{align}
Note that this transformation does not correspond to the symplectic transformation defined in Eq.~(\ref{Gaussian_evolution_general}), due to the effect of loss, which converts the pure input state into a mixed output state that remains Gaussian. In addition, the effect of losses can also be interpreted as a thermalization process, in which the virtual mode effectively acts as a thermal bath.

For any Gaussian state, an important result is that the quantum QFI matrix can be explicitly computed from the covariance matrix and displacement vector of the output state~\cite{Safranek2019},
\begin{equation}
    F_{ij} = \frac{1}{2} \lim_{\nu \rightarrow 1}  \text{vec}(\partial_{\lambda_i} \sigma_{\bm \lambda} )^\dagger \mathcal{M}^{-1} \text{vec}(\partial_{\lambda_j} \sigma_{\bm \lambda} ) + 2 (\partial_{\lambda_i} {\bf d}_{\bm \lambda} )^\dagger \sigma^{-1}_{\bm \lambda} (\partial_{\lambda_j} {\bf d}_{\bm \lambda}) ,
    \label{qfi_general_gaussian}
\end{equation}
$\text{vec}(\cdots)$ means the column vectorization of the corresponding matrix, and the following  $2M \times 2M$ complex matrix is introduced
\begin{equation}
    \mathcal{M} = \nu^2 \sigma^*_{\bm \lambda} \otimes \sigma_{\bm \lambda} - \Omega \otimes \Omega .
    \label{qfi_general_gaussian_matrixM}
\end{equation}
In the previous equation, $\Omega$ is the symplectic matrix defined in Eq.~(\ref{symplectic_matrix}), and $\nu$ is a regularization factor introduced to avoid singularities when the state is pure, i.e. $\eta \rightarrow 1$. This expression will be used in this entire section to calculate the  QFI matrix of the considered Gaussian states.

We begin by considering a \textit{single-mode Gaussian state} as the probe state. In this case, photons are injected only into the probe mode. The most general pure single-mode Gaussian state corresponds to a displaced squeezed state~\cite{BookSerafini}, denoted by $\ket{\psi^{(1)}_G}$, which is generated by the action of the squeezing operator followed by a displacement operator on the vacuum,
\begin{equation}
    | \psi^{(1)}_G \rangle = \hat{D}_1 (\alpha) \hat{S}_1 (r) | 0, 0 \rangle ,
    \label{gaussian_singlemode}
\end{equation}
with input beam splitter transmissivity to
$\tau_{\mathrm{in}} = 1$, as illustrated in Fig.~\ref{phase_loss_GeneralScheme}.(a).
Here, both the squeezing operator, defined in Eq.~(\ref{sm_operator_def}), and the displacement operator, defined in Eq.~(\ref{displacement_op_def}), act exclusively on the probe mode, in such a way that we do not have access to the ancillary mode, which is then represented by a vacuum state. We denote the average total photon number by $\aver{N} = \aver{n_r} + \aver{n_\alpha}$, where the squeezing contribution is $\aver{n_r} = \text{sinh}^2r$ and the coherent contribution is $\aver{n_\alpha} = |\alpha|^2$. Starting from the vacuum, the covariance matrix of the state $\ket{\psi^{(1)}_G}$ is obtained from the symplectic transformations defined in Eq.~(\ref{Gaussian_evolution_general}), by considering the symplectic matrix of the single-mode squeezing operator given in Eq.~(\ref{sm_squeezing_matrix}), resulting in
\begin{equation}
    \sigma^{(1)}_{\text{in}} = 
    \begin{pmatrix}
    \cosh (2r)  & 0 & -\sinh (2r) \hspace{0.5mm} \text{e}^{i \theta_1}  & 0 \\
    0 & 1 & 0 & 0 \\
    -\sinh (2r) \hspace{0.5mm} \text{e}^{-i \theta_1} & 0 & \cosh (2r) & 0 \\
    0 & 0 & 0 & 1
    \end{pmatrix} .
    \label{cov_singlemode}
\end{equation}
Since the squeezing operator $\hat{S}_1 (r) $ does not give any first moment contribution, i.e., $\aver{\hat{a}^\dagger_k}=0$ or $\aver{\hat{a}_k}=0$, the displacement vector is obtained directly from the transformation given in Eq.~(\ref{displacement_action}) applied in Eq.~(\ref{Gaussian_evolution_general}),
\begin{equation}
    {\bf d}^{(1)}_{\text{in}} =
    \begin{pmatrix}
    |\alpha| \hspace{0.5mm} \text{e}^{i \mu} &
    0 &
    |\alpha| \hspace{0.5mm} \text{e}^{-i \mu}  &
     0
    \end{pmatrix}^t ,
    \label{displacement_vector}
\end{equation}
where we denote $\alpha = |\alpha| \, \e^{i \mu}$. Note that these Gaussian parameters correspond to a vacuum state in the ancillary mode, represented by the coordinates $(\alpha_2, \alpha_2^*)$ Thus, the total state is, in fact, a single-mode Gaussian state.

In sequence, the covariance matrix and displacement vector at the output are obtained by replacing ${\bf d}^{(1)}_{\text{in}}$ and $\sigma^{(1)}_{\text{in}}$ into Eq.~(\ref{CovDisp_PhaseLoss}); and finally, the corresponding QFI matrix is obtained by inserting these results into Eq.~(\ref{qfi_general_gaussian}).
The final expressions are quite involved, thus we present here the asymptotic limits for large total photon number $\aver{N}$, while the complete expressions are presented in the Appendix~\ref{appendix:qfi_gaussian}. In order to determine the ratio of energy spent to displacement and squeezing, we introduce the coefficient $0 < p < 1$, which governs the energy contributions. We assume a regime dominated by displacement, which is experimentally more feasible. 
Thus, we set the displacement and squeezing contributions respectively as 
\begin{equation}
    \langle n_\alpha \rangle = \langle N \rangle - \langle N \rangle^{p}
    , \qquad
    \langle n_r \rangle = \langle N \rangle^{p}
    , \qquad
    0 < p < 1 .
    \label{singlemode_energy_disttrib}
\end{equation}
Finally, for each term of the probe incompatibility quantifier defined in Eq.~(\ref{probe_incomp}), we have the following asymptotic limits:
\begin{align}
    & \frac{F_{\varphi \varphi} (\rho^{(1)}_G)}{F^{\text{(max)}}_\varphi} \overset{\aver{N} \rightarrow \infty}{\longrightarrow} \sin^2\left(\frac{\theta_1-2\mu}{2} \right), \\
    & \frac{F_{\eta \eta} (\rho^{(1)}_G)}{F^{\text{(max)}}_\eta} \overset{\aver{N} \rightarrow \infty}{\longrightarrow} 
    \cos^2\left(\frac{\theta_1-2\mu}{2} \right) ,
    \label{qfi_sm_alpha}
\end{align}
which results in:
\begin{equation}
    \mathcal{F} (\rho^{(1)}_G) \overset{\aver{N} \rightarrow \infty}{\longrightarrow} \frac{1}{2} .
\end{equation}
Therefore, the incompatibility of employing a single-mode Gaussian state for the simultaneous estimation of phase and loss becomes evident, since $\mathcal{F}=1/2$ corresponds to the case in which only one parameter is estimated while the other is completely ignored \footnote{In Appendix~\ref{appendix:qfi_gaussian}, we also show that in the regime dominated by squeezing we also have $\mathcal{F} = 1/2$, indicating that probe incompatibility depends on the choice of the input state rather than on the energy contributions.}. Achieving the ultimate quantum precision bound for phase estimation requires the state being phase squeezed ($\theta_1 - 2 \mu = \pi$), which suppresses the loss estimation. In contrast, achieving the ultimate quantum precision bound for loss estimation requires the state being amplitude squeezed ($\theta_1 - 2 \mu = 0$), which suppresses the phase estimation.
In~\cite{Monras2007} the optimal phase choice for the loss estimation agrees with the Eq.~\eqref{qfi_sm_alpha}, however, there was no mention about the fundamental bound~\eqref{upper_loss}, since it was not known at the time.

As discussed on the beginning of this section, some intuition about probe incompatibility can be gained  by examining the photon number variance of the probe mode.
For the state considered here, defined in Eq.~\eqref{gaussian_singlemode}, we have:
\begin{equation}
    \Delta^2 n_1 = 2 \aver{n_r} (\aver{n_r} + \aver{n_\alpha}+ 1) + \aver{n_\alpha}- 2 \aver{n_\alpha}\sqrt{\aver{n_r} (\aver{n_r} + 1)} \cos \left( \theta - 2 \mu \right).
\end{equation}
According to Eq.~\eqref{qfi_sm_alpha}, in the strong displacement regime the optimal phase relation for phase estimation is $\theta - 2\mu = \pi$, which maximizes the photon number variance.
In contrast, the optimal phase relation for the loss estimation is $\theta - 2\mu = 0$, which minimizes the photon number variance. Note that, in the last case the photon number variance has a scaling $\Delta^2 n_1  \sim \aver{N}^{p/2}$, which cannot satisfy the necessary condition for probe compatibility, given by Eq.~(\ref{Conditions_SimultaneousEstimation}).
Additionally, the single-mode Gaussian state cannot reproduce the characteristic teeth-like structure of the optimal single-mode state $\ket{\psi^{(1)}_N}$, which is required for estimating loss in the simultaneous estimation scenario, as shown in Fig.~\ref{histogram_optimalstate}(b).

In addition, the correlation term of the QFI matrix is given by:
\begin{equation}
    F_{\varphi \eta} \big( \rho^{(1)}_G \big) =
    \frac{4  \eta \sin (\theta_1 -2 \mu )  \sqrt{\aver{n_r} (\aver{n_r}+1)} \aver{n_\alpha}}{4 (\eta -1) \eta  \aver{n_r}-1},
    \label{qfi_sm_correlation}
\end{equation}
which is zero in the two extremal cases $\theta_1 - 2 \mu = 0$ (squeezing and displacement in the same direction) or $\theta_1 - 2 \mu = \pi$ (squeezing and displacement in the opposite direction). In the strong displacement regime, each case corresponds to the optimal value for one of the QFIs, while the other QFI vanishes.
Furthermore, the maximum value of the correlation is attained when $\theta_1 - 2 \mu = \pi/2$, which corresponds to the case when both parameters are estimated with the same importance, resulting in $F_{\varphi \varphi} \big( \rho^{(1)}_G \big) = F_{\eta \eta} \big( \rho^{(1)}_G \big)$. 
We mention that Ref.~\cite{Pinel2013} also discusses parameter estimation using single-mode Gaussian states, 
including both phase and loss parameters in the single-parameter estimation scenario. However, that work did not discusses ultimate quantum limits or the asymptotic trade-offs presented here.

To conclude, the result obtained for the single-mode Gaussian state are consistent with the discussion presented in the previous section: either an ancillary mode is required, or a highly exotic single-mode state, such as the one shown in Fig.~\ref{histogram_optimalstate}(b). 

In the following, we consider different classes of \textit{two-mode Gaussian states}, as illustrated in Fig.\ref{phase_loss_GeneralScheme}(b). According to the previous discussion, a single-mode state exhibits probe incompatibility. 
To overcome this limitation, it is therefore necessary to exploit the ancillary.
From the single-mode and two-mode squeezing operators defined respectively in Eqs.~(\ref{sm_operator_def}) and ~(\ref{sm_operator_def}) we introduce a generalized two-mode squeezing operator as follows
\begin{equation}
    \hat{S}_{1,2} (r, \chi)
     = \text{exp} \left[
    r ~  \text{cos} \chi \left(
    \frac{
    \text{e}^{i \theta_1 } (\hat{a}^\dagger_1)^2 +
    \text{e}^{i \theta_2 } (\hat{a}^\dagger_2)^2
    }{2}
    \right) 
     + r ~ \text{sin} \chi \, \text{e}^{i \theta} \hat{a}^\dagger_1 \hat{a}^\dagger_2 - \text{h.c.} \right],
     \label{general_squeez}
\end{equation}
where here $\aver{n_r} = 2 ~ \sinh^2 r$  gives the average total photon number generated  in the system. The phase $\chi$ plays an important role in the preparation of the input state, as it determines when this operator generates two single-mode squeezed states ($\chi = 0$), a two-mode squeezed state ($\chi = \pi/2$), or a state between these two ($0 < \chi < \pi/2$). Such operator can be generated from the interference of two single-mode squeezed states in a beam splitter, where the phase $\chi$ is controlled by changing the transmissivity and reflectance phase of this beamsplitter \cite{Yeoman1993}. Thus, the general probe state is a two-mode Gaussian state starting from this generalized squeezed state, preceded by a displacement on the probe mode, and followed by the action of a beamsplitter with transmissivity $\tau_{in}$ coupling the probe and ancillary modes,
\begin{equation}
    | \psi^{(2)}_G (\chi) \rangle = \hat{B}(\tau_{\mathrm{in}}) \hat{D}_1 (\alpha) \hat{S}_{1,2} (r, \chi) | 0, 0 \rangle ,
    \label{gaussian_twomode}
\end{equation}
as illustrated in Fig.~\ref{phase_loss_GeneralScheme}.(b). The total photon number of such state is $\aver{N} = \aver{n_r} + \aver{n_\alpha} = 2 ~ \sinh^2 r + |\alpha|^2$.
In the previous equation, the displacement operator acts in exactly the same way as in the single-mode Gaussian state, and the beam splitter is represented by the already well-known matrix
\begin{equation}
    B(\tau_{\mathrm{in}})   =
    \begin{pmatrix}
    \sqrt{\tau_{\mathrm{in}}} & i \sqrt{1-\tau_{\mathrm{in}}}  \\
    i \sqrt{1-\tau_{\mathrm{in}}} & \sqrt{\tau_{\mathrm{in}}}
    \end{pmatrix}.
\end{equation}
Here, we assume that the displacement operator acts only on the first mode to ensure that, in principle, more photons are prepared into the probe mode rather than into the ancillary mode. 

Following the symplectic transformation shown in Eq.~(\ref{Gaussian_evolution_general}), we obtain the covariance matrix of this general two-mode Gaussian state in two steps:
(i) we first consider the action of the generalized two-mode squeezing operator $\hat{S}_{1,2}(r, \chi)$, using the corresponding generalized symplectic matrices from Eqs.~(\ref{sm_squeezing_matrix}) and~(\ref{tm_squeezing_matrix}); 
(ii) then we consider the action of the corresponding symplectic interferometer matrix to $B(\tau_{\mathrm{in}})$, presented to Eq.~(\ref{multiport_evolution_PhaseSpace}). This process results in the following covariance matrix
\begin{equation}
    \sigma^{(2)}_{\text{in}} = 
    \begin{pmatrix}
        B(\tau_{\mathrm{in}}) & 0_M \\
        0_M & B(\tau_{\mathrm{in}})^*
    \end{pmatrix}
    \begin{pmatrix}
        \cosh (2r) \, I_2  & 
        -\sinh (2r) \, R_\chi \\
        -\sinh (2r) \, R^*_\chi  & 
        \cosh (2r) \, I_2
    \end{pmatrix} 
        \begin{pmatrix}
        B(\tau_{\mathrm{in}})^\dagger & 0_M \\
        0_M & B(\tau_{\mathrm{in}})^t
    \end{pmatrix},
    \label{cov_twomode}
\end{equation}
where we introduce the following rotation matrix
\begin{equation}
    R_\chi = 
    \begin{pmatrix}
    \cos \chi \hspace{0.5mm} \text{e}^{i \theta_1}  & 
    \sin \chi \hspace{0.5mm} \text{e}^{i \theta} \\
    \sin \chi \hspace{0.5mm} \text{e}^{i \theta} & 
    \cos \chi \hspace{0.5mm} \text{e}^{i \theta_2}
    \end{pmatrix} ,
\end{equation}
which determines whether the transformation produces two independent single-mode squeezed states or a one two-mode squeezed state. 
The displacement vector is the same as for the single-mode case, 
\begin{equation}
    {\bf d}^{(2)}_{\text{in}} = {\bf d}^{(1)}_{\text{in}} ,
\end{equation}
since the displacement is applied only to the probe mode and the squeezing operator does not contribute to first moment averages.

In sequence, the covariance matrix and displacement vector at the output are obtained by substituting ${\bf d}^{(2)}_{\text{in}}$ and $\sigma^{(2)}_{\text{in}}$ into Eq.~(\ref{CovDisp_PhaseLoss}); and finally, the corresponding QFI matrix is obtained by inserting these results into Eq.~(\ref{qfi_general_gaussian}).
The final expressions are quite involved, thus we present here the asymptotic limits for large total photon number $\aver{N}$, while some of the complete expressions are presented in the Appendix~\ref{appendix:qfi_gaussian}.
Analogously to the single-mode Gaussian state, we introduce the coefficient $0 < p < 1$  that defines the energy contributions from each source.
From the entire class of two-mode Gaussian states $| \psi^{(2)}_G (\chi) \rangle$, defined in Eq.~(\ref{gaussian_twomode}), we classify the results into a few categories of states depending on the nature of the parameter $\chi$ and on the energy distribution. In the following, we highlight two main cases:

\begin{itemize}
    \item We first consider the case where we start with \textit{two independent single-mode squeezed states}, setting  $\chi=0$ in Eq.~(\ref{general_squeez}).
    In addition, we also assume the regime with \textit{an energy distribution dominated by displacement}, in the same way as considered for the single-mode state, as given by Eq.~(\ref{singlemode_energy_disttrib}). Such a state results in the following asymptotic limit for the probe incompatibility quantifier:
    \begin{equation}
        \mathcal{F} \overset{\aver{N} \rightarrow \infty}{\longrightarrow} \tau_{\mathrm{in}}.
        \label{assymptotic_tm_gauss}
    \end{equation}
    This result suggests that, as $\tau_{\text{in}} \rightarrow 1$, the incompatibility of the probe vanishes, as defined in Eq.~(\ref{probe_incomp}), motivating the introduction of a coefficient $0 < q < 1$ that governs how fast the transmissivity goes to one. 
    Note that, since we are considering displacement only in the first mode, the previous asymptotic limit gives $\langle n_1 \rangle \rightarrow \langle N \rangle$ in this strong displacement regime, which is precisely the first condition presented in Eq.~\ref{Conditions_SimultaneousEstimation}.
    With this parametrization, we prove analytically that the following state can asymptotically overcome probe incompatibility:
    \begin{equation}
        \ket{\psi^{(2)}_\alpha (0)} \equiv \Big\{ 
        \ket{\psi^{(2)}_G (\chi)} \big|
         ~
        \chi=0
        , ~
        \langle n_\alpha \rangle = \aver{N} - \aver{N}^{p}
        , ~
        \tau_{\mathrm{in}} = 1 - \frac{1}{\aver{N}^q}
        \Big\} .
        \label{main_state_chi0}
    \end{equation}
    In Appendix~\ref{appendix:qfi_gaussian}, the expressions for the phase and loss QFI of the state $\ket{\psi^{(2)}_\alpha(0)}$ are given respectively by Eqs.~(\ref{qfi_phase_chi0_alpha}) and~(\ref{qfi_loss_chi0_alpha}). Following these results, in the asymptotic limit, each term of the probe incompatibility quantifier defined in Eq.~(\ref{probe_incomp}) results:
    \begin{equation}
        \frac{F_{\varphi \varphi} ( \rho^{(2)}_\alpha (0) )}{F^{\text{(max)}}_\varphi} \overset{\aver{N} \rightarrow \infty}{\longrightarrow} 
        \begin{cases} 
        1 & \text{for } q < p, \\
            1-\frac{\eta  \cos ^2\left(\frac{\theta_1-2\mu}{2}\right)}{1 + (1-\eta ) \cos (\theta_1-\theta_2)} & \text{for } q = p, \\
        \sin^2\left(\frac{\theta_1-2\mu}{2} \right) & \text{for } q > p,
        \end{cases}
        \label{qfi_phase_sm}
    \end{equation}
    \begin{equation}
        \frac{F_{\eta \eta} ( \rho^{(2)}_\alpha (0) )}{F^{\text{(max)}}_\eta} \overset{\aver{N} \rightarrow \infty}{\longrightarrow} 
        \begin{cases} 
        1 & \text{for } q < p, \\
        1-\frac{\eta  \sin ^2\left(\frac{\theta1-2 \mu}{2})\right)}{1 + (1-\eta ) \cos (\theta1-\theta_2)} & \text{for } q = p, \\
        \cos^2\left(\frac{\theta_1-2\mu}{2} \right) & \text{for } q > p,
        \end{cases}
        \label{qfi_loss_sm}
    \end{equation}

    From the last two equations, we conclude that the ultimate quantum precision bounds for phase and loss are achieved simultaneously when $q<p$, which means that the transmissivity is converging sufficiently slowly to one, ensuring that the squeezing contributions from the two beams always interfere, i.e., $\lim_{\aver{N} \rightarrow \infty} (1-\tau_{\mathrm{in}})  \aver{n_r} = \infty$. In contrast, when $q>p$ the transmissivity is converging sufficiently fast to one in such a way that the squeezing contributions from the two beams do not interfere asymptotically, i.e., $\lim_{\aver{N} \rightarrow \infty} (1-\tau_{\mathrm{in}})  \aver{n_r} = 0$.
    This last case reproduces the results of the single-mode state given by Eq.~\eqref{qfi_sm_alpha} and then the same incompatibility problem appears. The probe incompatibility quantifier for this state is shown in blue at Fig.~\ref{gaussian_vs_optimization} (dashed lines). 
    Additionally, for $\aver{N} \gg 1$, the off-diagonal QFI matrix element of the state $\ket{\psi^{(2)}_\alpha (0)}$ can be approximated by:
    \begin{eqnarray}
        F_{\varphi \eta} \big( \rho^{(2)}_\alpha (0) \big)  
        &\approx&
        \frac{4 \eta \tau^2_{\mathrm{in}} \sin(\theta_1-2 \mu)  \sqrt{\aver{n_r}(\aver{n_r} + 1)} \aver{n_\alpha}}{\text{den}} +
        \label{qfi_correlation_sm}\\
        && - \frac{4 \eta \tau_{\mathrm{in}} (1 - \tau_{\mathrm{in}}) \sin(\theta_2-2 \mu) \sqrt{\aver{n_r}(\aver{n_r} + 1)} \aver{n_\alpha}}{\text{den}},
        \nonumber
    \end{eqnarray}
    where we introduce $\text{den} = 4 \eta  (1-\eta ) \aver{n_r}+8 (1-\eta)^2 (1-\tau_{\mathrm{in}}) \tau_{\mathrm{in}} \aver{n_r} (\aver{n_r}+1) \cos (\theta_1-\theta_2)+8 (1-\eta)^2 (1-\tau_{\mathrm{in}}) \tau_{\mathrm{in}} \aver{n_r}(\aver{n_r} + 1)+1 $.
    The previous equation is trivially zero when $\theta_1 - \mu = \theta_2 - \mu =0$, or asymptotically when $\tau_{\mathrm{in}} \rightarrow 1$ and $\theta_1 - \mu=0$. The results presented here for the state $\ket{\psi^{(2)}_\alpha (0)}$ have not yet been reported in the literature and provide new insights into the asymptotic conditions required to overcome such probe incompatibility.

    \item Next, we consider the case in which we start with \textit{one two-mode squeezed state}, by setting $\chi=\pi/2$ in Eq.~(\ref{general_squeez}).
    We also assume the regime of \textit{energy distribution dominated by displacement}. Here, the probe incompatibility also vanishes asymptotically for $\tau_{\mathrm{in}} = 1$, in the same way as presented in Eq.~(\ref{assymptotic_tm_gauss}).
    In contrast to the previous case, we can set $\tau_{\mathrm{in}} = 1$ from the beginning, since the two-mode squeezed state guarantees entanglement between the the probe and ancillary modes for any range of $\aver{N} > 1$. Thus, we define the following state:
    \begin{equation}
        \ket{\psi^{(2)}_\alpha (\pi/2)} \equiv \Big\{ 
        \ket{\psi^{(2)}_G (\chi)} \big|
        , ~
        \chi= \pi/2
        , ~
        \langle n_\alpha \rangle = \aver{N} - \aver{N}^{p}
        , ~
        \tau_{\mathrm{in}} = 1 
        \Big\} .
        \label{main_state_chipi2}
    \end{equation}
    In Appendix~\ref{appendix:qfi_gaussian}, the expressions for the phase and loss QFI of the state $\ket{\psi^{(2)}_\alpha(\pi/2)}$ are given respectively by Eqs.~(\ref{qfi_phase_chipi2_alpha}) and~(\ref{qfi_loss_chipi2_alpha}). Following these results, in the asymptotic limit, each term of the probe incompatibility quantifier defined in Eq.~(\ref{probe_incomp}) results:
    \begin{align}
        & \frac{F_{\varphi \varphi} ( \rho^{(2)}_\alpha (\pi/2) )}{F^{\text{(max)}}_\varphi} \overset{\aver{N} \rightarrow \infty}{\longrightarrow} 1 ,
        \\
        & \frac{F_{\eta \eta} ( \rho^{(2)}_\alpha (\pi/2) )}{F^{\text{(max)}}_\eta} \overset{\aver{N} \rightarrow \infty}{\longrightarrow} 1 .
    \end{align}
    We mention that after an appropriate reparametrization, the previous two limits can be derived from the results of previous works~\cite{Dowran2021,Woodworth2022}, which considered a bright two-mode squeezed state (displacement before squeezing) as the probe state. The probe incompatibility quantifier for this state is shown in blue at Fig.~\ref{gaussian_vs_optimization} (dotted lines). Finally, for $\aver{N} \gg 1$, the off-diagonal QFI matrix element of the state $\ket{\psi^{(2)}_\alpha (\pi/2)}$ can be approximated by:
    \begin{equation}
        F_{\varphi \eta} \big( \rho^{(2)}_\alpha (\pi/2) \big)  
        \approx
        \frac{8 \cos (\theta -2 \mu ) \eta \tau_{\mathrm{in}} \sqrt{(1-\tau_{\mathrm{in}}) \tau_{\mathrm{in}}} \sqrt{\aver{n_r}(\aver{n_r}+1)} \aver{n_\alpha}}{4 (1-\eta) \Big[ 4 (1-\eta) (1-\tau_{\mathrm{in}}) \tau_{\mathrm{in}}+(\eta -1) (1-2 \tau_{\mathrm{in}})^2 \aver{n_r} -1 \Big] \aver{n_r} -1
        }
        \label{qfi_correlation_tm}
    \end{equation}
    which is zero when the squeezing and displacement are performed in an orthogonal direction, $\theta - 2 \mu = \pm \pi/2$.

\end{itemize}

In agreement with the properties discussed In Eq.~(\ref{Conditions_SimultaneousEstimation}), these last two results reveal the essential structure that a Gaussian state must possess to overcome probe incompatibility and achieve asymptotically $\mathcal{F} \rightarrow 1$.  This structure comprises: (i) more photons entering the probe mode than the ancillary mode, i.e., $\tau_{\mathrm{in}} \rightarrow 1$, such that most of the energy is spent encoding the parameters of interest; (ii) an ancillary mode that continually interferes with the photons in the probe mode. In Fig.~\ref{gaussian_vs_optimization} is shown in blue all the two-mode Gaussian states that achieve the ultimate quantum precision limits for both parameter.

For completeness, an intermediate state between the two previous cases was considered, corresponding to $\ket{\psi^{(2)}_\alpha(\pi/4)}$, also with energy distribution dominated by the displacement. 
This state is also able to overcome probe incompatibility and exhibits a performance between $\ket{\psi^{(2)}_\alpha(0)}$ and $\ket{\psi^{(2)}_\alpha(\pi/2)}$, as shown in Fig.~\ref{gaussian_vs_optimization} (dashed-dotted lines). 
In this case, $\tau_{\mathrm{in}} = 1$ can be set from the beginning, since for $\chi \neq 0$ the squeezing guarantees entanglement between the probe and ancillary modes. It also provides new insights, since such states, whose statistical properties lie between those of two single-mode and one two-mode squeezed states, are generally not considered in quantum estimation.

Finally, states with energy distributions dominated by squeezing were also considered for different values of $\chi$. 
These states are denoted as
\begin{equation}
    \ket{\psi^{(2)}_r (\chi)} \equiv \Big\{ 
    \ket{\psi^{(2)}_G (\chi)} \big|
    , ~
    \chi \in [0, \pi/2]
    , ~
    \langle n_\alpha \rangle = \aver{N}^{p}
    , ~
    \tau_{\mathrm{in}} = 1 -\frac{1}{\aver{N}^q}
        \Big\} ,
\end{equation}
whith $0 < p,q < 1$. All cases of such states considered, converge to only half of the ultimate quantum precision, i.e., $\mathcal{F} ( \rho^{(2)}_r (\chi) ) \rightarrow 1/2$, as shown in green at Fig.~\ref{gaussian_vs_optimization}. Therefore, the class of states with stronger squeezing are not able to achieve the ultimate bound simultaneously for both parameters. This behavior can be attributed to the preparation of states with half of the photons in the probe mode and half in the ancillary mode, which clearly do not satisfy the first condition given in Eq.~(\ref{Conditions_SimultaneousEstimation}).

%--------------------------------------------------------------------------------------------------------------------------------------------------------------------------%

\section{Measurement incompatibility in the simultaneous estimation of phase and loss}
\label{sec:measurem_incomp}

In the previous section, we identified classes of Gaussian and non-Gaussian states capable of achieving the ultimate quantum precision limits for the simultaneous estimation of phase and loss. Thus, giving that optimal probe states, there are two remaining questions: (i) First, which measurement scheme is able to saturate the ultimate quantum CRB for each parameter? That is, which specific measurement strategies results in $\Delta^2 \lambda_i = 1 / F^{(\text{max})}$ for each parameter. (ii) Second, that measurement strategy is the same for both parameters? That is, even if there is \textit{no probe incompatibility}, is there \textit{measurement incompatibility}?

To begin, we recall the definition of the expectation value of the average of the commutators of the SLDs, given by Eq.~\eqref{incomp}.
When it is not zero, it implies trade-offs between the precision of different parameters, thus limiting the simultaneous estimation performance when using one single measurement strategy for both parameters.
First, for the probe state $|\psi^{(2)}_N\rangle$,
it was shown in the first work on the simultaneous estimation of phase and loss~\cite{Crowley2014} 
that the expectation value of such commutator is given by:
\begin{equation}
     I_{\varphi \eta} ( \rho^{(2)}_N ) =  \frac{i F_{\varphi \varphi} (\rho^{(2)}_N )}{2 \eta} .
     \label{incomp_psin}
\end{equation}
As a result, we have $I_{\varphi \eta} = 0$ only if $F_{\varphi \varphi} = 0$, which means that the measurement incompatibility can be overcome only at the cost of vanishing phase-estimation precision, 
a clearly undesirable situation. 

In a similar way as the QFI matrix in Eq.~(\ref{qfi_general_gaussian}), the expectation value $I_{\varphi \eta}$ of any Gaussian state can be calculated in terms of the covariance matrix $\sigma_{\bm \lambda}$ and displacement vector ${\bf d}_{\bm \lambda}$ as follows~\cite{Safranek2019}:
\begin{align}
    I_{ij} = & \, \lim_{\nu \rightarrow 1} \text{vec}(\partial_{\lambda_i} \sigma_{\bm \lambda} )^\dagger \mathcal{M}^{-1} 
    (\sigma^*_{\bm \lambda} \otimes \Omega - \Omega \otimes \sigma_{\bm \lambda})
    \mathcal{M}^{-1}\text{vec}(\partial_{\lambda_j} \sigma_{\bm \lambda} )  + 4 (\partial_{\lambda_i} {\bf d}_{\bm \lambda} )^\dagger \sigma^{-1}_{\bm \lambda} 
    \Omega \,
    \sigma^{-1}_{\bm \lambda} (\partial_{\lambda_j} {\bf d}_{\bm \lambda}) ,
\end{align}
where the matrix $\mathcal{M}$ was defined in Eq.~(\ref{qfi_general_gaussian_matrixM}) and $\Omega$ is the symplectic matrix in Eq.~(\ref{symplectic_matrix}). Here, we restrict our analysis to the Gaussian states that can achieve the ultimate quantum limit, 
namely, the states in the strong-displacement regime, 
$\ket{\psi^{(2)}_\alpha(0)}$ and $\ket{\psi^{(2)}_\alpha(\pi/2)}$, 
defined respectively in Eqs.~(\ref{main_state_chi0}) and~(\ref{main_state_chipi2}). For the first state, we have asymptotically that:
\begin{equation}
    \frac{I_{\varphi \eta} ( \rho^{(2)}_\alpha (0) )}{F^{\text{(max)}}_\varphi} \overset{N \rightarrow \infty}{\longrightarrow} 
    \begin{cases} 
     \frac{i}{\eta} & \text{for } q < p, \\
     \frac{i \eta \big[ (1-\eta ) \cos (\theta_1-\theta_2)+1 \big]}{(1-\eta ) \big[ \cos (\theta_1-\theta_2)+1 \big] } & \text{for } q = p, \\
     0 & \text{for } q > p,
    \end{cases}
    \label{incomp_sm}
\end{equation}
end for the second state:
\begin{equation}
    \frac{I_{\varphi \eta} ( \rho^{(2)}_\alpha (\pi/2) )}{F^{\text{(max)}}_\varphi} \overset{N \rightarrow \infty}{\longrightarrow} \frac{i}{\eta} .
    \label{incomp_tm}
\end{equation}

From the two previous equations, we conclude that the Gaussian states possess the same measurement incompatibility 
as the state $\ket{\psi^{(2)}_N}$. 
Therefore, the fact that $I_{\varphi \eta}$ cannot vanish in any of these cases reveals a fundamental measurement incompatibility, 
thus addressing question~(ii) stated at the beginning of this section.

\subsection{Measurement incompatibilities of practical optical detection schemes}

In this final section, we consider two common detection schemes: photon-number-resolving (PNR) and homodyne detection, to understand how the measurement incompatibilities discussed previously are manifested. 
To optimize the detection strategy, we introduce an additional beam splitter with transmissivity $\tau_{\mathrm{out}}$, 
placed right before detection, which transforms the output basis $\hat{b}_k$ into the detection basis $\hat{c}_k$, as follows:
\begin{equation}
     \begin{pmatrix}
    \hat{c}_1 \\
    \hat{c}_2
    \end{pmatrix} 
    =
    \begin{pmatrix}
    \sqrt{\tau_{\mathrm{out}}} & -i \sqrt{1-\tau_{\mathrm{out}}}  \\
    -i \sqrt{1-\tau_{\mathrm{out}}} & \sqrt{\tau_{\mathrm{out}}}
    \end{pmatrix}
     \begin{pmatrix}
    \hat{b}_1 \\
    \hat{b}_2
    \end{pmatrix} .
    \label{output_beams}
\end{equation}
Then, we select two observables of interest in each detection mode, $\hat{X}_1(\hat{c}_1, \hat{c}^\dagger_1)$ and $\hat{X}_2(\hat{c}_2, \hat{c}^\dagger_2)$, which are
the photon-number operator (in the case of PNR) and the quadrature operator (in the case of homodyne detection).
Thus, for each state considered $\rho_{\bm \lambda}$, the variances for the estimation of phase and loss are given by the generalization of the error propagation formula, given at Eq.~(\ref{error_prop}), as follows
\begin{equation}
    \Delta^2 \lambda_i (\rho_{\bm \lambda}) \equiv \left[ 
    \begin{pmatrix}
    \frac{d  \langle \hat{X}_1 \rangle}{d \lambda_i} & \frac{d  \langle \hat{X}_2 \rangle }{d \lambda_i}
    \end{pmatrix}
    \begin{pmatrix}
    \cov(\hat{X}_1, \hat{X}_1)  & \cov(\hat{X}_1, \hat{X}_2) \\
    \cov(\hat{X}_2, \hat{X}_1)  &  \cov(\hat{X}_2, \hat{X}_2) 
    \end{pmatrix}^{-1}
    \begin{pmatrix}
    \frac{d  \langle \hat{X}_1 \rangle}{d \lambda_i} \\ \frac{d  \langle \hat{X}_2 \rangle}{d \lambda_i}
    \end{pmatrix}
    \right]^{-1} ,
    \label{var}
\end{equation}
From the previous equation, we expect that maximizing the derivatives with respect to 
$\langle \hat{X}_1 \rangle$ and $\langle \hat{X}_2 \rangle$ results in a minimization of the corresponding variances. Notice that here we consider the practical approach in which the parameters are estimated from the mean values of these observables.  In order to understand how the incompatibility is manifested, we determine the conditions that gives the optimal variance obtained from single-parameter estimation and then check if these conditions coincides for both parameters. 
\textit{An optimal variance is achieved when it saturates the quantum CRB, resulting, at least asymptotically, in:}
\begin{equation}
    \Delta^2 \lambda_i (\rho_{\bm \lambda}) \, F^{\text{(max)}}_{\lambda_i}  \overset{N \rightarrow \infty}{\longrightarrow} 1
\end{equation}

Exploiting the full set of measurement outcomes, i.e. the projective measurement corresponding to the eigenstates of these observables, one 
may obtain more information about the parameters, as quantified by the classical Fisher information, at the expense of having to build a more complex estimator.

The first detection scheme considered is the PNR, by choosing $\hat{N}_j = \hat{c}^\dagger_j \hat{c}_j$ in Eq.~\eqref{var}.
The calculations are indeed much simpler in the basis of total photon number $\hat{N}_{+} = \hat{c}^\dagger_1 \hat{c}_1 + \hat{c}^\dagger_2 \hat{c}_2$ and photon number difference $\hat{N}_{-} = \hat{c}^\dagger_1 \hat{c}_1 - \hat{c}^\dagger_2 \hat{c}_2$. Considering the two-mode Gaussian states, for the phase estimation we have the following derivatives: 
\begin{align}
    &  \frac{\partial \langle \hat{N}_{+} \rangle}{\partial_\varphi}= 0 ,
    \nonumber\\
    & \frac{\partial \langle \hat{N}_{-} \rangle}{\partial_\varphi} = 4 \sqrt{\eta \tau_{\mathrm{in}} (1-\tau_{\mathrm{in}}) \tau_{\mathrm{out}} (1-\tau_{\mathrm{out}})} \aver{n_\alpha}  ,
\end{align}

and for the loss estimation we have: 
\begin{align}
    & \frac{\partial \langle \hat{N}_{+} \rangle}{\partial_\eta} = \tau_{\mathrm{in}} \aver{n_\alpha}  + \aver{n_r}  ,
    \nonumber\\
    & \frac{\partial \langle \hat{N}_{-} \rangle}{\partial_\eta} = (2 \tau_{\mathrm{out}}-1) \left( \tau_{\mathrm{in}} \aver{n_\alpha}  + \aver{n_r}  \right) .
\end{align}

\begin{figure}[t]
\includegraphics[width=1 \columnwidth]{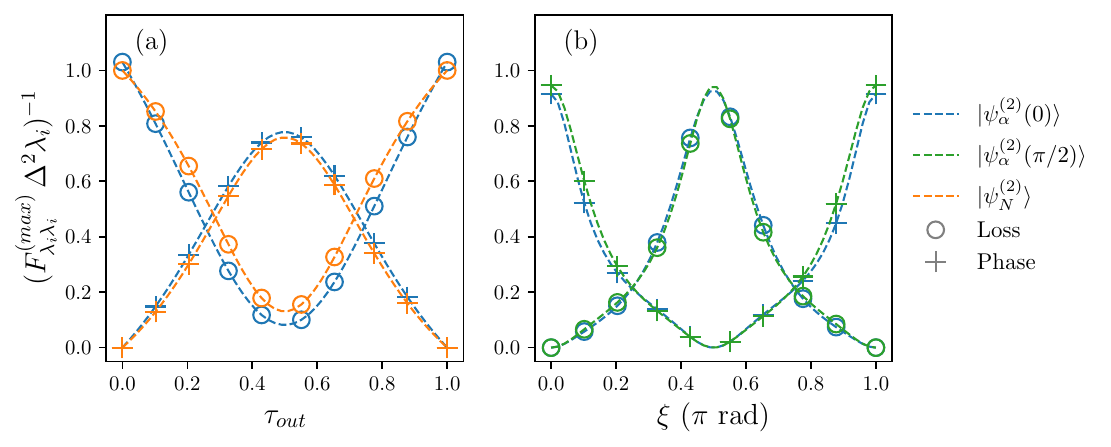}
\caption{
Panels (a) and (b) show the measurement incompatibility for the independent estimation with the figure of merit being $F^{-1}_{\varphi \varphi} / \Delta^2 \varphi$ for the phase estimation (crosses) and $F^{-1}_{\eta \eta} / \Delta^2 \eta$ for the loss estimation (circles), considering PNR in panel (a) and homodyne detection in panel (b).
In both graphs (a) and (b) we have $N=94$ and $\eta=0.1$.
In all the fours plots, we consider the energy distributions $\aver{n_\alpha} =\aver{N} - \sqrt{\aver{N}}$, 
$\aver{n_r} = \sqrt{\aver{N}}$ and $\mu=0$ for Gaussian states. 
\label{variances_vs_bound}
}
\end{figure}

We start our analysis considering the state $| \psi^{(2)}_\alpha (0) \rangle$ with $\tau_{\mathrm{in}} = 1 - 1/\aver{N}^q$ and $\theta_2 =  2 \mu$. To achieve the SQL scaling in the phase estimation, the squeezing and displacement should be oriented in the opposite direction: $\theta_1 - 2 \mu = \pi$, as well as a balanced beamsplitter at the output: $\tau_{\mathrm{out}}=1/2$. With these considerations, the phase variance achieves the ultimate precision limit when $q<p$, as given by:
\begin{equation}
    \left. \Delta^2 \varphi \left(
    \rho^{(2)}_\alpha (0) \right) \right|_{\tau_{\mathrm{out}}=1/2} F^{\text{(max)}}_\varphi \overset{N \rightarrow \infty}{\longrightarrow} 
    \begin{cases} 
     1 & \text{if } q < p, \\
     1 + \frac{1}{2 (1-\eta)} & \text{if } q = p, \\
     \infty & \text{if } q > p.
\end{cases}
\end{equation} 
In contrast, to achieve the SQL scaling in the loss estimation, the squeezing and displacement should be oriented in the same direction: $\theta_1 - 2 \mu = 0$ for any value of $\tau_{\mathrm{out}}$, which is optimal at $\tau_{\mathrm{out}}=1$, maximizing both $\partial_\eta \langle \hat{N}_{+} \rangle$ and $\partial_\eta \langle \hat{N}_{-} \rangle$. In that case, we have:
\begin{equation}
    \left. \Delta^2 \eta \left( 
    \rho^{(2)}_\alpha (0) \right)\right|_{\tau_{\mathrm{out}}=1} F^{\text{(max)}}_\eta \overset{N \rightarrow \infty}{\longrightarrow}
    \begin{cases} 
    1 + \frac{\eta}{1-\eta}
    & \text{if } q < p, \\
    1 + \frac{9 \eta}{10 (1 - \eta)}
    & \text{if } q = p, \\
    1 + \frac{\eta}{2 (1-\eta)}
    & \text{if } q > p.
\end{cases}
\end{equation}
which approaches the ultimate precision limit in the regime of large losses. Therefore, the measurement incompatibility arises from the different output transmissivity $\tau_{\mathrm{out}}$ required for each parameter. For that state,  Fig.~\ref{variances_vs_bound}(a) (in blue) presents the independent estimation variances for phase and loss as functions of the output transmissivity $\tau_{\mathrm{out}}$, clearly illustrating the tradeoff in choosing the measurement: the phase variance is minimized when the loss variance is maximized, and vice versa.

In sequence, we consider the state $| \psi^{(2)}_\alpha (\pi/2) \rangle$.
Now, we cannot set $\tau_{\mathrm{in}}=1$ as considered in the QFI, since it results in $\partial_\varphi \langle \hat{N}_{+} \rangle = \partial_\varphi \langle  \hat{N}_{-} \rangle  = 0$.
Therefore, to achieve the SQL for the phase estimation, we set $\tau_{\mathrm{out}}=1/2$, arriving at the following limit for the phase precision:
\begin{equation}
    \Delta^2 \varphi \left( 
    \rho^{(2)}_\alpha (\pi/2) )\right) F^{\text{(max)}}_\varphi \overset{N \rightarrow \infty}{\longrightarrow} 
    \frac{\eta + 1 - (1-\eta) \tau_{\mathrm{in}}}{(1-\eta) (1-\tau_{\mathrm{in}}) \tau_{\mathrm{in}}},
\end{equation}
which approaches the quantum CRB only when $\tau_{\mathrm{in}} \approx 1$ and $\eta \approx 0$.
In addition, from the previous equation, we conclude that even considering the input transmissivity in the form $\tau_{\mathrm{in}}=1-1/\aver{N}^q$ the phase variance diverges.
However, for loss estimation, this state cannot achieve the SQL scaling.
In order to achieve it, is necessary to choose an energy distribution in the form $\aver{n_\alpha}  = k \aver{N}$ and  $\aver{n_r}  = (1-k) \aver{N}$, in such a way that we achieve the SQL scaling only when $k=0$ (corresponding to a two-mode squeezed state, with $\Delta^2 \eta = 2 \eta (1-\eta)/\aver{N}$) or $k=1$ (corresponding to a coherent state, with $\Delta^2 \eta = \eta/\aver{N}$). 

Finally, Fig.~\ref{variances_vs_bound}(a) (in blue) also shows the phase and loss variances 
for the state $|\psi^{(2)}_N\rangle$ under the photon-counting strategy. 
This state performs similarly to the Gaussian state, 
where the measurement incompatibility is manifested by the need to choose different output transmissivities $\tau_{\text{out}}$ 
for the estimation of each parameter.

Finally, we consider the homodyne detection, by choosing $\hat{Q}_j = e^{i \xi_j} \hat{c}^\dagger_j + h.c.$, which are the quadrature operator introduced in Eq.~(\ref{quadrature_op}). In previous works, homodyne detection has been demonstrated as a promising detection strategy for Gaussian states. Here, we prove that the ultimate quantum precision limit can also be achieved asymptotically for each parameter; however, the measurement incompatibility is manifested in the fact that a different quadrature phase $\xi$ is optimal for each parameter.
This measurement strategy is implemented only for the Gaussian states, since the first moments of $\hat{Q}_j(\xi_j)$ vanish for the state with $N$ photons $| \psi^{(2)}_N \rangle$.

For two-mode Gaussian states, we have the following derivatives for phase estimation: 
\begin{align}
    & \frac{\partial \langle \hat{Q}_1 (\xi)\rangle }{\partial_\varphi} = 2 \sqrt{\eta  \tau_{\mathrm{in}} \tau_{\mathrm{out}}} \cos (\mu -\chi) \aver{n_\alpha}  ,
    \nonumber\\
    & \frac{\partial \langle \hat{Q}_2 (\xi) \rangle}{\partial_\varphi} = 2 \sqrt{\eta  \tau_{\mathrm{in}} (1-\tau_{\mathrm{out}})}  \sin (\mu -\chi ) \aver{n_\alpha}  ,
\end{align}
and for the loss estimation
\begin{align}    
    & \frac{\partial \langle \hat{Q}_1 (\xi) \rangle}{\partial_\eta} = \sqrt{\frac{\tau_{\mathrm{in}} \tau_{\mathrm{out}}}{\eta}}  \sin (\mu -\chi ) \aver{n_\alpha}   ,
    \nonumber\\
    & \frac{\partial \langle \hat{Q}_2 (\xi) \rangle}{\partial_\eta} = \sqrt{\frac{\tau_{\mathrm{in}} (1-\tau_{\mathrm{out}})}{\eta}} \cos (\mu -\chi ) \aver{n_\alpha}   .
\end{align}

Let us start by considering the state $| \psi^{(2)}_\alpha (\pi/2) \rangle$ with phase parameters $\theta_1 - 2 \xi = \theta_2 - 2 \xi = 0, \pi$, and transmissivity $\tau_{\mathrm{in}}=1-1/N^p$.
With this choice, the variance achieves the SQL scaling when $\tau_{\mathrm{out}}=1$, with the following asymptotic expressions:
\begin{equation}
    \Delta^2 \varphi \left(\rho^{(2)}_\alpha (0) \right) F^{\text{(max)}}_\varphi \overset{N \rightarrow \infty}{\longrightarrow} 
    \begin{cases} 
     \sec^2(\mu + \xi) & \text{if } q < p , \\
     \big[ 1 + \frac{\eta}{2 (1-\eta)} \big] \sec^2(\mu + \xi) & \text{if } q = p , \\
     \infty & \text{if } q > p ,
\end{cases}
\end{equation}
and the limit for the loss precision is:
\begin{equation}
    \Delta^2 \eta \left( \rho^{(2)}_\alpha (0)\right) F^{\text{(max)}}_\eta \overset{N \rightarrow \infty}{\longrightarrow}
    \begin{cases} 
    \csc^2(\mu + \xi) & \text{if } q < p, \\
    \big[ 1 + \frac{\eta}{2 (1-\eta)} \big] \csc^2(\mu + \xi) & \text{if } q = p \, . \\
    \infty & \text{if } q > p,
\end{cases}
\end{equation}

We proceed by considering the state $| \psi^{(2)}_\alpha (\pi/2) \rangle$ with $\tau_{\mathrm{in}}=1$.
In this case, the variances achieve the SQL scaling when the output transmissivity is $\tau_{\mathrm{out}}=1$ and the squeezing phase is $\theta - 2 \xi = \pm \pi/2$, resulting in the asymptotic expression for the phase variance:
\begin{equation}
    \Delta^2 \varphi \left(\rho^{(2)}_\alpha (\pi/2)\right) F^{\text{(max)}}_\varphi \overset{N \rightarrow \infty}{\longrightarrow} \sec^2(\mu + \xi) ,
\end{equation}
and for the loss variance:
\begin{equation}
    \Delta^2 \eta \left(\rho^{(2)}_\alpha (\pi/2)\right) F^{\text{(max)}}_\varphi \overset{N \rightarrow \infty}{\longrightarrow} \csc^2(\mu + \xi), 
\end{equation}
Therefore, for both states, the ultimate precision limit for the phase estimation is asymptotically achieved when  $\mu - \xi = 0$, maximizing the phase signal $\partial_\varphi \langle \hat{Q}_1 (\xi) \rangle$, but suppressing the loss signal.
In contrast, for loss estimation, the optimal phase is  $\mu - \xi = \pi/2$,  maximizing the loss signal $\partial_\eta \langle \hat{Q}_1 (\xi) \rangle$, but suppressing the phase signal. 
Additionally, this incompatibility is illustrated in Fig.~\ref{variances_vs_bound}(c), 
which shows each variance as a function of the quadrature phase $\xi$.

%%%%%%%%%%%%%%%%%%%%%%%%%%%%%%%%%%%%%%%%%%%%%%%%%%%%%%%%%%%%%%%%%%%%%%%%%%%%%%%%%%%%%%%%%%%%%%%%%%%%%%%%%%%%%%%%%%%%%%%%%%%%%%%%%%%%%%%%%%%%%%%%%%%%%%%%%%%%%%%%%%%%%%%%%%%%%%%%%%%%%%%%%%%%%%%%%%%%%%%%%%%%%%%%%%%%%%%%

\begin{center}
\myclearpage
\par\end{center}

\chapter{Conclusions}

In Chapter~\ref{chapter:suppression_laws}, we revealed the existence of entire families of suppression laws in the beamsplitter and tritter that cannot be explained by the permutation symmetry principle introduced in Refs.~\cite{Dittel1,Dittel2}. Our approach indicates that similar suppression laws, likewise not captured by permutation symmetry arguments, should arise for arbitrarily large multiports and for arbitrary total numbers of bosons. In principle, the generating-function framework developed here allows one to derive recurrence relations for suppression laws in multiports of any size. It should be emphasized, however, that this approach becomes impractical for sufficiently large interferometers, as it ultimately encounters the same computational hardness associated with the evaluation of matrix permanents.

In Chapter~\ref{chapter:GBS}, we developed a general framework to investigate the interference of single-mode squeezed states, explicitly incorporating the overlap of the internal states of the photons within a generating-function formalism. Using this framework, we examined two applications. The first considers a simplified model with homogeneous partial distinguishability, which is useful for analyzing distinguishability as a source of noise in Gaussian Boson Sampling protocols. The second consists in investigate how the partial  distinguishability degrades multiphoton interference in HOM-like scenarios, explicitly addressing the role of the mutual phases of the internal states using a realistic Gaussian model.

Finally, in Chapter~\ref{chapter:parameter_estimation}, we presented a comprehensive analysis of simultaneous phase and loss estimation in optical interferometry. We showed that probe incompatibility can be overcome either through careful engineering of single-mode non-Gaussian probe states or by exploiting mode entanglement with a lossless reference beam. Our results further provide strong evidence that measurement incompatibility cannot be eliminated, even asymptotically, for states capable of surpassing probe incompatibility. All findings were systematically benchmarked against the corresponding fundamental bounds, and we introduced powerful numerical optimization techniques.

%%%%%%%%%%%%%%%%%%%%%%%%%%%%%%%%%%%%%%%%%%%%%%%%%%%%%%%%%%%%%%%%%%%%%%%%%%%%%%%%%%%%%%%%%%%%%%%%%%%%%%%%%%%%%%%%%%%%%%%%%%%%%%%%%%%%%%%%%%%%%%%%%%%%%%%%%%%%%%%%%%%%%%%%%%%%%%%%%%%%%%%%%%%%%%%%%%%%%%%%%

\appendix

\chapter{\label{detailed_aver_coherent} Detailed calculations for the  distinguishability  as an average over displacements}

In this section we explain in details the derivation of Eqs.~(\ref{gen_func_aux2}), (\ref{Husimi_effective}), and
(\ref{rho_effective}) presented in the main text. To begin, it is convenient to rewrite Eq.~(\ref{gen_func_definition}) in terms of a Gaussian kernel $\mathcal{K}({\bm \alpha},H)$ as follows:
\begin{align}
    G({\bm \eta}) = \int d^2{\bm \alpha} \, Q({\bm \alpha}) \, \mathcal{K}({\bm \alpha},H) 
    ,
    \quad
    \mathcal{K}({\bm \alpha},H) = \frac{\text{exp}\left[- {\bm \alpha}^\dagger \left(H^{-1} - I_M \right) {\bm \alpha} \right]}{\text{det}(H)} .
    \label{kernel_def}
\end{align}

In the following, we analyze the decomposition of the photon-counting operator introduced in Eq.~(\ref{detec_oper_homog}). Since the operator $\hat{N}_l$ has already been explicitly expanded in the internal basis, we may directly apply the relation given in Eq.~(\ref{General_PhotonNumber_Output_1}) by setting $V_{ij}=1$, which results
\begin{align}
    \sum^M_{l=1} \hat{N}_l = {\bf \hat{a}}^\dagger_{\phi_0} \left(I_M - H_0 \right) {\bf \hat{a}}_{\phi_0}  + {\bf \hat{a}}^\dagger_{\phi_\perp} \left(I_M - H_\perp \right) {\bf \hat{a}}_{\phi_\perp} 
    \label{summation_Nl_model}
\end{align}
with $H_0 = U \Lambda U^\dagger$ and $H_\perp = \mathrm{diag}(H_0)$. In Eq.~(\ref{summation_Nl_model}), we introduce the annihilation-operator vectors associated with the common mode and the orthogonal modes, respectively, as ${\bf \hat{a}}_{\phi_0} = (\hat{a}_{1,\phi_0}, \ldots, \hat{a}_{M,\phi_0})$ 
and ${\bf \hat{a}}_{\phi_\perp} = (\hat{a}_{1,\phi^\perp_1}, \ldots, \hat{a}_{M,\phi^\perp_M})$. Substituting Eq.~(\ref{summation_Nl_model}) into the definition of the generating function given in Eq.~(\ref{vacuum_prob}), and making use of the normal-antinormal ordering identity in Eq.~(\ref{normal_antinormal}), we obtain
\begin{align}
    G_{\bf r}(\bm{\eta}, \epsilon)
    &= \left\langle 
    \mathcal{N} \left\{ 
    \exp \left[ {\bf \hat{a}}^\dagger_{\phi_0} \left(I_M - H_0 \right) {\bf \hat{a}}_{\phi_0} \right\} 
     \mathcal{N} \left\{  {\bf \hat{a}}^\dagger_{\phi_\perp} \left(I_M - H_\perp \right) {\bf \hat{a}}_{\phi_\perp}  \right]
    \right\} \right\rangle  
    \nonumber\\
    &= \left\langle 
    \frac{\mathcal{A} \left\{\exp \left[ {\bf \hat{a}}^\dagger_{\phi_0} \left(I_M - H_0 \right) {\bf \hat{a}}_{\phi_0} \right] \right\}}{H_0} 
    \frac{ \mathcal{A} \left\{\exp \left[{\bf \hat{a}}^\dagger_{\phi_\perp} \left(I_M - H_\perp \right) {\bf \hat{a}}_{\phi_\perp}  \right] \right\}}{H_\perp} 
    \right\rangle 
    \nonumber\\
    &= \int d^2 {\bm \alpha}_{\phi_0} 
    \, \mathcal{K}({\bm \alpha}_{\phi_0},H_0) \int d^2 {\bm \alpha}_{\phi_\perp}  
    \, \mathcal{K}({\bm \alpha}_{\phi_\perp},H_\perp)  
    \, Q_{\bf r}({\bm \alpha}_{\phi_0}, {\bm \alpha}_{\phi_\perp} )
    ,
    \label{gen_func_aux2_prove}
\end{align}
which recovers Eq.~(\ref{gen_func_aux2}) of the main text with the effective Husimi function defined as:
\begin{equation}
    Q_{\text{eff}}({\bm \alpha}_{\phi_0}) =   \int d^2 {\bm \alpha}_{\phi_\perp} 
    \, \mathcal{K}({\bm \alpha}_{\phi_\perp},H_\perp)  
    \, Q_{\bf r}({\bm \alpha}_{\phi_0}, {\bm \alpha}_{\phi_\perp} )
    .
    \label{Husimi_effective_def}
\end{equation}

In this way, by substituting the squeezed state, within the low-noise approximation given in Eq.~(\ref{smss_homog}), into Eq.~(\ref{Husimi_effective_def}), we obtain the Husimi function of the effective state,
\begin{align}
    &Q_{\bf r}({\bm \alpha}_{\phi_0}, {\bm \alpha}_{\phi_\perp} )
    \equiv \frac{1}{\pi^{2M}} \big| \langle {\bm \alpha}_{\phi_0}, {\bm \alpha}_{\phi_\perp} | {\bf r} (\epsilon) \rangle \big|^2
    \nonumber\\
    & \quad =  \frac{c_{\bf r}}{\pi^{2M}} \, \exp \left( - {\bm \alpha}^\dagger_{\phi_\perp} {\bm \alpha}_{\phi_\perp} \right) \, \langle {\bm \alpha}_{\phi_0}| 
    \, \prod_{k=1}^M \exp \Big[
    \frac{S_k}{2}(\hat{a}^\dagger_{k, \phi_0})^2  + 
    S_k \sqrt{\epsilon} \, \overline{\alpha}_{k, \phi^\perp_k} \, \hat{a}^\dagger_{k, \phi_0}
    \Big] |0 \rangle \langle 0 | \prod_{k=1}^M \exp \Big[
    ~\text{h.c.}~ \Big] | {\bm \alpha}_{\phi_0} \rangle
\end{align}
and replacing in Eq.~(\ref{Husimi_effective}) we have
\begin{align}
    Q_{\text{eff}}({\bm \alpha}_{\phi_0})
    &=\frac{c_{\bf r}}{\pi^M} \int \frac{d^2 {\bm \alpha}_{\phi_\perp} }{\pi^M} 
    \, \frac{\exp \left( -{\bm \alpha}^\dagger_{\phi_\perp} H_\perp {\bm \alpha}_{\phi_\perp} \right)}{H_\perp}  \, 
    \nonumber\\
    & \times \,\langle {\bm \alpha}_{\phi_0}| 
    \, \prod_{k=1}^M \exp \Big(
    \frac{S_k}{2}(\hat{a}^\dagger_{k, \phi_0})^2  + 
    S_k \sqrt{\epsilon} \, \overline{\alpha}_{k, \phi^\perp_k} \, \hat{a}^\dagger_{k, \phi_0}
    \Big) |0 \rangle \langle 0 | \prod_{k=1}^M \exp \Big(
    ~\text{h.c.}~ \Big) | {\bm \alpha}_{\phi_0} \rangle
    .
    \label{Husimi_effective_prove}
\end{align}
Therefore, we recover Eq.~(\ref{Husimi_effective}) by identifying $\hat{\rho}_\epsilon$ as the operator inside the integral of Eq.~(\ref{Husimi_effective_prove}) with the change of variables to the effective displacement ${\bm \beta}=\sqrt{\epsilon}\,\overline{D}_{\bf r}\,{\bm \alpha}_{\phi_\perp}$.

%----------------------------------------------------------------------------------------------------------------------------------------------------------------------------------------------------%

\chapter{Necessary conditions for probe-compatibility in phase and loss estimation}
\label{append:sufficient_cond}

Here, we provide the prove of the necessary conditions to achieve the upper bound of the probe incompatibility quantifier. During the proof wwe are also reviisiting some previous results developed in other works. That conditions where presented at Eq.~(\ref{Conditions_SimultaneousEstimation}) at the main text.
Given a quantum channel $\Lambda_{\bm \lambda}$ its Kraus representation 
\begin{equation}
\Lambda_{\bm \lambda}(\rho) = \sum_m K_{{\bm \lambda}, m} \rho K_{{\bm \lambda}, m}^\dagger   
\end{equation}
is not unique, and equivalent Kraus representations are connected with each other via a unitary matrix (more generally an isometry)~\cite{NielsenChuang}
\begin{equation}
\label{eq:Kraus_equiv}
    \tilde{K}_{{\bm \lambda}, m} = \sum_{m^\prime} u(\bm \lambda)_{m}^{m^\prime} K_{{\bm \lambda},m^\prime},
\end{equation}
where $u({\bm \lambda})$ is a unitary matrix, which importantly may depend on the estimated parameters.
In what follows we will drop explicit dependence of Kraus operators on $\blambda$ for conciseness.

In the single parameter case, one can show that the maximal achievable QFI for the output state of the channel, optimized over all input probe states, is upper bounded by~\cite{Fujiwara2008,Escher2011,Demkowicz2012}:
\begin{equation}
    \max_{\rho}F\left[\Lambda_{\lambda}(\rho)\right] \leq 4 \min_{\{K_{m}\}} \left\| \sum_m \partial_\lambda K_{m}^\dagger \partial_\lambda K_{m}\right\|,
\end{equation}
where minimization is performed over all equivalent Kraus representations of the channel and $\| \cdot \|$ is the operator norm.\footnote{The inequality becomes in fact equality, if one admits the possibility that the probe system may be entangled with a noiseless ancillary system on which the channel acts trivially.}
For finite-dimensional systems the above minimization may be performed efficiently, and cast in the form of a semidefinite program (SDP)~\cite{Demkowicz2012}.
Nevertheless, the bound is valid even if we do not perform full minimization over all Kraus representations and consider just a certain subclass. 

The above bound has been generalized in~\cite{Albarelli2022} to the multiparameter case, in order to upper bound the weighted sum of QFIs\footnote{\me{Similarly to the single-parameter case, if the probe state is allowed to be entangled with a noiseless ancillary system, this bound is saturable.}}, and hence it may be applied to upper bound the probe-incompatiblity measure 
\eqref{probe_incomp}:
\begin{equation}
\label{eq:probeincompbound}
\max_{\rho} \mathcal{F}[\Lambda_{\bm \lambda}(\rho)] \leq \frac{4}{d} \min_{\{K_{\lambda,m}\}}  \left\| \sum_{j=1}^d \frac{1}{ w_{\lambda_j}} \sum_{m} \partial_{\lambda_j} K_{m}^\dagger \partial_{\lambda_j} K_{m} \right\|, \ w_{\lambda_j} = F_{\lambda_j}^{\mathrm{(max)}},
\end{equation}
where $w_{\lambda_j}$ play the role of the weights.
Thus, if $\max_\rho \mathcal{F}[\Lambda_{\bm \lambda}(\rho)] < 1$ the model suffers from fundamental probe-incompatibility.
In contrast, the bound being equal to $1$ is a necessary condition for probe compatibility in the model~\cite{Albarelli2022}.

The above bound is expressed in terms of operator norms, well suited when we deal with small finite-dimensional spaces.
In the case of large or infinite-dimensional spaces, we usually impose some additional constraints on the states that are allowed in the problem.
In the case of optical interferometry, this usually amounts to a restriction on the mean photon number.
In such scenarios, it is more convenient to rewrite the above bounds, replacing the operator norms with expectation values on the states that are allowed in the problem.
For single-parameter bounds, this approach has been followed in~\cite{Escher2011}.\footnote{In \cite{Escher2011}, the bound includes also a second subtracted term, which, however, is irrelevant, as one may always choose such a Kraus representation, that the second term vanishes, see e.g. the discussion in \cite{KolodynskiPhd2014}, Appendix C.}
One can do the same in a straightforward way in case of multiparameter bound \eqref{eq:probeincompbound} and write:
\begin{equation}
\label{eq:probeincompboundexp}
\max_{\rho \in \mathcal{S}} \mathcal{F}[\Lambda_{\bm \lambda}(\rho)] \leq 
 \frac{4}{d} \min_{\{K_{m}\}} \max_{\rho \in \mathcal{S}}  \Tr \left[ \rho \sum_{j=1}^d \frac{1}{w_{\lambda_j}} \sum_{m} \partial_{\lambda_j} K_{,m}^\dagger \partial_{\lambda_j} K_{m} \right],
\end{equation}
where we restrict the set of input states to some subset $\mathcal{S}$. In sequence we will use this bound to obtain necessary conditions for probe-compatibility in simultaneous phase and loss estimation, expressed in terms of photon-number statistics properties that the state needs to satisfy, as shown in the main text in Eqs.~(\ref{Conditions_SimultaneousEstimation}). 

Following~\cite{Escher2011}, let us consider the following Kraus representation of the single mode channel $\Lambda_{\bm \lambda}$, ${\bm \lambda} =(\varphi, \eta)$
\begin{equation}
    \label{eq:KrausOps_alpha_beta}
K_{ m} = e^{-i \varphi \beta}\sqrt{\frac{(1-\eta)^m}{m!}}e^{i \varphi (\hat{n} - \alpha m)}\eta^{\frac{n}{2}} \hat{b}^m, \quad m=0,1,\dots, 
\end{equation}
where $\hat{n}=\hat{b}^\dagger \hat{b}$ is the photon number operator, while $\alpha$, $\beta$ are free real parameters that are used to obtain different equivalent Kraus representations.\footnote{Compared with \cite{Escher2011}, we introduced additional $\beta$ parameter, which is needed to compensate for the lack of additional subtracted term in the bound we are using compared to the one in \cite{Escher2011}.}
A Kraus operator $K_{m}$ represents an event where $m$ photons are lost from the mode.  

After some algebraic calculations, that make extensive use of commutation properties of annihilation operators, one arrives at: 
\begin{align}
\label{eq:alphaphi}
 \sum_{m=0}^\infty \partial_{\varphi} K_{m}^\dagger \partial_{\varphi}
 K_{m}  & %= \\ \nonumber
 =\left[\eta - \alpha(1-\eta) \right]^2  \hat{n}^2  +  \left[ \eta(1-\eta))(1+\alpha)^2 + 2\beta(\alpha(1-\eta) - \eta)\right] \hat{n} + \beta^2, \\
  \sum_{m=0}^\infty \partial_{\eta} K_{m}^\dagger \partial_{\eta}K_{m}
&=   \frac{1}{4 \eta (1-\eta)} \hat{n}.
\end{align}
We see that when we take the expectation values of the above operators with the state $\rho$, as in~\eqref{eq:probeincompboundexp}, we will obtain formulas that can be written in terms of first and second moments of the photon number operator $\langle \hat{n} \rangle$, $\langle \hat{n}^2 \rangle$, or equivalently first moment $\langle \hat{n} \rangle$ and the variance $\Delta^2 {n} = \langle \hat{n}^2 \rangle - \langle \hat{n} \rangle^2$.

To compute a useful bound, we need to minimize the expression in \eqref{eq:probeincompboundexp} over different Kraus representation, which in our case amount to minimization over $\alpha$ and $\beta$. Note that only part involving derivatives over $\varphi$ \eqref{eq:alphaphi} depends on $\alpha$, $\beta$. The dependence is quadratic and hence we can minimize the expectation value of this term over $\alpha$ and $\beta$ explicitly and obtain (see also \cite{Escher2011}):
\begin{equation}
    \min_{\alpha,\beta} \Tr \left[ \rho \sum_{m=0}^\infty \partial_{\varphi} K_{m}^\dagger \partial_{\varphi}
 K_{m}  \right] =  \frac{\eta \langle{ \hat{n}\rangle} }{1 - \eta + \frac{\langle \hat{n} \rangle}{\Delta^2 n} \eta},
\end{equation}
where the optimal choice of parameters corresponds to 
\begin{equation}
\alpha = \frac{\eta(\Delta^2 n - \langle \hat{n} \rangle )}{\Delta^2 n (1-\eta) + \langle \hat{n} \rangle \eta},\quad \beta = \frac{\langle \hat{n} \rangle^2 \eta}{\Delta^2 n (1-\eta)+ \langle \hat{n} \rangle \eta}.
\end{equation}
The probe-incompatibility bound \eqref{eq:probeincompboundexp}, therefore, reads: 
\begin{equation}
    \max_{\rho} \mathcal{F}[\Lambda_{\blambda}(\rho)] \leq \frac{1}{2} \left( \frac{1}{F_{\varphi}^{\mathrm{(max),(1)}}}  \frac{4 \eta \langle{ \hat{ n} \rangle} }{1 - \eta + \frac{\langle \hat{n} \rangle}{\Delta^2 n} \eta} +  \frac{1}{F_\eta^{\mathrm{(max)}}}   \frac{ \langle \hat{n} \rangle}{\eta (1-\eta)}  \right).  
\end{equation}
If we now substitute the formulas for asymptotically saturable single parameter bounds $F_{\varphi}^{\mathrm{(max)}}$ \eqref{upper_phase} and $F_\eta^{\mathrm{(max)}}$\eqref{upper_loss}, we get:  
\begin{equation}
 \max_{\rho} \mathcal{F}[\Lambda_{\blambda}(\rho)] \leq \frac{1}{2} \frac{\langle \hat{n} \rangle}{N} 
 \left( \frac{1}{1 + \frac{\langle \hat{n} \rangle}{\Delta^2 n} \frac{\eta}{1-\eta}} + 1 \right),
\end{equation}
where $N$ is the maximal number of photons allowed.  We see that, in order to have the upper bound equal to 1 (and hence satisfy necessary condition for probe compatibility), we need to satisfy two conditions:
\begin{equation}
\label{eq:neccondprobe}
    \frac{\langle \hat{n} \rangle}{N} \rightarrow 1, \quad \frac{\langle \hat{n} \rangle}{\Delta^2 n} \rightarrow 0.
\end{equation}

%----------------------------------------------------------------------------------------------------------------------------------------------------------------------------------------------------%

\chapter{\label{appendix:qfi_gaussian} Quantum Fisher Information for the selected Gaussian states}

Following this, we denote the state with strong squeezing as $| \psi^{(1)}_r \rangle$, that is defined as the state in Eq.~\eqref{gaussian_singlemode} with $\aver{n_r} = \aver{N} - \aver{N}^p$ and $\aver{n_\alpha}= \aver{N}^p$. For this state, we have:
\begin{equation}
    \frac{F_{\varphi \varphi} (\rho^{(1)}_r)}{F^{\text{(max)}}_\varphi} \overset{\aver{N} \rightarrow \infty}{\longrightarrow} 1 ,
    \qquad
    \frac{F_{\eta \eta} (\rho^{(1)}_r)}{F^{\text{(max)}}_\eta} \overset{\aver{N} \rightarrow \infty}{\longrightarrow} 0 ,
    \label{qfi_sm_r}
\end{equation}
which achieves the ultimate quantum precision bound for the phase estimation but fails to achieve even the SQL scaling for the loss estimation.

Here, we focus on two-mode Gaussian states that can attain the ultimate precision limit for both parameters. Assuming the regime of strong displacement, we retain only the displacement contribution in Eq.~\eqref{qfi_general_gaussian}.
Under these conditions, from Eqs.~\eqref{cov_twomode}, ~\eqref{displacement_vector},~\eqref{qfi_general_gaussian} and considering $\aver{n_r}  \gg 1$, the phase and loss QFIs for the state $| \psi^{(2)}_\alpha (0) \rangle$ are given, respectively, by the following expressions:
\begin{align}
F_{\varphi \varphi} ( \rho^{(2)}_\alpha (0) )  & \approx \frac{ 4 \eta \tau_{\mathrm{in}} \Big[ 1 + \gamma (1-\eta ) (1-\tau_{\mathrm{in}}) \tau_{\mathrm{in}}  \aver{N}^2_r \Big] \aver{n_\alpha} 
    }{
    (1-\eta ) \Big[ \gamma (1-\eta ) (1-\tau_{\mathrm{in}}) \tau_{\mathrm{in}}  \aver{N}^2_r + 4 \eta  \aver{n_r}  \Big]+1
    } + \nonumber \\ 
& + \frac{8 \eta^2 \tau_{\mathrm{in}} \Big[ 1 -
    \tau_{\mathrm{in}} \cos (\theta_1-2 \mu )+(1-\tau_{\mathrm{in}}) \cos(\theta_2-2 \mu )
    \Big] \aver{n_r}  \aver{n_\alpha} 
    }{ 
    (1-\eta ) \Big[ \gamma (1-\eta ) (1-\tau_{\mathrm{in}}) \tau_{\mathrm{in}}  \aver{N}^2_r + 4 \eta  \aver{n_r}  \Big]+1
    } ,
    \label{qfi_phase_chi0_alpha}
\end{align}
\begin{align}
F_{\eta \eta} ( \rho^{(2)}_\alpha (0) ) & \approx \frac{ \tau_{\mathrm{in}} \Big[ 1 +
    \gamma (1-\eta ) (1-\tau_{\mathrm{in}}) \tau_{\mathrm{in}}  \aver{N}^2_r \Big] \aver{n_\alpha} 
    }{
    \eta (1-\eta ) \Big[ \gamma (1-\eta ) (1-\tau_{\mathrm{in}}) \tau_{\mathrm{in}}   \aver{N}^2_r+4 \eta  \aver{n_r}  \Big]+\eta 
    } + \nonumber \\ 
& + \frac{
    2 \eta \Big[1 +\tau_{\mathrm{in}} \cos (\theta_1-2 \mu )-(1-\tau_{\mathrm{in}}) \cos (\theta_2-2 \mu ) \Big] \aver{n_r}  \aver{n_\alpha} 
    }{
    \eta (1-\eta ) \Big[ \gamma (1-\eta ) (1-\tau_{\mathrm{in}}) \tau_{\mathrm{in}}   \aver{N}^2_r+4 \eta  \aver{n_r}  \Big]+\eta 
    \label{qfi_loss_chi0_alpha}
    } .
\end{align}
where we introduce the coefficient $\gamma = 16 \cos ^2\left(\frac{\theta_1-\theta_2}{2}\right)$. Similarly, the phase and loss QFIs for the state $| \psi^{(2)}_\alpha (\pi/2) \rangle$ are given, respectively, by the following expressions:
\begin{align}
F_{\varphi \varphi} ( \rho^{(2)}_\alpha (\pi/2) )  & \approx \frac{
    4 \eta  \tau_{\mathrm{in}} \Big[ 1 + 4 (1-\eta ) (2 \tau_{\mathrm{in}}-1)^2 \aver{N}^2_r \Big] \aver{n_\alpha} 
    }{
    1 + 4 (1-\eta ) \Big[ 1 -4 (1-\eta ) (1-\tau_{\mathrm{in}}) \tau_{\mathrm{in}}+(1-\eta ) (1-2 \tau_{\mathrm{in}})^2 \aver{n_r}  \Big] \aver{n_r} 
    } + \nonumber \\ 
& + \frac{
    8 \eta  \tau_{\mathrm{in}} \Big[2 -\eta +2 \eta  \sqrt{(1-\tau_{\mathrm{in}}) \tau_{\mathrm{in}}} \sin (\theta -2 \mu )-8 (1-\eta ) (1-\tau_{\mathrm{in}}) \tau_{\mathrm{in}} \Big]  \aver{n_r}  \aver{n_\alpha} 
    }{
    1 + 4 (1-\eta ) \Big[ 1 -4 (1-\eta ) (1-\tau_{\mathrm{in}}) \tau_{\mathrm{in}}+(1-\eta ) (1-2 \tau_{\mathrm{in}})^2 \aver{n_r}  \Big] \aver{n_r} 
    } ,
    \label{qfi_phase_chipi2_alpha}
\end{align}
\begin{align}
F_{\eta \eta} ( \rho^{(2)}_\alpha (\pi/2) ) & \approx \frac{
    \tau_{\mathrm{in}} \Big[ 1-4 \eta  \sqrt{(1-\tau_{\mathrm{in}}) \tau_{\mathrm{in}}} \sin (\theta -2 \mu ) \aver{n_r} \Big] \aver{n_\alpha} 
    }{
    \eta +4 (1-\eta ) \eta  \Big[1-4 (1-\eta ) (1-\tau_{\mathrm{in}}) \tau_{\mathrm{in}}+(1-\eta ) (1-2 \tau_{\mathrm{in}})^2 \aver{n_r} \Big] \aver{n_r} 
    } + \nonumber \\ 
& + \frac{
    4 \tau_{\mathrm{in}} (1-\eta) (1-2 \tau_{\mathrm{in}})^2 \aver{N}^2_r  \aver{n_\alpha} 
    }{
    \eta +4 (1-\eta ) \eta  \Big[1-4 (1-\eta ) (1-\tau_{\mathrm{in}}) \tau_{\mathrm{in}}+(1-\eta ) (1-2 \tau_{\mathrm{in}})^2 \aver{n_r} \Big] \aver{n_r}  
    } .
    \label{qfi_loss_chipi2_alpha}
\end{align}

%%%%%%%%%%%%%%%%%%%%%%%%%%%%%%%%%%%%%%%%%%%%%%%%%%%%%%%%%%%%%%%%%%%%%%%%%%%%%%%%%%%%%%%%%%%%%%%%%%%%%%%%%%%%%%%%%%%%%%%%%%%%%%%%%%%%%%%%%%%%%%%%%%%%%%%%%%%%%%%%%%%%%%%%%%%%%%%%%%%%%%%%%%%%%%%%%%%%%%%%%

\bibliographystyle{unsrt}
\bibliography{tese_Matheus}

\end{document}